\documentclass[aps,pra,10pt,showpacs,amsmath,twocolumn,floatfix,superscriptaddress,eqsecnum,longbibliography]{revtex4-1}
\usepackage[dvipsnames]{xcolor}
\usepackage[colorlinks=true,bookmarks=false,linkcolor=NavyBlue,urlcolor=NavyBlue,citecolor=NavyBlue,breaklinks]{hyperref}
\usepackage{amsmath}
\usepackage{amsfonts}
\usepackage{graphicx}
\usepackage{times}
\usepackage{braket}

\newcommand{\real}{\operatorname{Re}}

\newcommand{\parti}[2]{\frac{\partial #1}{\partial #2}}

\newcommand{\intall}{\int_{-\infty}^{\infty}}

\newcommand{\sinc}{\operatorname{sinc}}

\newcommand{\abs}[1]{\left|#1\right|}
\newcommand{\bk}[1]{\left(#1\right)}
\newcommand{\Bk}[1]{\left[#1\right]}

\newcommand{\trace}{\operatorname{tr}}

\newcommand{\expect}{\mathbb E}

\begin{document}

\title{Quantum Theory of Superresolution for Two Incoherent Optical
  Point Sources}

\author{Mankei Tsang}
\email{mankei@nus.edu.sg}
\affiliation{Department of Electrical and Computer Engineering,
  National University of Singapore, 4 Engineering Drive 3, Singapore
  117583}

\affiliation{Department of Physics, National University of Singapore,
  2 Science Drive 3, Singapore 117551}

\author{Ranjith Nair}
\affiliation{Department of Electrical and Computer Engineering,
  National University of Singapore, 4 Engineering Drive 3, Singapore
  117583}

\author{Xiao-Ming Lu}
\affiliation{Department of Electrical and Computer Engineering,
  National University of Singapore, 4 Engineering Drive 3, Singapore
  117583}

\date{\today}

\begin{abstract}
  Rayleigh's criterion for resolving two incoherent point sources has
  been the most influential measure of optical imaging resolution for
  over a century. In the context of statistical image processing,
  violation of the criterion is especially detrimental to the
  estimation of the separation between the sources, and modern
  farfield superresolution techniques rely on suppressing the emission
  of close sources to enhance the localization precision.  Using
  quantum optics, quantum metrology, and statistical analysis, here we
  show that, even if two close incoherent sources emit simultaneously,
  measurements with linear optics and photon counting can estimate
  their separation from the far field almost as precisely as
  conventional methods do for isolated sources, rendering Rayleigh's
  criterion irrelevant to the problem. Our results demonstrate that
  superresolution can be achieved not only for fluorophores but also
  for stars.
\end{abstract}

\maketitle

\section{Introduction}
Rayleigh's criterion for resolving two incoherent point sources,
requiring them to be separated at least by a diffraction-limited spot
size on the image plane \cite{rayleigh,born_wolf}, has been the most
influential measure of optical imaging resolution for over a
century. More recently, insights from quantum optics \cite{hell94} and
statistics \cite{betzig95} have led to revolutions in farfield
superresolution techniques \cite{hell,betzig,moerner} beyond his
criterion. The techniques proposed in Refs.~\cite{hell94,betzig95}
rely on locating a point source when no other nearby sources are
radiating in the same optical mode. While such techniques have
achieved spectacular success in microscopy, they require sophisticated
control of the emission of special fluorophores and are irrelevant to
astronomy and remote sensing.

For two sources with overlapping radiations on the image plane,
studies have found that signal processing of the imaging data can
still determine their locations, although the precision in the
presence of photon shot noise quickly deteriorates when Rayleigh's
criterion is violated \cite{bettens,vanaert,ram}. The precision
degradation is mandated by the Cram\'er-Rao lower error bound
\cite{vantrees}, suggesting that the degradation is fundamental to
direct imaging.  Given such prior work, conventional wisdom thus
suggests that the positions of two incoherent sources should become
harder to estimate when their radiations overlap, a statistical
phenomenon we call \emph{Rayleigh's curse}.

Since photon shot noise is now the dominant noise source in
fluorescence microscopy \cite{ram,pawley} as well as stellar imaging
\cite{zmuidzinas03,labeyrie,howell06,huber}, it is timely to inquire
whether a quantum treatment can lead to new insights. Here we attack
the problem from the perpsective of quantum metrology, a branch of
quantum information theory relevant to sensing and imaging
\cite{helstrom,glm2011}.  To be specific, we derive the fundamental
quantum limit to the precision of locating two weak thermal optical
point sources in the form of the quantum Cram\'er-Rao bound (QCRB)
proposed by Helstrom \cite{helstrom}.  Surprisingly, we find that the
QCRB maintains a fairly constant value for any separation and shows no
sign of Rayleigh's curse. This behavior is in stark contrast to the
QCRB for in-phase coherent sources, in which case Rayleigh's curse is
fundamental \cite{localization}.

It is known mathematically that there exists a measurement scheme to
attain the QCRB for one parameter asymptotically
\cite{hayashi05,fujiwara2006}. For a more concrete experimental
implementation, here we propose the method of SPAtial-mode
DEmultiplexing (SPADE). We show that SPADE can ideally estimate the
separation between the two sources with a quantum-optimal Fisher
information, and we also propose linear optical system designs that
can implement the measurement. Direct imaging is poor at localization
of two close sources precisely because it estimates their separation
poorly, and SPADE is able to overcome this problem and Rayleigh's
curse via further linear optical processing before photon counting.

The subject of quantum imaging has been extensively studied; see
Appendix~\ref{review} for a literature review. Most prior proposals
rely on nonclassical sources or multiphoton coincidence measurements,
however, making them difficult and inefficient to use in
practice. Incoherent sources, such as fluorophores and stars, are of
course much more common, and linear optical methods to enhance the
localization precision for close incoherent sources will be of
monumental interest to both localization microscopy
\cite{betzig95,betzig,moerner} and astrometry
\cite{howell06,huber}. The most relevant prior work remains the
pioneering studies by Helstrom on thermal sources \cite{helstrom}, yet
he studied two sources only in the context of binary hypothesis
testing and assumed a given separation in the two-source hypothesis
\cite{helstrom73b}.  As the separation is usually unknown and needs to
be estimated in the first place
\cite{betzig95,betzig,moerner,bettens,vanaert,ram,howell06,huber}, our
parameter-estimation framework should be more useful.

\section{\label{optics}Quantum optics for weak thermal sources}
To illustrate the essential physics, we follow Lord Rayleigh's lead
\cite{rayleigh} and assume quasi-monochromatic scalar paraxial waves
and one spatial dimension on the object and image planes. Within each
short coherence time interval for a thermal source at an optical
frequency, it is standard
\cite{goodman_stat,mandel,mandel59,labeyrie,zmuidzinas03,gottesman,stellar}
to assume that the average photon number $\epsilon$ arriving on the
image plane is much smaller than $1$, and useful information is
obtained only after many photons have been measured over many such
intervals. This means that the quantum density operator for the
optical fields on the image plane in each coherence time interval can
be well approximated as
\begin{align}
\rho = (1-\epsilon)\rho_0 +\epsilon \rho_1 + O(\epsilon^2),
\label{rho}
\end{align}
where $\rho_0 = \ket{\textrm{vac}}\bra{\textrm{vac}}$ is the
zero-photon state, $\rho_1$ is a one-photon state, and $O(\epsilon^2)$
denotes terms on the order of $\epsilon^2$; see Appendix~\ref{state}
for a detailed derivation.  For the rest of the paper, we neglect the
$O(\epsilon^2)$ terms and use the $\approx$ sign to denote the
first-order approximations. Similar approximations were also used
earlier to study stellar interferometry \cite{gottesman,stellar}.

A connection with classical statistical optics can be made by
observing that $\rho_1$ is related to the mutual coherence of the
optical fields with respect to the Sudarshan-Glauber distribution. As
shown in Appendix~\ref{state}, the one-photon state for two incoherent
point sources and a diffraction-limited imaging system can be taken as
\begin{align}
\rho_1 &\approx \frac{1}{2}
\bk{\ket{\psi_1}\bra{\psi_1}+\ket{\psi_2}\bra{\psi_2}},
\label{rho1}
\\
\ket{\psi_s} &= \intall dx \psi_s(x)\ket{x}, \quad s = 1,2,
\label{psis}
\end{align}
where $x$ is the image-plane coordinate normalized with respect to the
magnification factor of the imaging system \cite{goodman},
$\ket{x} = a^\dagger(x)\ket{\textrm{vac}}$ is the photon image-plane position
eigenket defined with respect to annihilation and creation operators
that obey $[a(x),a^\dagger(x')] = \delta(x-x')$
\cite{yuen_shapiro1,shapiro09}, and $\psi_s(x)$ is the image-plane
wavefunction from each source.

\begin{figure}[htbp!]
\includegraphics[width=0.45\textwidth]{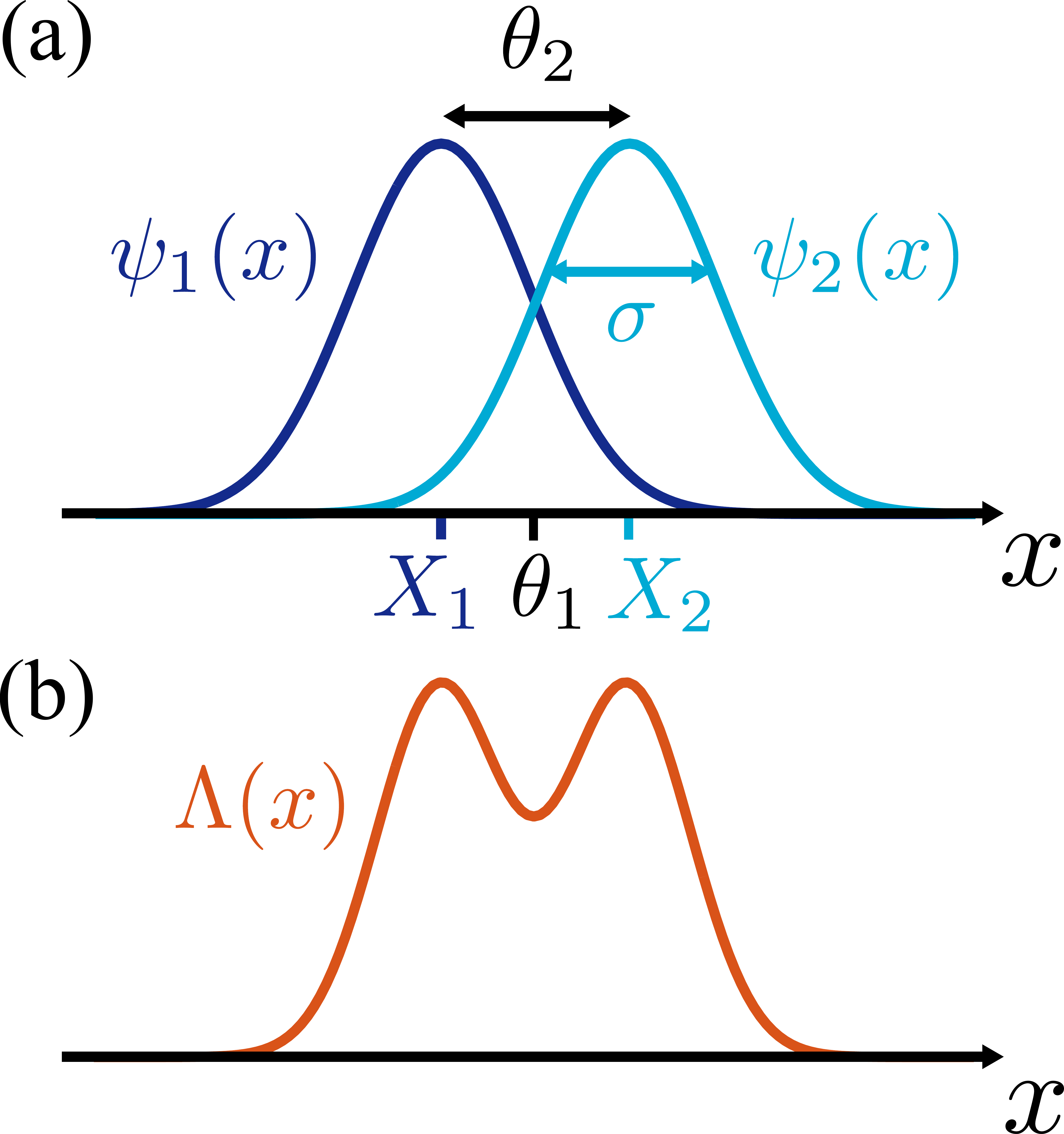}
\caption{\label{psf}(a) Two photonic wavefunctions on the
  image plane, each coming from a point source. $X_1$ and $X_2$ are
  the point-source positions, $\theta_1$ is the centroid, $\theta_2$
  is the separation, and $\sigma$ is the width of the point-spread
  function. (b) If photon counting is performed on the image plane,
  the statistics are Poisson with a mean intensity proportional to
  $\Lambda(x) = [|\psi_1(x)|^2+|\psi_2(x)|^2]/2$.}
\end{figure}

We can reproduce the standard Poisson model of direct image-plane
photon counting
\cite{ram,ober,deschout,chao16,pawley,labeyrie,zmuidzinas03} by
considering the $1-\epsilon$ probability of no photon count and the
$\epsilon \ll 1$ probability of measuring a photon. If a photon is
detected, the probability density of the photon position $x$ is
\begin{align}
\Lambda(x) &= \frac{1}{2}
\bk{\abs{\braket{x|\psi_1}}^2+\abs{\braket{x|\psi_2}}^2}
\nonumber\\
&= \frac{1}{2}\Bk{\abs{\psi_1(x)}^2+\abs{\psi_2(x)}^2}.
\label{intensity}
\end{align}
With $\epsilon \ll 1$, the photon count at each pixel with width $dx$
can be approximated as Poisson with a mean given by
$\epsilon\Lambda(x) dx$.  The total photon count over $M$ coherence
time intervals then remains approximately Poisson with a mean
$M\epsilon \Lambda(x) dx = N\Lambda(x) dx$, where $N \equiv M\epsilon$
is the average photon number collected over the $M$ intervals and
$\Lambda(x)$ becomes the mean intensity profile.  To illustrate,
Fig.~\ref{psf} depicts the wavefunctions and the mean intensity for a
typical imaging system.  Note the crucial point that
$\braket{\psi_1|\psi_2} = \intall dx \psi_{1}^*(x) \psi_{2}(x) \neq
0$, and the spatial modes excited by the two sources are in general
not orthogonal, especially when Rayleigh's criterion is violated.
This overlap underlies all the physical and mathematical difficulties
with the resolution problem, as it implies on a fundamental level that
the two modes cannot be separated for independent measurements.

\section{\label{sec_crb}Classical and quantum Cram\'er-Rao bounds}
To investigate the impact of measurement noise on parameter
estimation, suppose that $\rho$ depends on a set of unknown parameters
denoted by $\{\theta_\mu; \mu = 1,2,\dots\}$ \cite{implicit}, and a
quantum measurement is made on the image plane over the $M$ intervals
to estimate $\theta$. Any quantum measurement can be mathematically
described by a positive operator-valued measure (POVM) $E(\mathcal Y)$
\cite{helstrom}, such that the probability distribution of measurement
outcome $\mathcal Y$ is
$P(\mathcal Y) = \trace E(\mathcal Y)\rho^{\otimes M}$, with $\trace$
denoting the operator trace and $\rho^{\otimes M}$ denoting a tensor
product of $M$ density operators.  Let $\check\theta_\mu(\mathcal Y)$
be an estimator and
\begin{align}
\Sigma_{\mu\nu} \equiv \int d\mathcal Y P(\mathcal Y)
\Bk{\check\theta_\mu(\mathcal Y)-\theta_\mu}
\Bk{\check\theta_\nu(\mathcal Y)-\theta_\nu}
\end{align}
be the error covariance matrix. For any unbiased estimator, the
Cram\'er-Rao bound is given by
\begin{align}
\Sigma_{\mu\mu} \ge \bk{\mathcal J^{-1}}_{\mu\mu},
\end{align}
where 
\begin{align}
\mathcal J_{\mu\nu} &\equiv \int d\mathcal Y 
\frac{1}{P(\mathcal Y)}\parti{P(\mathcal Y)}{\theta_\mu}
\parti{P(\mathcal Y)}{\theta_\nu}
\end{align}
is the Fisher information matrix with respect to $P(\mathcal Y)$
\cite{vantrees}.

For the Poisson model of direct imaging
\cite{ram,ober,deschout,chao16},
\begin{align}
\mathcal J_{\mu\nu}^{(\textrm{direct})} =
N \intall dx \frac{1}{\Lambda(x)}
\parti{\Lambda(x)}{\theta_\mu}\parti{\Lambda(x)}{\theta_\nu}.
\label{Jdirect}
\end{align}
Alternatively, the same result can be derived without the Poisson
approximation by considering the one-photon distribution given by
Eq.~(\ref{intensity}) and no multiphoton coincidence.  As the
Cram\'er-Rao bound is asymptotically achievable \cite{vantrees}, the
Fisher information has become the standard precision measure in modern
fluorescence microscopy \cite{ram,ober,deschout,chao16} as well as
astronomy \cite{lindegren78,king83,zmuidzinas03,huber}.

Direct imaging, though standard, is but one of the infinite
measurement methods permitted by quantum mechanics. The ultimate
performance of any quantum measurement and any unbiased estimator can
be quantified using the quantum Cram\'er-Rao bound
\begin{align}
\Sigma_{\mu\mu} \ge \bk{\mathcal J^{-1}}_{\mu\mu} \ge 
\bk{\mathcal K^{-1}}_{\mu\mu},
\end{align}
where $\mathcal K$ is the quantum Fisher
information matrix in terms of $\rho^{\otimes M}$ \cite{helstrom}. To compute
$\mathcal K$ analytically, we assume a spatially invariant imaging
system with $\psi_s(x) =\psi(x-X_s)$, where $\psi(x)$ is the
point-spread function of the imaging system and $X_s$ is the unknown
position of each source \cite{goodman}. Both
$\mathcal J^{(\textrm{direct})}$ and $\mathcal K$ turn out to be
diagonal if we redefine the parameters of interest as the centroid
\begin{align}
\theta_1 &= \frac{X_1+X_2}{2}
\end{align}
and the separation 
\begin{align}
\theta_2 = X_2- X_1,
\end{align}
as depicted in Fig.~\ref{psf}.  We also assume, with little loss of
generality, that the point-spread function has a constant
$x$-independent phase, which can be easily implemented by a two-lens
system \cite{goodman}. The phase is then irrelevant to $\rho_1$ in
Eq.~(\ref{rho1}) and $\psi(x)$ can be taken as real.

The computation of $\mathcal K$ is described in
Appendix~\ref{metrology}; the result is
\begin{align}
\mathcal K_{11} &\approx 4N(\Delta k^2-\gamma^2),
&
\mathcal K_{22} &\approx N\Delta k^2,
\label{K}
\end{align}
with $\mathcal K_{12} = \mathcal K_{21} \approx 0$, where
\begin{align}
\Delta k^2 \equiv \intall dx \Bk{\parti{\psi(x)}{x}}^2
\label{dk2}
\end{align}
is the spatial-frequency variance of the real point-spread function
set by the diffraction limit and
\begin{align}
\gamma \equiv 
\intall dx \parti{\psi(x)}{x}\psi(x-\theta_2)
\label{gamma}
\end{align}
is a parameter that depends on $\theta_2$.  The prefactor $N$
indicates a shot-noise scaling with respect to the average photon
number, as expected from classical sources
\cite{glm2011,localization}.  For $\theta_2 \to \infty$,
$\gamma^2 \to 0$, and we recover the standard shot-noise limit to the
localization of isolated sources.

\begin{figure}[htbp!]
\includegraphics[width=0.45\textwidth]{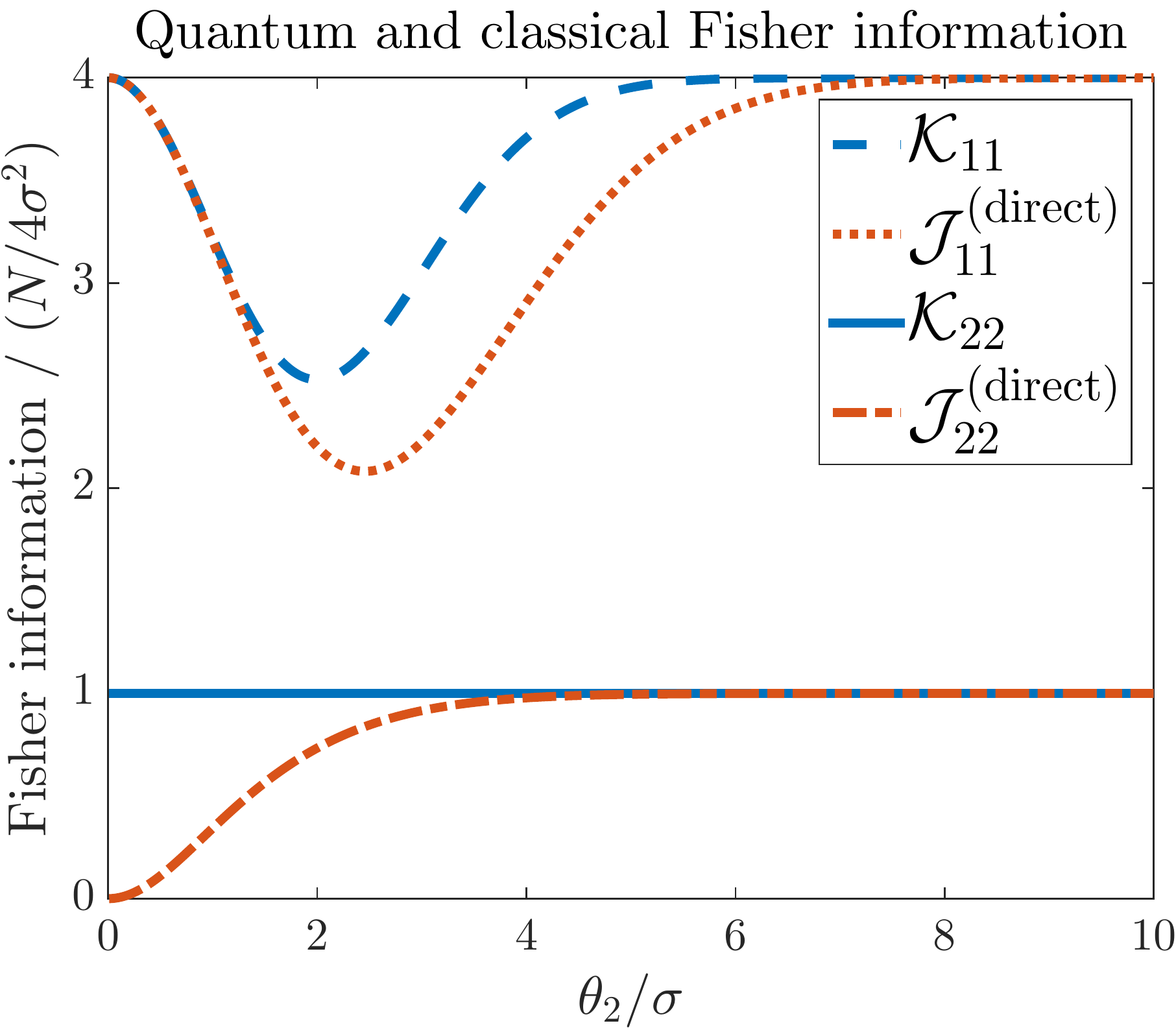}
\caption{\label{fisher}Plots of Fisher information versus the
  separation for a Gaussian point-spread function.  $\mathcal K_{11}$
  and $\mathcal K_{22}$ are the quantum values for the estimation of
  the centroid $\theta_1 = (X_1+X_2)/2$ and the separation
  $\theta_2 = X_2- X_1$, respectively, while
  $\mathcal J_{11}^{(\rm direct)}$ and
  $\mathcal J_{22}^{(\rm direct)}$ are the corresponding classical
  values for direct imaging. The horizontal axis is normalized with
  respect to the point-spread function width $\sigma$, while the
  vertical axis is normalized with respect to $N/(4\sigma^2)$, the
  value of $\mathcal K_{22}$.}
\end{figure}

To compare the quantum Fisher information with the classical
information for direct imaging, Fig.~\ref{fisher} plots the diagonal
elements of $\mathcal J^{(\rm direct)}$ and $\mathcal K$, assuming a
Gaussian point-spread function \cite{pawley}
$\psi(x) = (2\pi\sigma^2)^{-1/4}\exp[-x^2/(4\sigma^2)]$, where
$\sigma = 1/(2\Delta k) = \lambda/(2\pi\textrm{NA})$, $\lambda$ is the
free-space wavelength, and $\textrm{NA}$ is the effective numerical
aperture.  The constant $\mathcal K_{22}$ in particular becomes
\begin{align}
  \mathcal K_{22} &\approx \frac{N}{4\sigma^2}.
\label{K22_gauss}
\end{align}
The Gaussian case is representative and the same qualitative behaviors
can be observed for other common point-spread functions.  For the
centroid, both the classical and quantum information is within a
factor of $2$ of the standard limit $N/\sigma^2$.
$\mathcal J_{11}^{(\rm direct)} \le \mathcal K_{11}$ as it should, but
the small gap between the two means that there is little room for
improvement.

The difference between the separation information quantities
$\mathcal K_{22}$ and $\mathcal J_{22}^{(\rm direct)}$ in
Fig.~\ref{fisher} is much more dramatic.  Both quantities approach the
same limit $N/(4\sigma^2)$ as $\theta_2 \to\infty$, implying that
direct imaging is quantum-optimal for well-separated sources. For
$\theta_2/\sigma \to 0$, however, the classical information
$\mathcal J_{22}^{(\rm direct)}$ decreases to zero. This means that
direct imaging is progressively worse at estimating the separation for
closer sources, to the point that the information vanishes and the
Cram\'er-Rao bound diverges at $\theta_2 = 0$. We call this divergent
behavior due to overlapping wavefunctions Rayleigh's curse, as it
implies a severe penalty on the localization precision when the
intensity profiles overlap significantly and Rayleigh's criterion is
violated for a given $N$.

\begin{figure}[htbp!]
\includegraphics[width=0.45\textwidth]{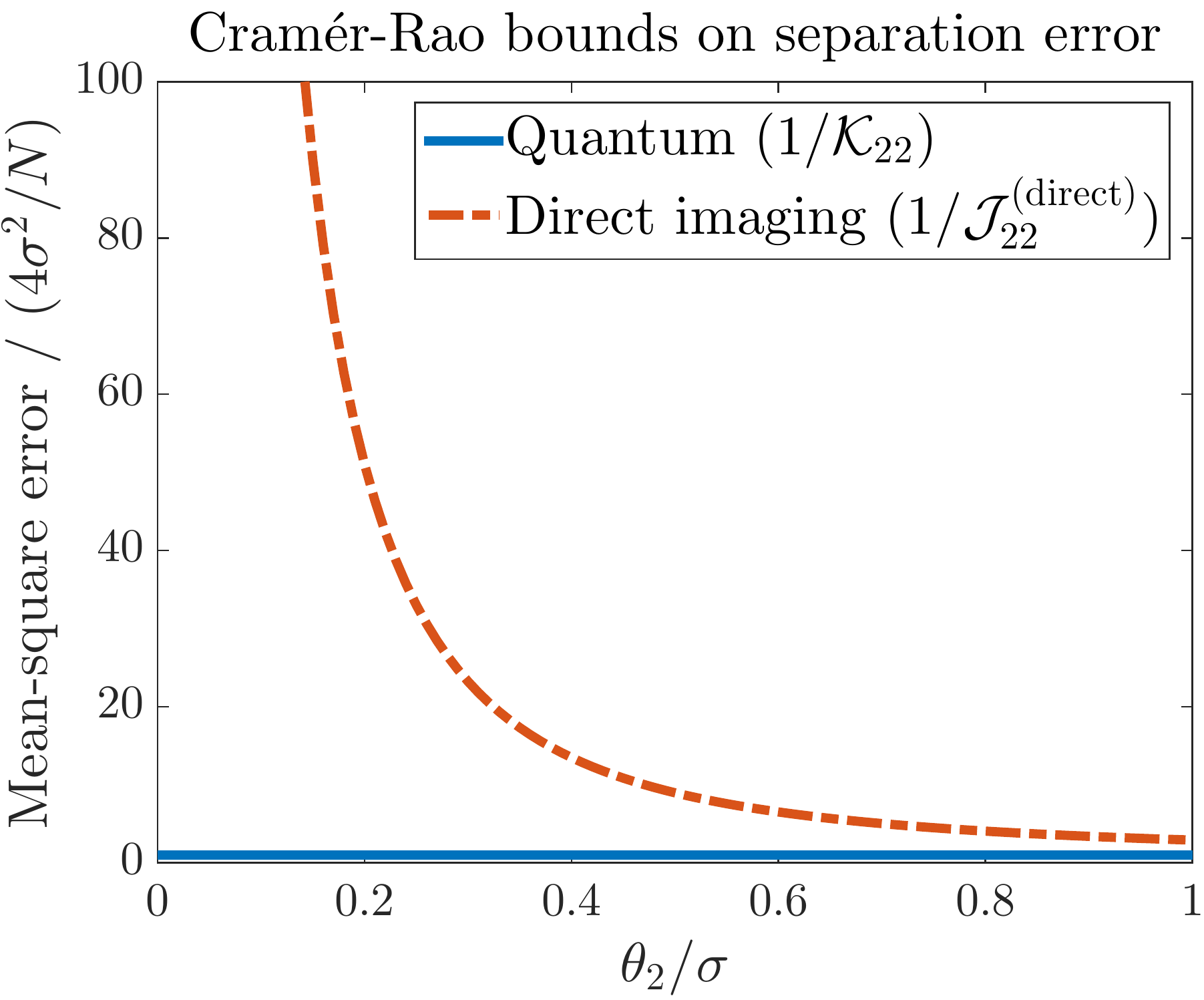}
\caption{\label{crb}The quantum Cram\'er-Rao bound
  ($1/\mathcal K_{22}$) and the classical bound for direct imaging
  ($1/\mathcal J_{22}^{(\rm direct)}$) on the error of separation
  estimation. The bounds are normalized with respect to the quantum
  value $4\sigma^2/N$. Rayleigh's curse refers to the divergence of
  the classical bound when $\theta_2 \lesssim \sigma$, as discovered
  by Refs.~\cite{bettens,vanaert,ram}.}
\end{figure}

Direct imaging suffers from Rayleigh's curse for any point-spread
function, as $\partial \Lambda(x)/\partial \theta_2$ vanishes at
$\theta_2 = 0$ while $\Lambda(x)$ remains nonzero in regions of $x$
where the derivative vanishes, causing
$\mathcal J_{22}^{(\rm direct)}$ to vanish via Eq.~(\ref{Jdirect}).
This is the reason why the Cram\'er-Rao bounds derived in
Refs.~\cite{bettens,vanaert,ram} for separation estimation all diverge
when Rayleigh's criterion is violated.  Remarkably, the quantum
information $\mathcal K_{22}$ in Eq.~(\ref{K}) stays constant
regardless of the separation. If the centroid $\theta_1$ is known,
there exists a POVM with error $\Sigma_{22}$ asymptotically attaining
the single-parameter QCRB \cite{hayashi05,fujiwara2006}, viz.,
\begin{align}
\Sigma_{22} \to \frac{1}{\mathcal K_{22}}
\approx \frac{1}{N\Delta k^2},
\quad
N \to \infty.
\end{align}
This means that Rayleigh's curse can be avoided for separation
estimation, and considerable improvements can be obtained, if the
optimal quantum measurement can be implemented.  To expound the issue,
Fig.~\ref{crb} plots the quantum and classical Cram\'er-Rao bounds
$1/\mathcal K_{22}$ and $1/\mathcal J_{22}^{(\rm direct)}$,
demonstrating more dramatically the divergent error in the classical
case and the substantial room for improvement offered by quantum
mechanics.

\section{\label{sec_spade}Spatial-mode demultiplexing (SPADE)}
Instead of measuring the position of each photon in the direct imaging
method, we propose a discrimination in terms of the Hermite-Gaussian
spatial modes \cite{yariv} to estimate the separation. 
Consider the basis $\{\ket{\phi_q}; q = 0,1,\dots\}$ with
eigenkets given by
\begin{align}
\ket{\phi_q} &= \intall dx \phi_q(x)\ket{x},
\quad
q = 0,1,\dots
\\
\phi_q(x) &= \bk{\frac{1}{2\pi\sigma^2}}^{1/4}\frac{1}{\sqrt{2^q q!}}
H_q\bk{\frac{x}{\sqrt{2}\sigma}}\exp\bk{-\frac{x^2}{4\sigma^2}},
\end{align}
where $H_q$ is the Hermite polynomial \cite{yariv}. The POVM for each
coherence time interval can be expressed as projections
\begin{align}
E_0 &= \ket{\textrm{vac}}\bra{\textrm{vac}},
&
E_1(q) &= \ket{\phi_q}\bra{\phi_q}.
\label{POVM_HG}
\end{align}
Conditioned on a detection event, the probability of detecting the
photon in the $q$th mode becomes
\begin{align}
P_1(q) &\approx \frac{1}{2}
\bk{\abs{\braket{\phi_q|\psi_1}}^2+\abs{\braket{\phi_q|\psi_2}}^2}.
\label{P1q}
\end{align}
Similar to direct imaging, $\epsilon \ll 1$ implies that, over $M$
intervals, the total photon count $m_q$ in each Hermite-Gaussian mode
can be approximated as Poisson with a mean given by $N P_1(q)$.

To proceed further, we assume that the centroid $\theta_1$ is known,
and only $\theta_2$ is to be estimated.  Since centroid estimation
using direct imaging is relatively insensitive to the separation, the
assumption of an accurately known centroid is not difficult to
satisfy; see Appendix~\ref{misalignment} for a detailed
discussion. Under this assumption, we can assume $\theta_1 = 0$
without loss of generality, and the wavefunctions become
$\psi_1(x) = \psi\bk{x+\theta_2/2}$, and
$\psi_2(x) = \psi\bk{x-\theta_2/2}$.  For simple analytic results, we
further assume that the point-spread function is Gaussian.  The
overlap factors in Eq.~(\ref{P1q}) can then be evaluated by
recognizing that $\ket{\phi_q}$ is mathematically equivalent to an
energy eigenstate of a harmonic oscillator (in the configuration space
of the photon), and $\ket{\psi_1}$ and $\ket{\psi_2}$ are equivalent
to configuration-space coherent states with displacements
$\pm\theta_2/(4\sigma)$. The result is
\begin{align}
P_1(q) &\approx \abs{\braket{\phi_q|\psi_1}}^2 = 
\abs{\braket{\phi_q|\psi_2}}^2 = 
\exp\bk{-Q}\frac{Q^q}{q!},
\nonumber\\
Q &\equiv \frac{\theta_2^2}{16\sigma^2}.
\label{Q}
\end{align}
This formula is valid even if the two sources have unequal intensities
and $\rho_1$ is any mixture of $\ket{\psi_1}\bra{\psi_1}$ and
$\ket{\psi_2}\bra{\psi_2}$.  The classical Fisher information for the
Hermite-Gaussian-basis measurement over $M$ intervals becomes
\begin{align}
\mathcal J_{22}^{(\rm HG)} &\approx 
N \sum_{q=0}^\infty P_1(q)\Bk{\parti{}{\theta_2} \ln P_1(q)}^2
\approx \frac{N}{4\sigma^2},
\label{J22_HG}
\end{align}
which is equal to the quantum information given by
Eq.~(\ref{K22_gauss}) and also free of Rayleigh's curse.

\begin{figure}[htbp!]
\includegraphics[width=0.45\textwidth]{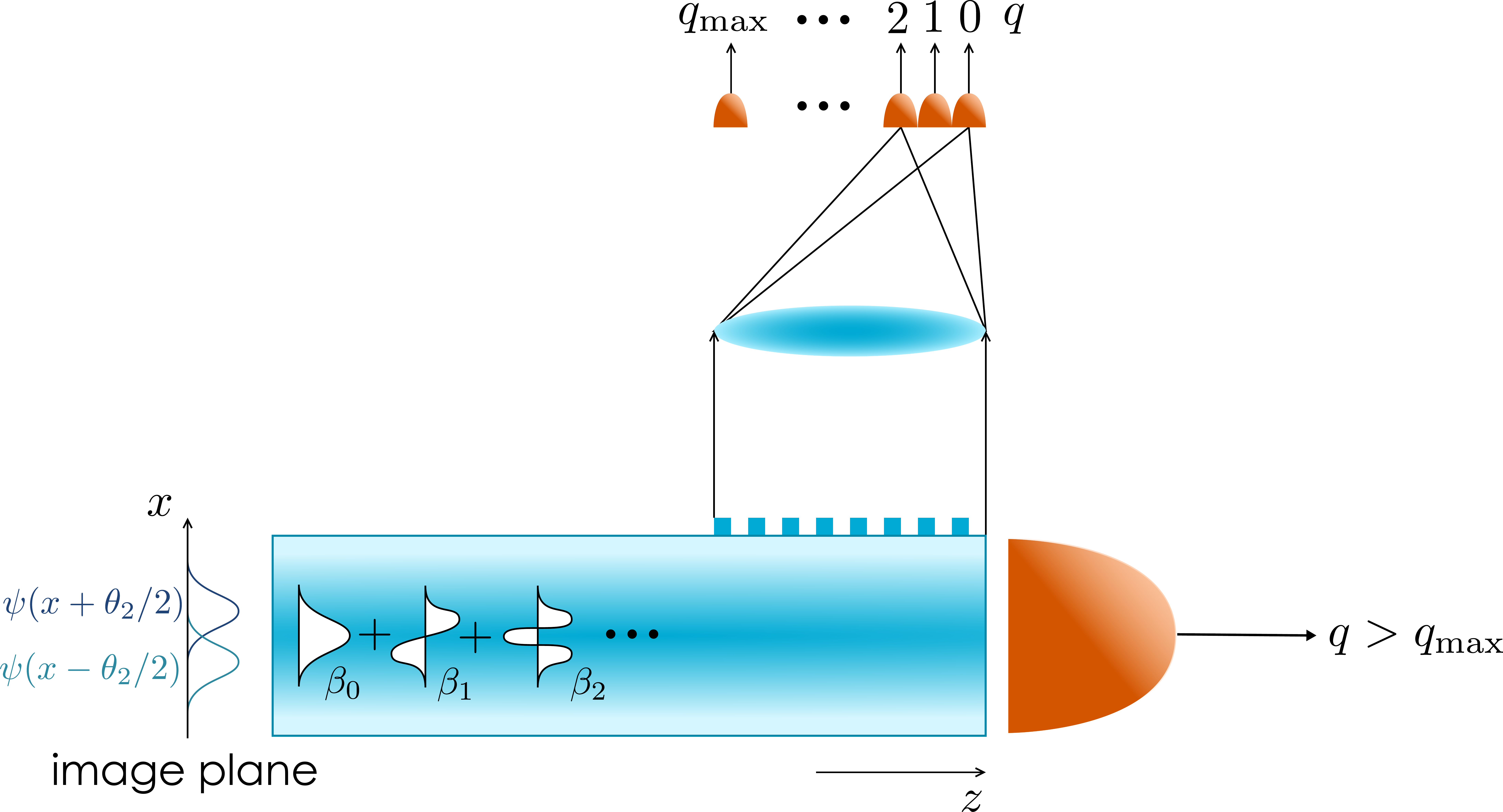}
\caption{\label{mms} A multimode-waveguide SPADE with a grating
  output coupler and farfield photon counting. The photon counter at
  the end of the multimode waveguide captures any remaining photon in
  the higher-order or leaky modes.}
\end{figure}

To measure in the Hermite-Gaussian basis, one needs to demultiplex the
image-plane field in terms of the desired spatial modes before
determining the outcome based on the mode in which the photon is
detected.  To do so with a high information-extraction efficiency, one
should perform a one-to-one conversion of the Hermite-Gaussian modes
into modes in a more accessible degree of freedom with minimal loss
and measurements that capture as many photons as possible.  For
example, we can take advantage of the fact that the Hermite-Gaussian
modes are waveguide modes of a quadratic-index waveguide \cite{yariv}.
Suppose that we couple the image-plane optical field into such a
highly multimode waveguide centered at the centroid position, as shown
in Fig.~\ref{mms}. Each mode with index $q$ acquires a different
propagation constant $\beta_q$ along the longitudinal direction $z$.
If a grating coupler \cite{yariv_yeh} with spatial frequency $\kappa$
is then used to couple all the modes into free space, each mode will
be coupled to a plane wave with a different spatial frequency
$\beta_q - \kappa$ along the $z$ direction in free space, and a
Fourier-transform lens can be used to focus the different plane waves
onto different spots of a photon-counting array in the far field.

\begin{figure}[htbp!]
\includegraphics[width=0.45\textwidth]{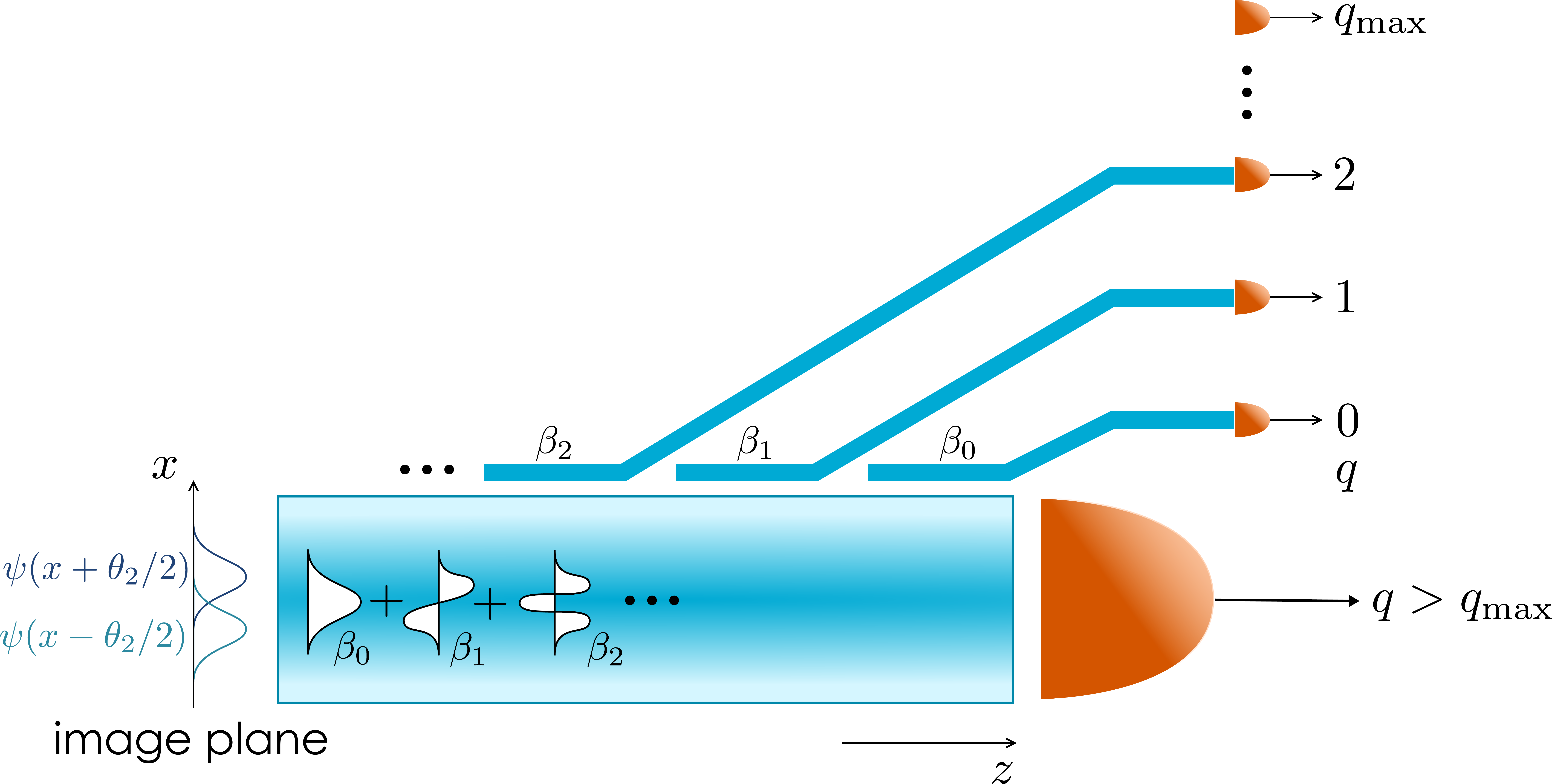}
\caption{\label{mms2} An alternative design, with evanescent coupling
  to single-mode waveguides with different propagation constants for
  phase matching. The photon counter at the end of the multimode
  waveguide captures any remaining photon in the higher-order or leaky
  modes.}
\end{figure}

An alternative is to use evanescent coupling with different
single-mode waveguides \cite{sorin}, as depicted in
Fig.~\ref{mms2}. If each single-mode waveguide is fabricated to have a
propagation constant equal to a different value of $\beta_q$, the
phase-matching condition will cause each mode in the multimode
waveguide to be coupled to a specific fiber.

Given these physical setups, we can now explain the operation of SPADE
in a more intuitive semiclassical optics language: it is based on the
exquisite sensitivity of the mode-coupling efficiencies to the offset
of the wavefunctions from the centroid. The incoherent sources are
literally blinking on the fundamental coherence time scale, causing
each image-plane photon to have a wavefunction given randomly by
$\psi_1(x)$ or $\psi_2(x)$. Either wavefunction can excite the
waveguide modes coherently with the same excitation probabilities,
causing the final photon counts to be as sensitive to the offset for
two sources as it is for one. Put another way, the incoherence between
the two sources implies a random relative phase between the two fields
and enables coupling into the first-order odd mode, which is the main
spatial mode responsible for the high sensitivity to small offsets.


The use of photon counting is essential here to discriminate against
the abundant but uninformative zero-photon events. If homodyne or
heterodyne methods were used instead, they would suffer from excess
vacuum fluctuations when no photon arrives. The poor performance of
heterodyne methods for weak thermal sources is also known in the
context of stellar interferometry \cite{townes,stellar}. The situation
is different from measurements of coherent light, the density operator
of which contains off-diagonal terms with respect to the photon-number
basis and the probabilistic photon picture is less adequate. Our
preliminary calculations \cite{nair_unpublished} confirm this
expectation and suggest that heterodyne detection of the spatial modes
still suffers from Rayleigh's curse.

Suppose that a total of $L$ photons are detected over the $M$ trials.
A record of the modes for the $L$ photons $(q_1,\dots,q_L)$ can be
obtained, but in fact a time-resolved record is not necessary, as
$\sum_l q_l$ is a sufficient statistic for estimating $Q$ and
$\theta_2$ \cite{wasserman}, meaning that the set of photon numbers
$\{m_q = \sum_l \delta_{q q_l}; q = 0,1,\dots\}$ detected in different
modes are also sufficient. The maximum-likelihood estimator becomes
\begin{align}
\check{Q}_{\rm ML} &= \frac{1}{L} \sum_{q} q  m_q,
&
\check\theta_{2\rm ML} &= 4\sigma\sqrt{\check{Q}_{\rm ML}},
\label{QML}
\end{align}
which is straightforward to implement computationally.  For $L = 0$,
one can set $\check\theta_{2}$ to a constant value; the $L = 0$
probability $(1-\epsilon)^M \approx \exp(-N)$ is in any case
negligible for large $N$. Maximum-likelihood estimation can
asymptotically saturate the Cram\'er-Rao bound
$\Sigma_{22} \ge 1/\mathcal J_{22}^{(\rm HG)}$ for large $M$
\cite{wasserman}. With
$\mathcal J_{22}^{(\rm HG)} \approx \mathcal K_{22}$, the QCRB is
asymptotically attainable as well. Appendix~\ref{monte_carlo} reports
a Monte Carlo analysis of the maximum-likelihood estimator for SPADE,
confirming that the Cram\'er-Rao bound remains close to the estimation
error for finite photon numbers.

\section{\label{sec_bspade} Binary SPADE}
\begin{figure}[htbp!]
\includegraphics[width=0.45\textwidth]{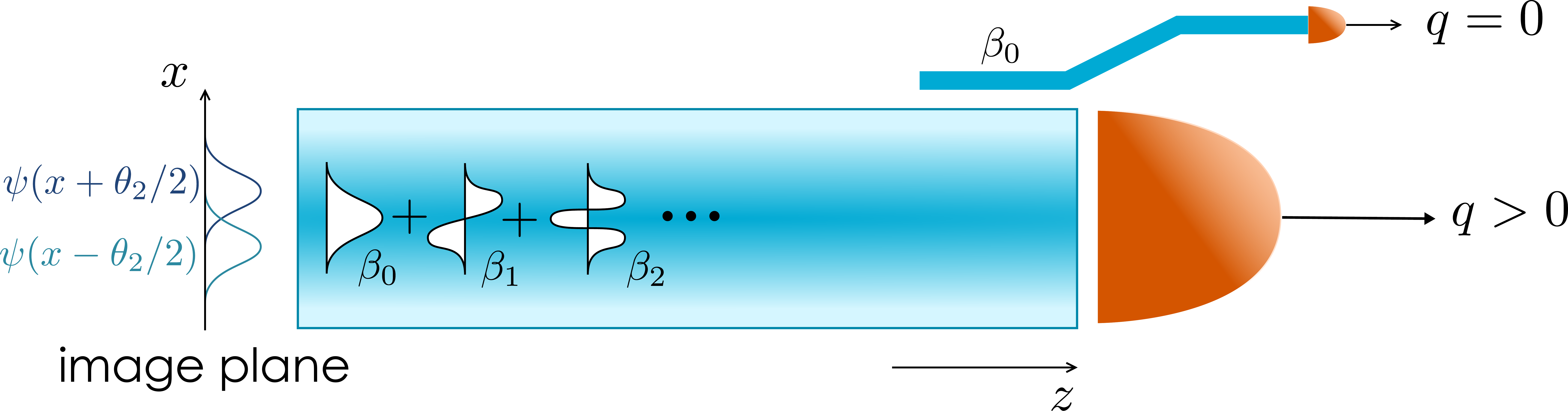}
\caption{\label{sms}Binary SPADE with evanescent coupling to only
  one single-mode waveguide.}
\end{figure}

\begin{figure}[htbp!]
\includegraphics[width=0.45\textwidth]{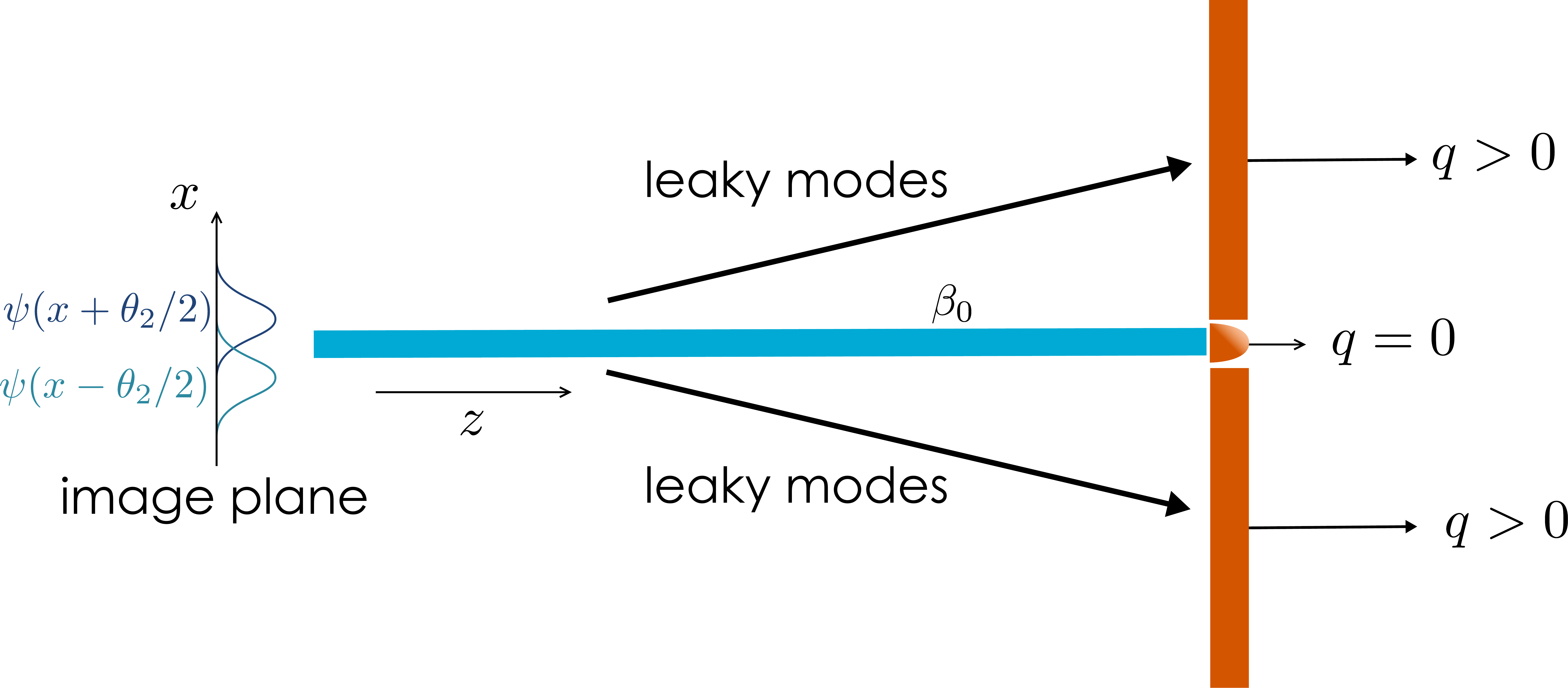}
\caption{\label{leaky_sms}An alternative design of binary SPADE with a
  single-mode waveguide and leaky-mode detection.}
\end{figure}

Since direct imaging has trouble estimating the separation only when
$\theta_2/\sigma$ is small, and only low-order Hermite-Gaussian modes
in SPADE are excited significantly in that case, we can focus on the
discrimination of low-order modes to simplify the SPADE design.  One
such design is depicted in Fig.~\ref{sms}, where only the $q = 0$
component is coupled into the single-mode waveguide, while any photon
in the higher-order modes remains in the multimode waveguide for
subsequent detection. An alternative design is depicted in
Fig.~\ref{leaky_sms}: the $q = 0$ mode is coupled to a single-mode
waveguide, while higher-order modes are necessarily coupled to the
leaky modes of the waveguide, which are also measured.

\begin{figure}[htbp!]
\includegraphics[width=0.45\textwidth]{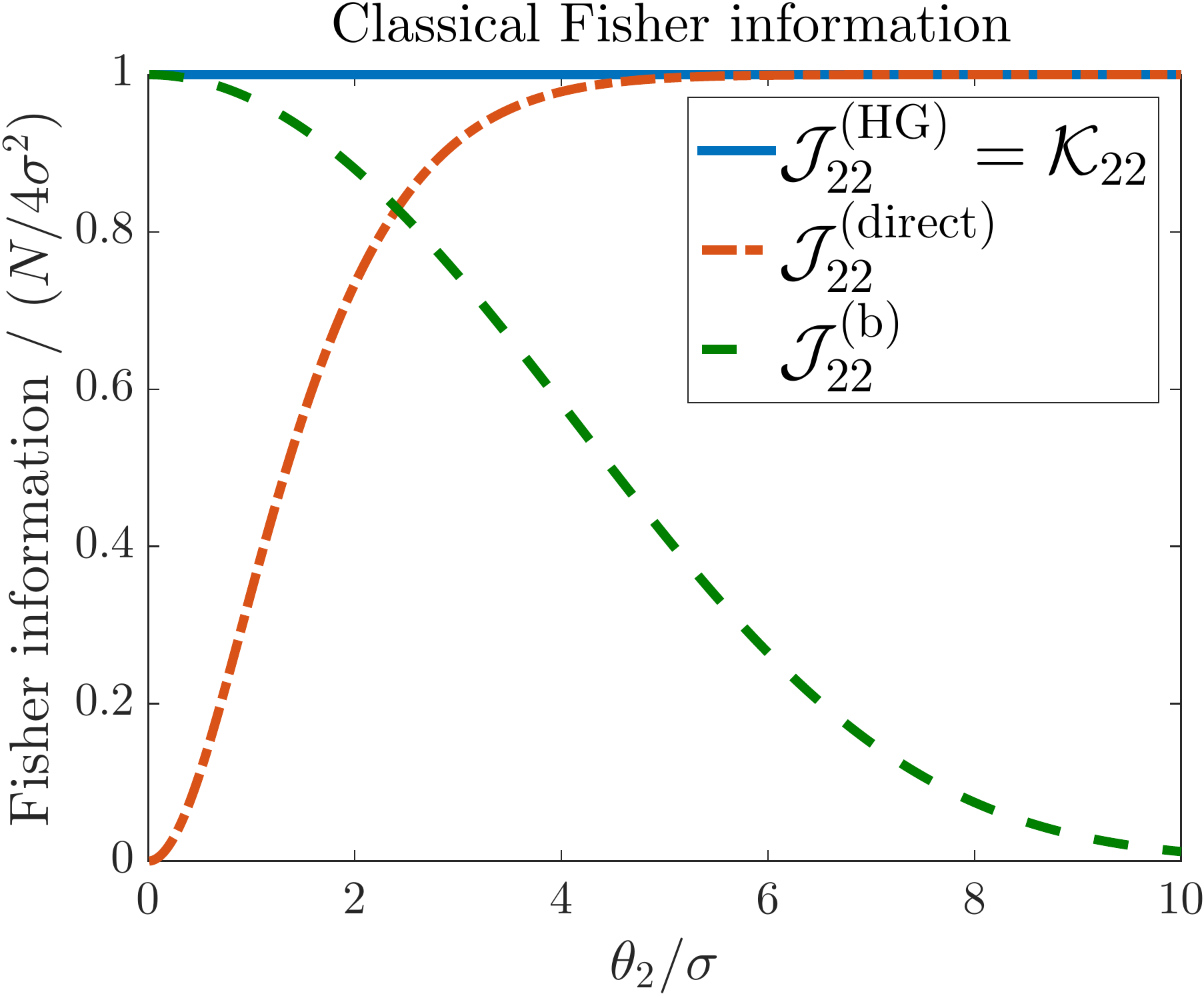}
\caption{\label{bspade}Fisher information for separation estimation
  versus normalized separation $\theta_2/\sigma$ for a Gaussian
  point-spread function.  $\mathcal J_{22}^{(\rm HG)}$ is the
  information for the ideal Hermite-Gaussian-basis measurement, which
  is equal to the quantum value $\mathcal K_{22}$,
  $\mathcal J_{22}^{(\rm direct)}$ is for direct imaging, and
  $\mathcal J_{22}^{(\rm b)}$ is for binary SPADE. The vertical axis
  is normalized with respect to
  $\mathcal J_{22}^{(\rm HG)} = \mathcal K_{22} = N/(4\sigma^2)$.}
\end{figure}

Conditioned on a detection event, the probability of detecting the
photon in the $q = 0$ mode remains
\begin{align}
P_1(q = 0) &\approx \exp\bk{-Q},
\label{P1q0}
\end{align}
but now the higher-order modes
cannot be discriminated, and the probability of detecting a photon in
any higher-order mode becomes
\begin{align}
P_1(q > 0) &= 1-P_1(q=0) \approx  1-\exp\bk{-Q}.
\label{P1qhigh}
\end{align}
The Fisher information for this scheme is hence
\begin{align}
\mathcal J_{22}^{(\rm b)} &\approx
 \frac{N}{4\sigma^2}
 \frac{Q\exp(-Q)}{1-\exp(-Q)}.
\label{fi_bspade}
\end{align}
Figure~\ref{bspade} compares $\mathcal J_{22}^{(\rm b)}$ with the
optimal value $\mathcal J_{22}^{(\rm HG)}\approx \mathcal K_{22}$ as
well as $\mathcal J_{22}^{(\rm direct)}$ for direct imaging. It can be
seen that binary SPADE gives significant information for small
$\theta_2/\sigma$, which happens to be the regime where direct imaging
performs poorly. Binary SPADE actually works less well when the
sources are far apart, and the two methods can complement each other
to enhance the localization precision, as shown in
Appendix~\ref{misalignment}.

For a total of $L$ detected photons, $L$ and $m_0$, the number of
photons detected in the $q=0$ mode, are sufficient statistics for
estimating $Q$ and $\theta_2$, and $m_0$ follows the binomial
distribution for $L$ trials and success probability $\exp(-Q)$
\cite{wasserman}.  The maximum-likelihood estimator becomes
\begin{align}
\check Q_{\rm ML}^{(\rm b)} &=-\ln \frac{m_0}{L},
&
\check\theta_{2\rm ML}^{(\rm b)}
&= 4\sigma \sqrt{\check Q_{\rm ML}^{(\rm b)}}.
\label{ML_bspade}
\end{align}
For $L = 0$ or $m_0 = 0$, one can select finite values for
$\check\theta_2$ to regularize the
estimator. Appendix~\ref{monte_carlo} reports a Monte Carlo analysis
of the resulting estimation error, confirming that it remains close to
the Cram\'er-Rao bound for finite photon numbers.

Compared with the large amount of data generated by direct imaging and
the complex algorithms needed to process them, only two photon numbers
are needed by binary SPADE to estimate the separation precisely. The
highly compressed measurement output and computationally simple
estimators, enabled by the coherent optical processing, come as
bonuses with our schemes.

\section{\label{other}Other point-spread functions}
\begin{figure}[htbp!]
\includegraphics[width=0.45\textwidth]{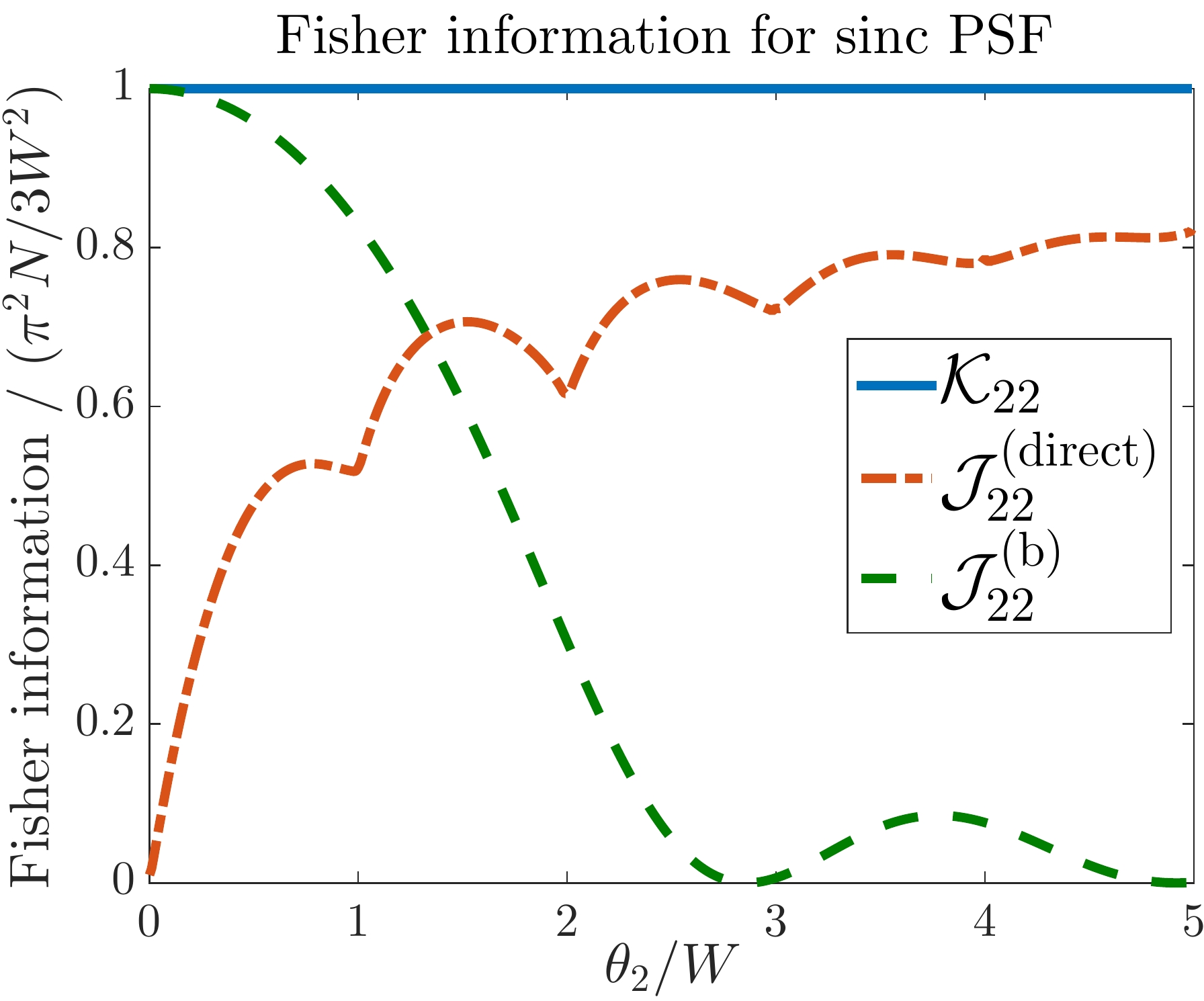}
\caption{\label{fisher_sinc}Fisher information for separation
  estimation versus normalized separation $\theta_2/W$ for the sinc
  point-spread function.  $\mathcal K_{22}$ is the quantum value,
  $\mathcal J_{22}^{(\rm direct)}$ is the numerically computed value
  for direct imaging, and $\mathcal J_{22}^{(\rm b)}$ is that for
  binary SPADE tailored for the sinc function. The vertical axis is
  normalized with respect to $\mathcal K_{22} = \pi^2N/(3W^2)$.}
\end{figure}

Our analysis of SPADE so far relies on the assumption of a Gaussian
point-spread function. For other point-spread functions, it is
nontrivial to find a suitable basis of spatial modes, although we can
still rely on the mathematical existence of a quantum-optimal
measurement \cite{hayashi05,fujiwara2006} to be sure that the QCRB can
be saturated. For a more concrete method, the analysis of the binary
SPADE schemes is fortunately still tractable, if we assume a
single-mode waveguide with a mode profile that matches the
point-spread function $\psi(x)$ centered at the centroid position.
Define $\ket{\psi} = \intall dx \psi(x)\ket{x}$ as the state of one
photon in the waveguide mode.  The efficiency of coupling a photon in
state $\ket{\psi_1}$ or $\ket{\psi_2}$ into the waveguide mode becomes
\begin{align}
\abs{\braket{\psi|\psi_1}}^2 &= \abs{\braket{\psi|\psi_2}}^2
=\abs{\intall dx \psi^*(x)\psi\bk{x+\frac{\theta_2}{2}}}^2
\nonumber\\
&= \abs{\intall dk \abs{\Psi(k)}^2\exp\bk{\frac{ik\theta_2}{2}}}^2
\equiv \Upsilon(\theta_2),
\label{f}
\end{align}
where $\Upsilon(\theta_2)$ is the mode overlap factor and
$\Psi(k) \equiv (2\pi)^{-1/2}\intall dx\psi(x)\exp(-ikx)$ is the
optical transfer function of the imaging system before the image plane
\cite{goodman}.  For the density operator in Eqs.~(\ref{rho}),
(\ref{rho1}), and (\ref{psis}), or in fact any mixture of
$\ket{\psi_1}\bra{\psi_1}$ and $\ket{\psi_2}\bra{\psi_2}$, the
probability of finding a photon in the waveguide mode becomes
$P(\psi) \approx \epsilon \Upsilon$, and the probability of finding a
photon in any other mode is
$P(\bar\psi) \approx \epsilon\bk{1-\Upsilon}$.  The Fisher information
over $M$ intervals is then
\begin{align}
\mathcal J_{22}^{(\rm b)}
\approx
\frac{N}{\Upsilon(1-\Upsilon)}\bk{\parti{\Upsilon}{\theta_2}}^2.
\end{align}
To study its behavior for small $\theta_2$, expand
$\Upsilon(\theta_2)$ in Eq.~(\ref{f}) as
$\Upsilon(\theta_2) = 1 - \Delta k^2 \theta_2^2/4 + O(\theta_2^4)$
with
$\Delta k^2 = \intall dk |\Psi(k)|^2 k^2 -[\intall dk |\Psi(k)|^2
k]^2$, giving
\begin{align}
\mathcal J_{22}^{(\rm b)}(\theta_2 = 0) &\approx N \Delta k^2.
\end{align}
$\mathcal J_{22}^{(\rm b)}$ can hence reach the quantum information
$\mathcal K_{22}= N\Delta k^2$ at $\theta_2 = 0$, precisely where
$\mathcal J_{22}^{(\rm direct)}$ vanishes and Rayleigh's curse is at
its worst. For larger $\theta_2$, $\mathcal J_{22}^{(\rm b)}$ is
expected to decrease, as the scheme is unable to discriminate the
higher-order modes that become more likely to be occupied.
Figure~\ref{fisher_sinc} plots $\mathcal K_{22}$, the numerically
computed $\mathcal J_{22}^{(\rm direct)}$, and
$\mathcal J_{22}^{(\rm b)}$ for the sinc point-spread function
\cite{rayleigh} $\psi(x) = (1/\sqrt{W})\sinc(x/W)$, where
$W = \lambda/(2\textrm{NA})$, $\sinc u \equiv \sin(\pi u)/(\pi u)$ for
$u \neq 0$ and $\sinc(0) \equiv 1$.  The information quantities
demonstrate behaviors similar to the Gaussian case.

\section{\label{sec_2d}Two-dimensional imaging}
The essential physics remains unchanged when we consider
two-dimensional imaging, and we discuss the generalization only
briefly here; the details are given elsewhere \cite{ant}.  The
single-photon ket in Eq.~(\ref{psis}) should now be expressed as
$\ket{\psi_s} = \intall dx \intall dy \psi_s(x,y)\ket{x,y}$, where
$\braket{x,y|x',y'} = \delta(x-x')\delta(y-y')$ and $\psi_s(x,y)$ is a
two-dimensional wavefunction \cite{yuen_shapiro1,shapiro09}. In terms
of a point-spread function $\psi(x,y)$ and unknown positions
$(X_1,Y_1)$ and $(X_2,Y_2)$, $\psi_s(x,y) = \psi(x-X_s,y-Y_s)$, and we
can define the four centroid and separation parameters as
$\theta_{1} = (X_1 + X_2)/2$, $\theta_{2} = X_2 - X_1$,
$\theta_{3} = (Y_1 + Y_2)/2$, and $\theta_{4} = Y_2 - Y_1$.
$\mathcal J^{(\rm direct)}$ for the estimation of $\theta_{2}$ and
$\theta_{4}$ decreases to zero when the sources are close, and
Rayleigh's curse still exists for direct imaging
\cite{vanaert,ram}. On the other hand, the quantum Fisher information
matrix, to be reported in Ref.~\cite{ant}, again shows no sign of
Rayleigh's curse for two-dimensional separation estimation.

For SPADE, we can use the two-dimensional Hermite-Gaussian basis
\cite{yariv}. Assuming a Gaussian point-spread function and a known
centroid, it is straightforward to show that a measurement of each
photon in the Hermite-Gaussian basis with mode indices $q$ and $p$
obeys a two-variable Poisson distribution, and the classical Fisher
information with respect to $\theta_{2}$ and $\theta_{4}$ remains a
constant and free of Rayleigh's curse, similar to the one-dimensional
case. For other point-spread functions, such as the Airy disk
\cite{born_wolf,goodman}, binary SPADE with a matching mode profile
can estimate the separation without Rayleigh's curse for small
separations in the same way as the one-dimensional case, but
information about the direction of the separation is lost.  To obtain
directional information, one needs to discriminate at least some of
the higher-order modes in different directions.

A quadratic-index optical fiber can support two-dimensional
Hermite-Gaussian modes, while a weakly guiding step-index fiber also
has modes closely resembling the Hermite-Gaussian modes
\cite{yariv_yeh}.  A complication arises for cylindrically symmetric
fibers, as modes with the same total order $q+p$ will have a
degenerate propagation constant, causing multiple modes to satisfy the
same phase-matching conditions in grating or evanescent coupling and
preventing discrimination of modes with the same order.  The net
result is that directional information is compromised. One solution is
to turn the point-spread function into an elliptic one with asymmetric
widths and use an elliptic fiber to break the degeneracy.

\section{Conclusion}
We have presented two important results in this paper: the fundamental
quantum limit to locating two incoherent optical point sources and the
SPADE measurement schemes for quantum-optimal separation estimation.
Our quantum bound sets the ultimate limit to localization precision in
accordance with the fundamental laws of quantum mechanics, while SPADE
can extract the full information offered by quantum mechanics
concerning the separation parameter via linear photonics. The proposed
SPADE schemes work well for close sources with significant overlap in
their wavefunctions, avoiding Rayleigh's curse and the divergent error
that plagues direct imaging. The computational simplicity of the
estimators is an additional advantage.  Foreseeable applications
include binary-star astrometry \cite{howell06,huber,becker} and
single-molecule imaging \cite{moerner}, either as a replacement of
techniques based on fluorescence resonant energy transfer
\cite{pawley,michalet06} or as an enhancement of localization
microscopy \cite{ram,moerner,hell,betzig,pawley} to provide
complementary information about close pairs of fluorophores.

\emph{Note added.}---Subsequent to the completion of this work (the
first version of this manuscript was submitted to the arXiv preprint
server on Nov 2, 2015 \cite{tnlv1}), we have developed a semiclassical
but less general theory to explain our results here for pedagogy
\cite{tnl2}, discovered an alternative scheme called
Super-Localization via Image-inVERsion interferometry (SLIVER) that
can overcome Rayleigh's curse without the need to tailor the device to
the point-spread function \cite{sliver,ant}, derived the QCRB for
thermal sources without the $\epsilon \ll 1$ approximation using an
alternative approach \cite{nair_tsang16}, and shown that variations of
SPADE and SLIVER can attain the bound for arbitrary $\epsilon$
\cite{nair_tsang16}, validating the results here. A generalization of
our theory presented here to two dimensions, with similar conclusions,
is described in detail elsewhere \cite{ant}.  Following
Ref.~\cite{tnlv1}, Lupo and Pirandola have derived the ultimate
quantum Fisher information for separation estimation with arbitrary
quantum sources \cite{lupo}, including our independent result on
thermal sources \cite{nair_tsang16} as a special case. Experiments
inspired by our theory have been reported in
Refs.~\cite{tang16,lvovsky16,steinberg16,paur16};
Refs.~\cite{lvovsky16,steinberg16,paur16} also propose variations of
SPADE that are easier to implement experimentally.


\section*{Author contributions}
M.~T.\ conceived the idea of applying quantum metrology to
two-incoherent-source localization and developed the quantum optics
formalism in Sec.~\ref{optics} and Appendix~\ref{state}. R.~N.\
derived the quantum Fisher information in Sec.~\ref{sec_crb} and
Appendix~\ref{metrology} with X.-M.~L.\ and M.~T.'s inputs and checks.
M.~T.\ invented and analyzed the SPADE measurement schemes described
in Secs.~\ref{sec_spade}--\ref{sec_2d}, while R.~N.\ first recognized
the importance of the first-order Hermite-Gaussian mode to SPADE, as
explained in Sec.~\ref{sec_spade}. M.~T.\ supervised the project and
wrote the paper. All authors discussed extensively during the course
of this work.

\section*{Acknowledgments}
We acknowledge useful discussions with Shan Zheng Ang and Shilin
Ng. This work is supported by the Singapore National
Research Foundation under NRF Grant No.~NRF-NRFF2011-07 and the
Singapore Ministry of Education Academic Research Fund Tier 1 Project
R-263-000-C06-112.  

\appendix
\section{\label{review}Quantum-imaging literature review}
Helstrom pioneered the application of his quantum estimation and
detection bounds to optical imaging problems
\cite{helstrom70,helstrom73b,helstrom}, focusing on coherent and
thermal sources. In particular, the now well-known expression for the
shot-noise-limited localization error for one classical source can be
found in Ref.~\cite{helstrom70}; similar expressions in the context of
direct imaging were later reported in
Refs.~\cite{lindegren78,king83,bobroff}. For more recent studies of
quantum metrology for coherent-state or nonclassical-state imaging,
see, for example, Refs.~\cite{delaubert,nair_yen,perez12}. For studies
on the use of squeezed light for single-object localization, see, for
example,
Refs.~\cite{fabre,barnett,treps,taylor2013,localization}. None of
these studies considered the problem of locating two close incoherent
sources.

The standard quantum model of paraxial imaging and the use of
nonclassical light for that purpose were proposed by Yuen and Shapiro
\cite{yuen_shapiro1}. This topic has been further investigated most
notably by Kolobov and co-workers \cite{kolobov,kolobov07}, who
focused on coherent or squeezed light, homodyne detection, and field
fluctuations. Such models are irrelevant to incoherent sources such as
stars and fluorophores, for which the mean field is zero and photon
counting is the more relevant method to minimize vacuum noise; to
quote Helstrom \cite{helstrom70b},

\begin{quotation}
With such incoherently
illuminated or radiating objects, it is not the field of the light
that is of interest, for that field is best described as a random
process having zero mean value and a most erratic spatiotemporal
variation. Rather it is the mean-square value of the field, averaged
over many cycles of the dominant temporal frequency, that
characterizes the object in the most informative way.
\end{quotation}

Subsequent work by Kolobov and co-workers
\cite{kolobov_fabre,beskrovnyy05,beskrovny08,piche} considered the
squeezing and measurement of the eigenmodes of an imaging system for
image-reconstruction superresolution. Again, these studies focused on
coherent or squeezed light only.
%
%
The ``Rayleigh resolution limit''
mentioned by many of these papers is a misnomer, as the resolution
limit for coherent imaging should be attributed to Abbe, while
Rayleigh's criterion is defined for two incoherent sources
\cite{rayleigh,born_wolf} and ill-suited to coherent imaging
\cite{horstmeyer}. Moreover, the imaging-system eigenmodes they
studied have no relation to the spatial modes we propose for the
two-source localization problem and they did not use the more rigorous
framework of statistical parameter estimation.

We can consider the schemes proposed in
Refs.~\cite{centroid,glm_imaging,shin,rozema,oppel,schwartz13,cui13,monticone}
as another class of superresolution imaging protocols, which require
coherent or nonclassical sources and multiphoton coincidence
measurements and do not consider statistical inference. It is well
known in statistical optics that a multiphoton coincidence
measurement, such as the obsolete Hanbury Brown-Twiss interferometry,
fundamentally has a much poorer signal-to-noise ratio than amplitude
interferometry because multiphoton coincidence events are rare for
thermal optical sources \cite{goodman_stat,stellar}. The actual
statistical resolution of this class of protocols is thus
questionable, especially for weak optical sources, without further
proofs in the context of inference accuracy.  In recent years, there
has also been significant interest in quantum lithography
\cite{boto,shih07,boyd2012,hemmer12} and ghost imaging
\cite{pittman,gatti,shih07,erkmen10}, although their applications are
clearly different from our purpose and will not be elaborated here.

The relative neglect of incoherent sources in the quantum-imaging
literature, despite their obvious importance, may be due to a lack of
appreciation that quantum mechanics can be relevant to such highly
classical light. Our work thus showcases quantum metrology as a
powerful tool to discover the ultimate performance of sensing and
imaging even for classical sources, providing not only rigorous
quantum limits but also pleasant surprises for one of the most
important applications in optics.

\section{\label{state}Quantum optics: derivation of
  Eqs.~(\ref{rho})--(\ref{psis})}
Define $\alpha = (\alpha_1,\dots,\alpha_J)^\top$ as a column vector of
complex field amplitudes for $J$ optical spatial modes on the image
plane and $\ket{\alpha}$ as a multimode coherent state with amplitude
$\alpha$. Any quantum state can be expressed as
\begin{align}
\rho &= \int D\alpha \Phi(\alpha)\ket{\alpha}\bra{\alpha},
\label{glauber}
\end{align}
where $\Phi(\alpha)$ is the Sudarshan-Glauber representation and
$D\alpha$ is an appropriate measure \cite{mandel}. For thermal
sources, it is standard \cite{mandel} to assume $\Phi$ to be a
zero-mean complex Gaussian given by
\begin{align}
\Phi(\alpha) &= \frac{1}{\det(\pi \Gamma)}
\exp\bk{-\alpha^\dagger  \Gamma^{-1}\alpha},
\end{align}
where $\alpha^\dagger = (\alpha_1^*,\dots,\alpha_J^*)$ denotes the
complex transpose of $\alpha$,
\begin{align}
 \Gamma &= \expect\bk{\alpha\alpha^\dagger}
\label{Gamma}
\end{align}
is the image-plane mutual coherence matrix, and
$\expect\Bk{f(\alpha)} \equiv \int D\alpha \Phi(\alpha) f(\alpha)$
denotes the expectation of any function $f$ with respect to the $\Phi$
distribution.  Writing the coherent state in terms of a superposition
of Fock states and applying the Gaussian moment theorem \cite{mandel}
to Eq.~(\ref{glauber}), we can express $\rho$ as the incoherent
mixture
\begin{align}
\rho &= \sum_{n=0}^\infty \pi_n \rho_n,
&
n &\equiv \sum_j n_j,
\label{rho_full}
\end{align}
where $\pi_n$ is the probability of having $n$ total photons in the
state and $\rho_n$ is an $n$-photon multimode Fock state.

At optical frequencies or beyond, it is standard
\cite{goodman_stat,mandel,mandel59,labeyrie,zmuidzinas03,gottesman,stellar}
to assume that, within the short coherence time of a source, the
average photon number arriving at the imaging device is much smaller
than $1$. We will make the same assumption for two sources, viz.,
\begin{align}
\epsilon \equiv \sum_j \trace \rho a_j^\dagger a_j = 
\expect\bk{\alpha^\dagger\alpha} =\sum_j  \Gamma_{jj} \ll 1,
\end{align} 
where $\trace$ denotes the operator trace, $a_j$ is the annihilation
operator for the $j$th mode, and $a_j^\dagger$ is the creation
operator. For example, a star with sun-like temperature
$6000~\textrm{K}$ emits $\sim 10^{-2}$ photon on average per mode at
wavelength $500~\textrm{nm}$, while the limited fraction of the
coherence area captured by the telescope aperture further reduces the
received photon number \cite{goodman_stat}.  In microscopy, a typical
fluorophore emits $< 10^7$ photons per second \cite{ober} with
coherence time $< 50~\textrm{fs}$ \cite{pawley}, leading to
$\epsilon < 10^{-6}$ for two sources.  The zero-photon probability
given by
\begin{align}
\pi_0 &= \expect\bk{e^{-\alpha^\dagger\alpha}} = 1-\epsilon + O(\epsilon^2)
\end{align}
is then the highest, the one-photon probability given by
\begin{align}
\pi_1 &= \expect\bk{e^{-\alpha^\dagger\alpha}\alpha^\dagger\alpha} = \epsilon
+O(\epsilon^2)
\end{align}
is $\epsilon$ to the first order, and the multiphoton
probability
\begin{align}
\sum_{n=2}^\infty \pi_n &= 1-\pi_0-\pi_1 = O(\epsilon^2)
\label{pn2}
\end{align}
is in the second order, leading to Eq.~(\ref{rho}).  As the vacuum
state provides no information and multiphoton events are rare, we will
focus on the one-photon state $\rho_1$. This focus also makes our
formalism applicable to inefficient single-photon emitters, which may
have non-Poissonian multiphoton statistics but rare multiphoton
events, and electron microscopy \cite{vanaert}.

The negligence of the $O(\epsilon^2)$ multiphoton probability leads to
a Poisson photon-counting distribution \cite{goodman_stat}, which
ignores bunching or antibunching effects but remains an excellent
empirical model for both astronomical optical sources
\cite{goodman_stat,mandel,mandel59,labeyrie,zmuidzinas03} and
fluorophores \cite{ram,ober,deschout,chao16,pawley} by virtue of the
$\epsilon \ll 1$ condition. To quote Mandel \cite{mandel59},

\begin{quotation}
  The light from these sources is always so weak that
  $\bar n\xi/T \ll 1$ [$\epsilon$ in our terminology] and the
degeneracy is unlikely to be detected in measurements on a single
beam. The situation is, of course, improved when correlation
measurements are undertaken on two or more coherent beams (Hanbury
Brown and Twiss 1956), since these measurements single out the
degenerate photons (Mandel 1958). Even so it is unlikely that any
faint stars could be studied in this way.
\end{quotation}

Similarly, Goodman states that
\cite{goodman_stat}

\begin{quotation}
  If the count degeneracy parameter [$\epsilon$ in our terminology]
is much less than 1, it is highly probable that there will be either
zero or one counts in each separate coherence interval of the incident
classical wave. In such a case the classical intensity fluctuations
have a negligible "bunching" effect on the photo-events, for (with
high probability) the light is simply too weak to generate multiple
events in a single coherence cell. If negligible bunching of the
events takes place, the count statistics will be indistinguishable
from those produced by stabilized single-mode laser radiation, for
which no bunching occurs.
\end{quotation}

A more recent work by Zmuidzinas
\cite{zmuidzinas03} also states that

\begin{quotation}
It is well established that the photon counts registered by
  the detectors in an optical instrument follow statistically
  independent Poisson distributions, so that the fluctuations of the
  counts in different detectors are uncorrelated. To be more precise,
  this situation holds for the case of thermal emission (from the
  source, the atmosphere, the telescope, etc.) in which the mean
  photon occupation numbers of the modes incident on the detectors are
  low, $n \ll 1$ [$\epsilon$ in our terminology]. In
  the high occupancy limit, $n \gg 1$, photon bunching becomes
  important in that it changes the counting statistics and can
  introduce correlations among the detectors. We will discuss only the
  first case, $n \ll 1$, which applies to most astronomical
  observations at optical and infrared wavelengths.
\end{quotation}

Define $\ket{j} = a_j^\dagger\ket{\textrm{vac}}$ as the ket with one
photon only in the $j$th mode. Consider the one-photon matrix elements
\begin{align}
\bra{j}\rho\ket{k} &= \expect\bk{e^{-\alpha^\dagger\alpha} \alpha_j \alpha_k^*}
=  \Gamma_{jk} + O(\epsilon^2).
\end{align}
We can then assume, to the first order of $\epsilon$,
\begin{align}
\pi_1 &\approx \epsilon,
&
\rho_1 &\approx \frac{1}{\epsilon}\sum_{j,k} \Gamma_{jk} \ket{j}\bra{k}.
\label{rho1_discrete}
\end{align}
Similar approximations were also used in
Refs.~\cite{stellar,gottesman}.  To derive $\Gamma$, let
$\beta \equiv (\beta_1,\dots,\beta_K)^\top$ be the field amplitudes
for optical modes on the object plane, and consider the field
propagation rule $\alpha = S \beta$ for a linear optical system, where
$S$ is the field scattering matrix.  The image-plane mutual coherence
$\Gamma$ is then related to the object-plane mutual coherence matrix
$\Gamma^{({\rm o})}$ by
\begin{align}
 \Gamma &= S  \Gamma^{({\rm o})} S^\dagger.
\label{wolf}
\end{align}
This propagation rule is a basic principle in both classical and
quantum statistical optics \cite{mandel}.

In the paraxial regime, we can use localized wavepacket modes as a
basis \cite{yuen_shapiro1,shapiro09}. Let $u$ be the position index
for a wavepacket mode on the one-dimensional object plane and consider
two incoherent sources with equal intensities at positions $u = u_1$
and $u = u_2$. The fields are uncorrelated at different points on the
object plane, with nonzero intensities only at the sources. Then
\begin{align}
  \Gamma_{uv}^{({\rm o})} = \epsilon_0 \delta_{uv}
\bk{\delta_{u u_1}+\delta_{u u_2}},
\end{align}
where $\epsilon_0$ is the average photon number from each source.  On
the image plane, the mutual coherence becomes
\begin{align}
\Gamma_{jk} = \epsilon_0\bk{S_{j u_1} S^*_{k u_1} 
+ S_{j u_2}S^*_{k u_2}},
\label{Gamma2}
\end{align}
and the average photon number can be expressed as
$\epsilon = 2\epsilon_0 \eta$, where
$\eta \equiv \sum_j \abs{S_{j u_s}}^2$ is the quantum efficiency of
the imaging system and we have made the reasonable assumption that the
efficiency is the same for both
sources. Equation~(\ref{rho1_discrete}) can then be expressed as
Eq.~(\ref{rho1}) if we define single-photon kets
$\ket{\psi_s} \equiv \sum_j \psi(j,u_s)\ket{j}$ with normalized
wavefunctions $\psi(j,u_s) = S_{j u_s}/\sqrt{\eta}$. Assuming
image-plane wavepacket positions $x_j = x_0 + jdx$, position eigenkets
$\ket{x_j} = \ket{j}/\sqrt{dx}$, and wavefunctions
$\psi_s(x_j) = \psi(j,u_s)/\sqrt{dx}$, we arrive at Eq.~(\ref{psis})
by taking the continuous-space limit with infinitesimal $dx$
\cite{yuen_shapiro1,shapiro09}.

\section{\label{metrology}Quantum metrology:
derivation of Eqs.~(\ref{K})}
The quantum Fisher information matrix with respect to
$\rho^{\otimes M}$ proposed by Helstrom \cite{helstrom} is defined as
\begin{align}
\mathcal K_{\mu\nu}(\rho^{\otimes M}) &= M\mathcal K_{\mu\nu}(\rho) = 
M \real \trace \mathcal L_\mu(\rho) \mathcal L_\nu(\rho) \rho,
\label{qfi}
\end{align}
where $\mathcal L_\mu(\rho)$ is a symmetric logarithmic derivative
(SLD) of $\rho$.  Writing $\rho$ in its eigenbasis as
\begin{align}
\rho &= \sum_j  D_j \ket{e_j}\bra{e_j},
\end{align}
$\mathcal L_\mu(\rho)$ can be expressed as
\cite{helstrom,braunstein,paris}
\begin{align}
\mathcal L_\mu(\rho) &= \sum_{j,k; D_j + D_k \neq 0}
\frac{2}{D_j+D_k}
\bra{e_j}\parti{\rho}{\theta_\mu}\ket{e_k}
\ket{e_j}\bra{e_k}.
\label{sld}
\end{align}
Given this definition and Eq.~(\ref{rho_full}), it can be shown that
\begin{align}
\mathcal K(\rho) &= \sum_n \pi_n \mathcal K(\rho_n) \ge \pi_1 \mathcal K(\rho_1),
\end{align}
as each $\rho_n$ is in an orthogonal subspace.  Since the vacuum state
$\rho_0 = \ket{\textrm{vac}}\bra{\textrm{vac}}$ contains no
information and multiphoton events are rare, the total information
will be dominated by that from the one-photon state $\rho_1$. We will
therefore focus on the one-photon component $\pi_1\mathcal K(\rho_1)$
as a tight lower bound on the quantum information and assume in the
following
\begin{align}
\mathcal K(\rho) &\approx \pi_1\mathcal K(\rho_1).
\label{K1}
\end{align}
With $\pi_1 \approx \epsilon$ and the probability of multiphoton
events being $O(\epsilon^2)$ according to Eq.~(\ref{pn2}), this
approximation is accurate to the first order of $\epsilon$.

To compute the quantum Fisher information matrix $\mathcal K(\rho_1)$
according to Eqs.~(\ref{qfi})--(\ref{sld}), we first need to
diagonalize the $\rho_1$ in Eqs.~(\ref{rho1}) and (\ref{psis}), noting
that the eigenvectors should span the supports of $\rho_1$ and
$\partial\rho_1/\partial\theta_\mu$.  The partial derivative of
$\rho_1$ with respect to $X_\mu$ can be expressed as
\begin{align}
\parti{\rho_1}{X_\mu} &= \parti{D_1}{X_\mu}\ket{e_1}\bra{e_1}
+\parti{D_2}{X_\mu}\ket{e_2}\bra{e_2}
\nonumber\\&\quad
+
\Bk{D_1 \parti{\ket{e_1}}{X_\mu}\bra{e_1} + 
D_2 \parti{\ket{e_2}}{X_\mu}\bra{e_2} + \textrm{H.c.}},
\label{drho}
\end{align}
where H.c.\ denotes the Hermitian conjugate. In addition to the
support of $\rho_1$ spanned by $\ket{e_1}$ and
$\ket{e_2}$, we also need to find more eignevectors that span the
support of $\partial\ket{e_1}/\partial X_\mu$ and
$\partial\ket{e_2}/\partial X_\mu$.

Assuming that $\psi_\mu(x) = \psi(x-X_\mu)$ and the point-spread
function $\psi(x)$ has an $x$-independent phase, we can take $\psi(x)$
to be real without loss of generality and choose the following
orthonormal set of eigenvectors:
\begin{align}
\ket{e_1} &= \frac{1}{\sqrt{2(1-\delta)}}
\bk{\ket{\psi_1} - \ket{\psi_2}},
\nonumber\\
\ket{e_2} &= \frac{1}{\sqrt{2(1+\delta)}}
\bk{\ket{\psi_1} + \ket{\psi_2}},
\nonumber\\
\ket{e_3} &= \frac{1}{c_3}
\Bk{\frac{\Delta k}{\sqrt{2}}\bk{\ket{\psi_{11}} + \ket{\psi_{22}}} 
- \frac{\gamma}{\sqrt{1-\delta}}\ket{e_1}},
\nonumber\\
\ket{e_4} &= \frac{1}{c_4}
\Bk{\frac{\Delta k}{\sqrt{2}}\bk{\ket{\psi_{11}} - \ket{\psi_{22}}} + 
\frac{\gamma}{\sqrt{1+\delta}}\ket{e_2}},
\end{align}
where $\Delta k^2$ and $\gamma$ are given by Eqs.~(\ref{dk2})
and (\ref{gamma}), respectively,
\begin{align}
\ket{\psi_{11}} &\equiv \frac{1}{\Delta k}\intall dx \parti{\psi(x-X_1)}{X_1}\ket{x},
\nonumber\\
\ket{\psi_{22}} &\equiv \frac{1}{\Delta k}\intall dx \parti{\psi(x-X_2)}{X_2}\ket{x},
\nonumber\\
c_3 &\equiv \bk{\Delta k^2 + b^2 - \frac{\gamma^2}{1-\delta}}^{1/2},
\nonumber\\
c_4 &\equiv \bk{\Delta k^2 - b^2 - \frac{\gamma^2}{1+\delta}}^{1/2},
\nonumber\\
b^2 &\equiv \intall dx \parti{\psi(x-X_1)}{X_1}\parti{\psi(x-X_2)}{X_2},
\nonumber\\
\delta &\equiv \intall dx \psi(x-X_1)\psi(x-X_2),
\end{align}
and the eigenvalues of $\rho_1$ are
\begin{align}
D_1 &= \frac{1-\delta}{2},
&
D_2 &= \frac{1+\delta}{2},
&
D_3 &= D_4 = 0.
\label{eigenvalues}
\end{align}
After more algebra, the SLD in Eq.~(\ref{sld}) with respect to the
derivative in Eq.~(\ref{drho}) can be expressed as
\begin{align}
\mathcal L_\mu^{(X)} &= \sum_{j,k} \mathcal L_{\mu,jk}^{(X)}\ket{e_j}\bra{e_k}
\label{LX}
\end{align}
with a real and symmetric matrix
$\mathcal L_{\mu,jk}^{(X)} = \mathcal L_{\mu,kj}^{(X)}$, the nonzero
and unique elements of which are found to be
\begin{align}
\mathcal L_{1,11}^{(X)} &=  -\mathcal L_{2,11}^{(X)} = \frac{\gamma}{1-\delta},
&
\mathcal L_{1,12}^{(X)} &= \mathcal L_{2,12}^{(X)} = 
\frac{\gamma\delta}{\sqrt{1-\delta^2}},
\nonumber\\
\mathcal L_{1,13}^{(X)} &= -\mathcal L_{2,13}^{(X)} = \frac{c_3}{\sqrt{1-\delta}},
&
\mathcal L_{1,14}^{(X)} &= \mathcal L_{2,14}^{(X)} = \frac{c_4}{\sqrt{1-\delta}},
\nonumber\\
\mathcal L_{1,22}^{(X)} &=  -\mathcal L_{2,22}^{(X)} = \frac{-\gamma}{1+\delta},
&
\mathcal L_{1,23}^{(X)} &= \mathcal L_{2,23}^{(X)} = \frac{c_3}{\sqrt{1+\delta}},
\nonumber\\
\mathcal L_{1,24}^{(X)} &= -\mathcal L_{2,24}^{(X)} = \frac{c_4}{\sqrt{1+\delta}}.
\end{align}
In terms of the centroid and separation parameters given by
$\theta_1 = (X_1+X_2)/2$ and $\theta_2 = X_2 - X_1$, the SLDs become
\begin{align}
\mathcal L_1 &= \mathcal L_1^{(X)} + \mathcal L_2^{(X)},
&
\mathcal L_2 &= \frac{\mathcal L_2^{(X)} - \mathcal L_1^{(X)}}{2}.
\label{L12}
\end{align}
We can now substitute Eqs.~(\ref{eigenvalues})--(\ref{L12}) into
Eq.~(\ref{qfi}) to compute the quantum Fisher information matrix
$\mathcal K(\rho_1)$. The final result, assuming
$M\pi_1 \approx M\epsilon = N$, is given by Eqs.~(\ref{K}) with
zero off-diagonal terms.

\section{\label{misalignment}Unknown centroid
and misalignment}
Our analysis of SPADE in Sec.~\ref{sec_spade} and \ref{sec_bspade}
assumes that the centroid of the two sources is known exactly and the
device is optimally aligned with the centroid. For astronomy, it is
reasonable to assume that the centroid is known accurately, as
extensive telescopic data on stellar objects should be readily
available and conventional imaging is accurate in estimating the
centroid. Even if the centroid is unknown, stellar objects usually
shine long enough for one to collect ample prior information before
aligning the SPADE device.  For microscopy, however, biological
samples may drift more quickly and fluorophores can bleach, giving
little time and few photons for one to estimate both parameters.  One
option, to be explored in future work, is to scan the SPADE device
across the image plane in a manner similar to the operation of a
confocal microscope \cite{pawley}.

\begin{figure}[htbp!]
\includegraphics[width=0.45\textwidth]{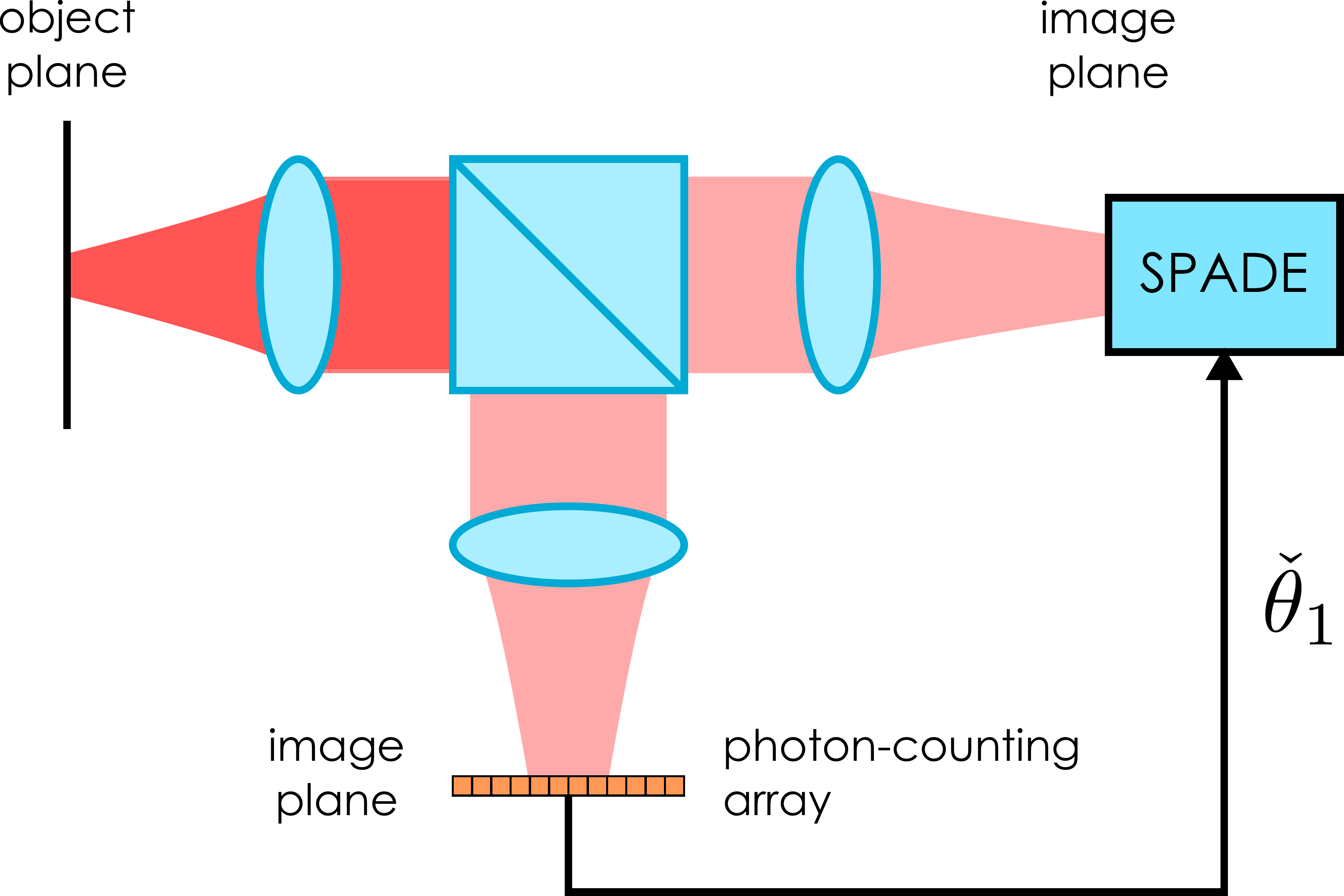}
\caption{\label{hybrid_scheme}A hybrid measurement scheme that splits
  the optical field by a beam splitter, measures one output port by a
  photon-counting array, and use the centroid estimate
  $\check\theta_1$ to align SPADE at the other port.}
\end{figure}

Another option, illustrated in Fig.~\ref{hybrid_scheme}, is to split
the optical field by a beam splitter, measure one output port by
direct imaging, and use the centroid estimate to align SPADE at the
other port in a hybrid scheme. As the overall optical system is linear
with photon counting, the output statistics remain Poisson for
$\epsilon \ll 1$, meaning that the statistics of the measurements are
independent and can be analyzed separately. The penalty of
beam-splitting with the classical sources is simply a reduction of
photon number at each port. With direct imaging offering little
information about $\theta_2$ when Rayleigh's criterion is violated,
the additional information offered by SPADE for a reduced photon
number can still be helpful.  The outstanding issues are then the
robustness of SPADE to the misalignment due to imperfect centroid
estimation, and the overhead resources of photons needed to achieve
satisfactory alignment.

\begin{figure}[htbp!]
\includegraphics[width=0.45\textwidth]{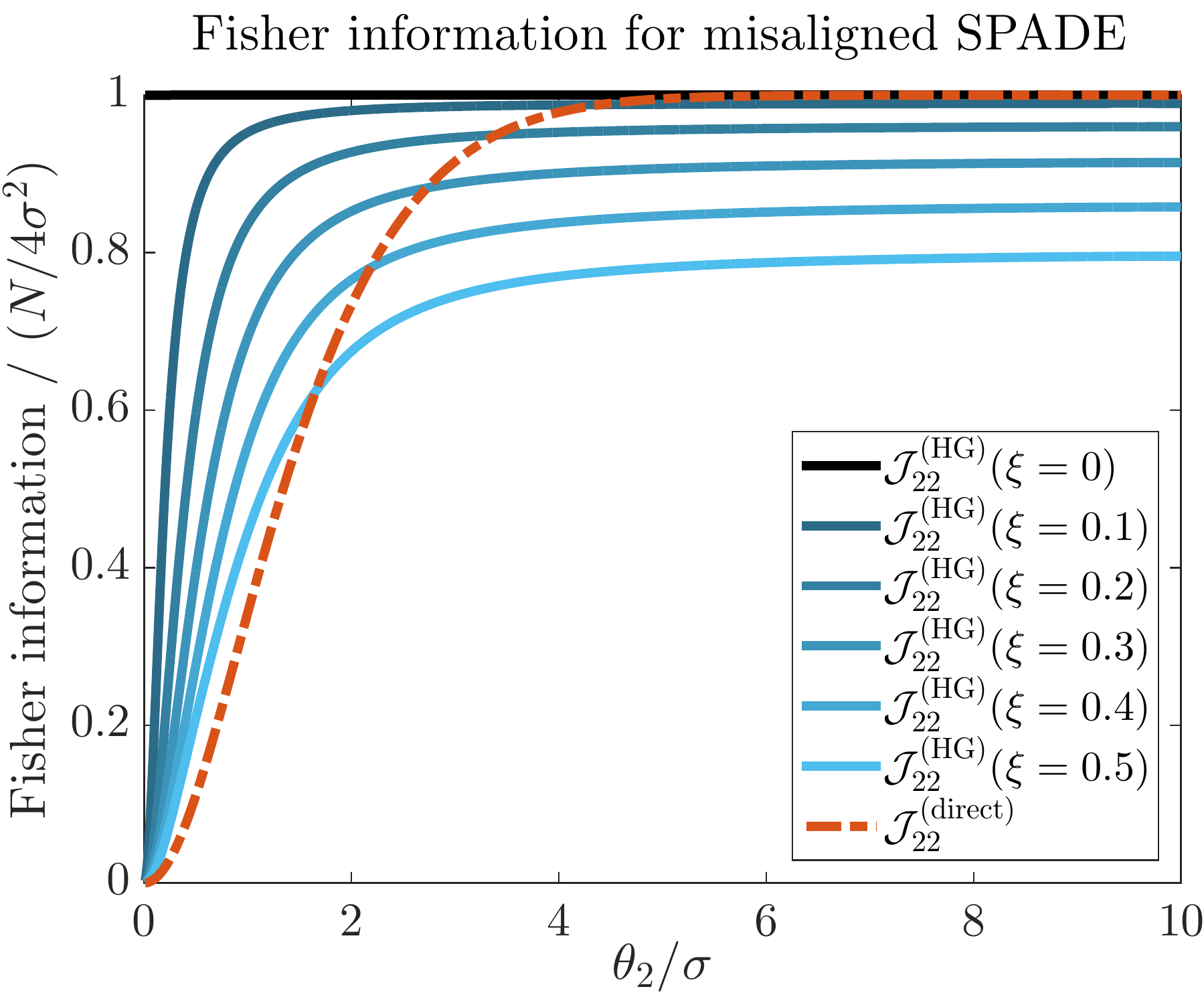}
\caption{\label{misalign}Fisher information for separation estimation
  with SPADE with misalignment levels
  $\xi = 0,0.1,\dots,0.5$ (solid curves) and direct imaging
  (dash-dotted curve). The different solid curves can be
  distinguished by their decreasing values with larger misalignments.}
\end{figure}

Let the center of a SPADE device be $\check \theta_1$ and consider
$\theta_1 \neq \check \theta_1$ due to misalignment.  For a Gaussian point-spread
function and the Hermite-Gaussian-basis measurement, Eq.~(\ref{P1q})
should be generalized to
\begin{align}
P_1(q) &\approx
\frac{1}{2}\Bk{\exp(-Q_1)\frac{Q_1^q}{q!}+\exp(-Q_2)\frac{Q_2^q}{q!}},
\nonumber\\
Q_1 &\equiv \frac{1}{4\sigma^2}\bk{\check \theta_1- \theta_1 + \frac{\theta_2}{2}}^2,
\nonumber\\
Q_2 &\equiv \frac{1}{4\sigma^2}\bk{\check \theta_1 - \theta_1 - \frac{\theta_2}{2}}^2.
\end{align}
Define the level of misalignment as
\begin{align}
\xi &\equiv \frac{|\check \theta_1-\theta_1|}{\sigma}.
      \label{xi}
\end{align}
We treat $\xi$ as a systematic error and $\theta_2$ as the parameter
of interest for SPADE. Figure~\ref{misalign} plots the resulting
Fisher information for several levels of misalignment. It can be seen
that the information degrades with misalignment, but appreciable
enhancements over direct imaging are still present even if
$\theta_2 \ll \sigma$ and the wavefunction overlap is significant, as
long as $\xi \ll 1$. Appendix~\ref{monte_carlo} confirms this
conclusion numerically for finite photon numbers.

To attain a tolerable level of misalignment, $\theta_1$ first needs to
be estimated and $\check \theta_1$ should be aligned with the
estimate. With conventional imaging, the centroid estimation error is
near-optimal and on the order of $\sigma/\sqrt{N}$ in terms of the
root-mean-square value, meaning that the number of extra photons $N_1$
needed to attain $\xi$ is roughly
\begin{align}
N_1 &\sim \frac{1}{\xi^2}.
\label{N1}
\end{align}
An even more realistic analysis would take $\check \theta_1$ to be a stochastic
waveform determined by the centroid measurements and the adaptive
alignment control \cite{wiseman_milburn}.

\begin{figure}[htbp!]
\includegraphics[width=0.45\textwidth]{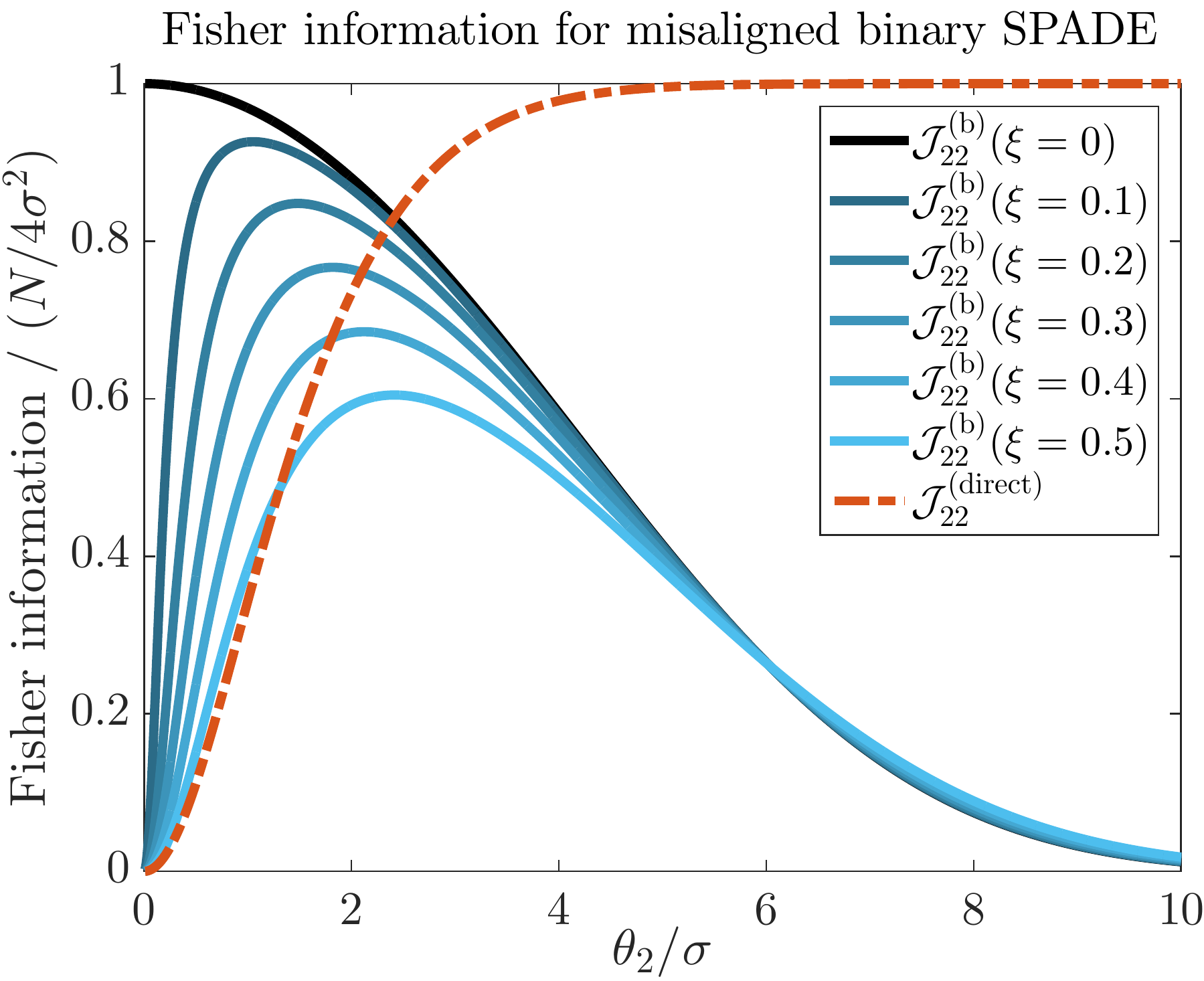}
\caption{\label{misalign_bin}Fisher information for separation
  estimation with binary SPADE with misalignment
  levels $\xi = 0,0.1,\dots,0.5$ (solid curves) and direct imaging
  (dash-dotted curve). The different solid curves can be distinguished
  by their decreasing values with larger misalignments.}
\end{figure}

For binary SPADE, Eqs.~(\ref{P1q0}) and (\ref{P1qhigh}) should be
generalized to
\begin{align}
P_1(q= 0) &\approx \frac{1}{2} \Bk{\exp(-Q_1)+\exp(-Q_2)},
\\
P_1(q> 0) &\approx 1-\frac{1}{2}\exp(-Q_1)-\frac{1}{2}\exp(-Q_2).
\end{align}
Figure~\ref{misalign_bin} plots the Fisher information for misaligned
binary SPADE, showing a similar degradation behavior to that in
Fig.~\ref{misalign} for nonzero $\xi$. Significant improvements over
direct imaging are still possible for small separations and
$\xi \ll 1$. 

\begin{figure}[htbp!]
\includegraphics[width=0.45\textwidth]{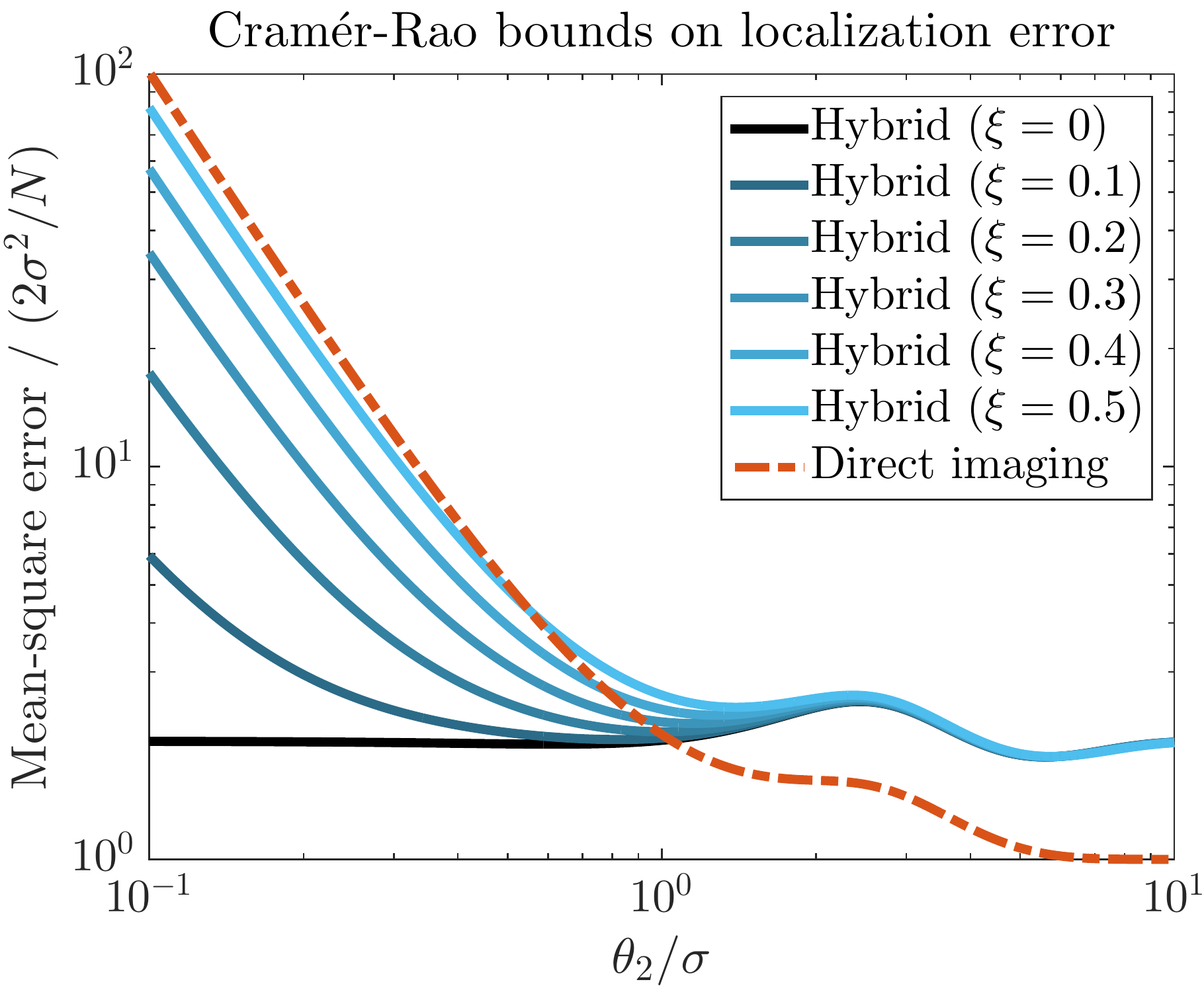}
\caption{\label{hybrid_crb}Cram\'er-Rao bounds on the mean-square
  error of estimating $X_1$ or $X_2$ for a 50-50 hybrid scheme (solid)
  and direct imaging (dash-dotted). Note that the log-log scale is
  used here for clarity, unlike all the other plots in this paper. The
  vertical axis is normalized with respect to the error of locating an
  isolated source with direct imaging.}
\end{figure}

For two-parameter estimation, consider the hybrid scheme in
Fig.~\ref{hybrid_scheme}, assuming 50-50 beam-splitting and binary
SPADE for example. For simplicity, assume that the binary-SPADE output
is used only for separation estimation, such that the total
information matrix with respect to $\theta_1$ and $\theta_2$ remains
diagonal.  Compared with direct imaging, the centroid information for
the hybrid scheme is halved, viz.,
\begin{align}
\mathcal J_{11}^{(\rm hybrid)} &= \frac{\mathcal J_{11}^{(\rm direct)}}{2},
\end{align}
but the separation information gained by binary SPADE can be
appreciable, with
\begin{align}
\mathcal J_{22}^{(\rm hybrid)}  &= 
\frac{\mathcal J_{22}^{(\rm direct)}}{2} + 
\frac{\mathcal J_{22}^{(\rm b)}}{2}.
\end{align}
The net performance of the hybrid scheme can be quantified in terms of
the Cram\'er-Rao bounds for locating $X_1$ and $X_2$. For a diagonal
information matrix $\mathcal J$ with respect to $\theta_1$ and
$\theta_2$, the bound on the mean-square error $\Sigma^{(X)}$ of
estimating either $X_1 = \theta_1-\theta_2/2$ or
$X_2 = \theta_1+\theta_2/2$ is simply
\begin{align}
\Sigma_{ss}^{(X)} &\ge \frac{1}{\mathcal J_{11}}+
\frac{1}{4\mathcal J_{22}},
\quad
s = 1,2,
\end{align}
which demonstrates the detrimental effect of small $\mathcal J_{22}$
for localization. Figure~\ref{hybrid_crb} compares the localization
bounds for the hybrid scheme and direct imaging in log-log scale.  For
small separations, it can be seen that the increased separation
information in the hybrid scheme more than compensates for the reduced
centroid information and allows localization errors substantially
lower than those for direct imaging. With a higher $N_1 = N/2$, more
accurate centroid information from the imaging port can be used to
reduce the misalignment at the SPADE port, and performance converging
to the ideal $\xi = 0$ case in Fig.~\ref{hybrid_crb} can be expected
for high $N$.

\section{\label{monte_carlo}Monte Carlo analysis}
To confirm that the classical Cram\'er-Rao bounds satisfactorily
represent the actual performance of SPADE for finite photon numbers,
here we simulate the device output data numerically, apply
maximum-likelihood estimation, and investigate the resulting error.
To refine our error analysis, we condition our results on the total
number of detected photons $L$, which is obtained after an experiment,
rather than the average photon number $N$ \cite{bettens}. It is not
difficult to show that, conditioned on $L$, the classical and quantum
Fisher information retain their expressions except that $N$ is
replaced by $L$. The error bounds become
\begin{align}
\frac{1}{\mathcal J_{22}'^{(\rm HG)}} 
&\approx \frac{1}{\mathcal K_{22}'}
\approx \frac{4\sigma^2}{L}.
\label{crb_L}
\end{align}
It can also be shown that the sufficient statistic $\sum_q q m_q$ in
the maximum-likelihood estimatior for SPADE in Eq.~(\ref{QML}) is
Poisson with mean $LQ$, so it is simple to generate samples of the
maximum-likelihood estimates $\check Q_{\rm ML}$ and
$\check \theta_{2\rm ML}$ according to Eq.~(\ref{QML}).  

\begin{figure}[htbp!]
\includegraphics[width=0.45\textwidth]{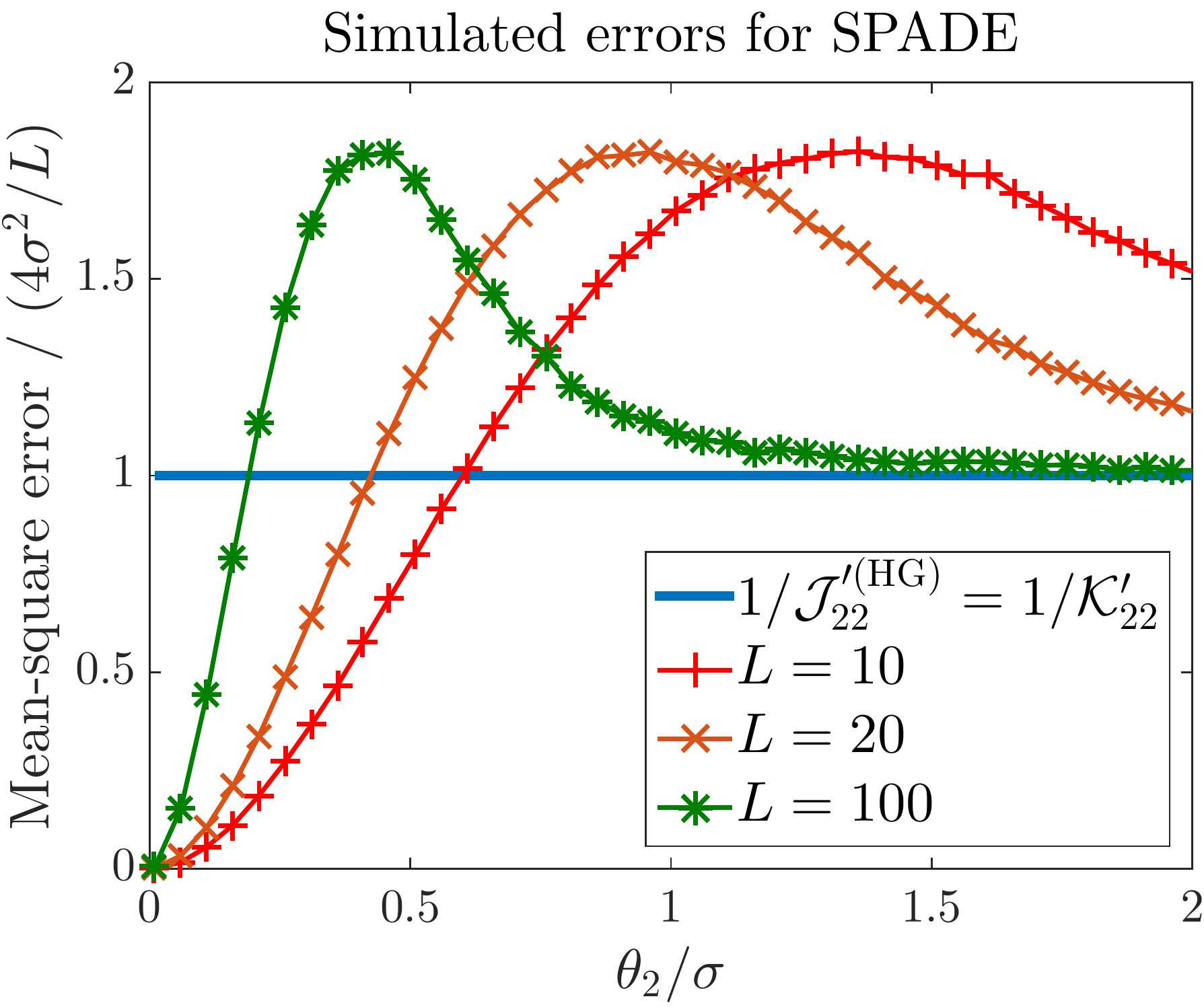}
\caption{\label{mc_errors}Simulated mean-square errors for SPADE with
  maximum-likelihood estimation, conditioned on $L$ detected photons.
  Note that the vertical axis is normalized with respect to the
  Cram\'er-Rao bounds $4\sigma^2/L$, so the plotted values are the
  actual errors magnified by $L/(4\sigma^2)$. Each error is computed
  by averaging $10^5$ simulations, and the lines connecting the data
  points are guides for eyes.}
\end{figure}

Figure~\ref{mc_errors} plots the simulated mean-square errors,
normalized with respect to Eq.~(\ref{crb_L}), for several values of
$L$.  It is intriguing to see that, as $\theta_2/\sigma \to 0$, the
errors go below the bounds. This is a well known statistical
phenomenon called superefficiency \cite{cox79,vaart97}, as the
maximum-likelihood estimator here is actually biased for finite
samples, and the simple Cram\'er-Rao bounds considered here need not
apply. In asymptotic frequentist statistics, superefficiency is not
regarded as an important idea \cite{cox79}, because a superefficient
estimator can beat the Cram\'er-Rao bound only on a set of points with
zero measure in the asymptotic limit \cite{cox79,vaart97}, suggesting
that any region of superefficiency should shrink for larger samples,
as also shown in Fig.~\ref{mc_errors}, and its usefulness is
increasingly limited.  A Bayesian version of the Cram\'er-Rao bound
\cite{vantrees} can also be used to bound the global or minimax error
of any biased or unbiased estimator; the Fisher information still
plays a decisive role in the Bayesian bound and its significance as a
precision measure remains strong in Bayesian and minimax statistics
\cite{tsang16}.

For our present purpose, the main point of Fig.~\ref{mc_errors} is
that the errors remain less than twice the Cram\'er-Rao bound \emph{at
  worst} and even offer the pleasant surprise of superefficiency for
small separations. The overall closeness of the errors to the
Cram\'er-Rao bounds supports our use of the Fisher information to
represent the performance of SPADE.

\begin{figure}[htbp!]
\includegraphics[width=0.45\textwidth]{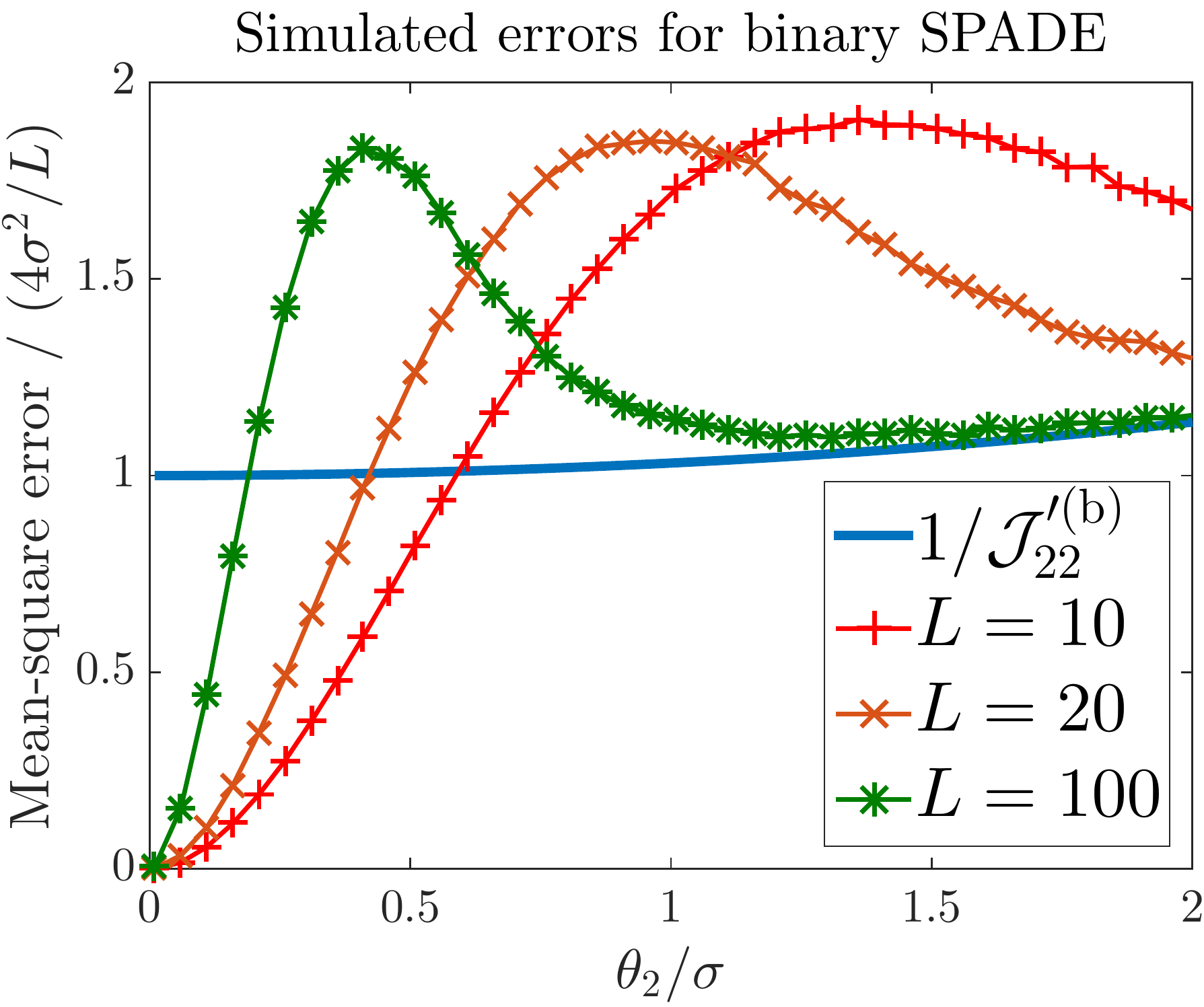}
\caption{\label{mc_errors_bspade}Simulated mean-square errors for
  binary SPADE with maximum-likelihood estimation, conditioned on $L$
  detected photons.  Note that the vertical axis is normalized with
  respect to $4\sigma^2/L$, so the plotted values are the actual
  errors magnified by $L/(4\sigma^2)$. Each error is computed by
  averaging $10^5$ simulations, and the lines connecting the data
  points are guides for eyes.}
\end{figure}

For binary SPADE, the Fisher information conditioned on $L$ has the
same form as Eq.~(\ref{fi_bspade}), and the Cram\'er-Rao bound can be
expressed as
\begin{align}
\frac{1}{\mathcal J_{22}'^{(\rm b)}} &\approx \frac{4\sigma^2}{L}
\frac{1-\exp(-Q)}{Q\exp(-Q)}.
\label{crb_bspade}
\end{align}
The sufficient statistic $m_0$ in $\check\theta_{2\rm ML}^{(\rm b)}$
given by Eq.~(\ref{ML_bspade}) is binomial and also simple to
generate. In case $m_0 = 0$, we set $\check\theta_2 = 2\sigma$, the
maximum of our considered range of
$\theta_2$. Figure~\ref{mc_errors_bspade} plots the simulated
mean-square errors for binary SPADE with otherwise the same parameters
as those for Fig.~\ref{mc_errors}. For small $\theta_2/\sigma$, the
errors follow very similar trends as their counterparts in
Fig.~\ref{mc_errors}, and for larger $\theta_2/\sigma$ the errors
begin to follow the rising Cram\'er-Rao bound according to
Eq.~(\ref{crb_bspade}). This supports our use of the Fisher
information to represent the performance of binary SPADE.

To investigate the effect of misalignment described in
Appendix~\ref{misalignment}, Fig.~\ref{misalign_mc} plots the
simulated errors for binary SPADE with a misalignment level defined in
Eq.~(\ref{xi}) given by $\xi = 0.1$. The overhead photon number
required to achieve $\xi = 0.1$ is $N_1\sim 100$ according to
Eq.~(\ref{N1}) and negligible if $L\gg N_1$. Since $\xi$ is unknown in
reality, the maximum-likelihood estimator used in the simulations
assumes zero misalignment for simplicity. Despite the model mismatch,
the errors remain close to the Cram\'er-Rao bound, especially for
larger $L$, and substantially below the bound for direct imaging.

\begin{figure}[htbp!]
\includegraphics[width=0.45\textwidth]{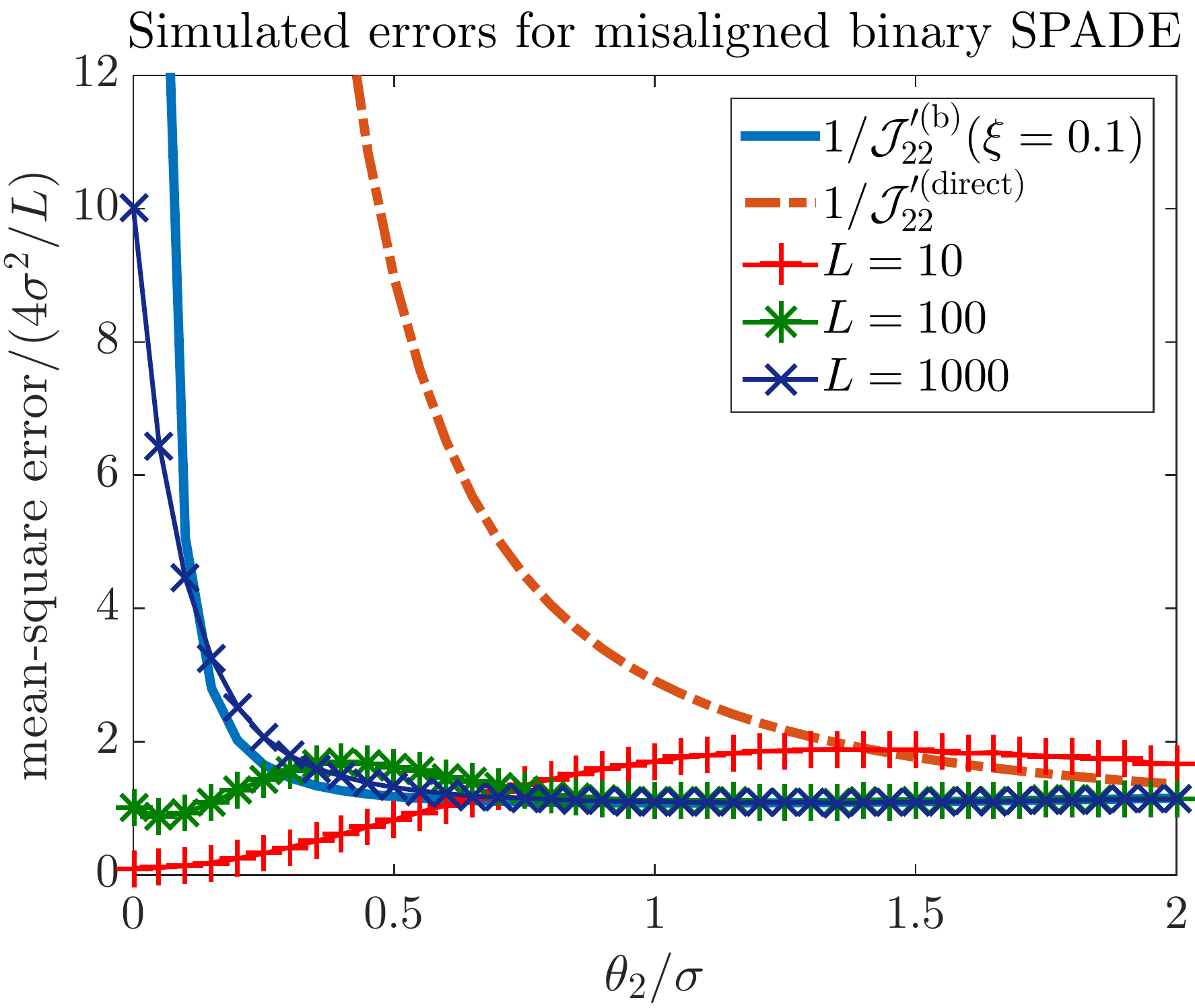}
\caption{\label{misalign_mc}Simulated mean-square errors of binary
  SPADE with misalignment level $\xi = 0.1$, conditioned on $L$
  detected photons. The maximum-likelihood estimator that assumes no
  misalignment is used. Note that the vertical axis is normalized with
  respect to $4\sigma^2/L$. Each error is computed by averaging
  $10^5$ simulations, and the lines connecting the data points are
  guides for eyes. The errors are substantially below the Cram\'er-Rao
  bound for direct imaging (dash-dotted curve).}
\end{figure}

For a given $N$, $L$ has a mean $M\epsilon = N$ and standard deviation
$\sqrt{M\epsilon(1-\epsilon)} \approx \sqrt{N}$. This means that the
distribution of $L$ becomes increasingly sharp around the mean at
$L = N$ for large $N$, and we can expect the performance for a given
$L = N$ to be an increasingly accurate approximation of the average
performance in the given-$N$, random-$L$ scenario.



\bibliography{research}

\begin{thebibliography}{92}%
\makeatletter
\providecommand \@ifxundefined [1]{%
 \@ifx{#1\undefined}
}%
\providecommand \@ifnum [1]{%
 \ifnum #1\expandafter \@firstoftwo
 \else \expandafter \@secondoftwo
 \fi
}%
\providecommand \@ifx [1]{%
 \ifx #1\expandafter \@firstoftwo
 \else \expandafter \@secondoftwo
 \fi
}%
\providecommand \natexlab [1]{#1}%
\providecommand \enquote  [1]{``#1''}%
\providecommand \bibnamefont  [1]{#1}%
\providecommand \bibfnamefont [1]{#1}%
\providecommand \citenamefont [1]{#1}%
\providecommand \href@noop [0]{\@secondoftwo}%
\providecommand \href [0]{\begingroup \@sanitize@url \@href}%
\providecommand \@href[1]{\@@startlink{#1}\@@href}%
\providecommand \@@href[1]{\endgroup#1\@@endlink}%
\providecommand \@sanitize@url [0]{\catcode `\\12\catcode `\$12\catcode
  `\&12\catcode `\#12\catcode `\^12\catcode `\_12\catcode `\%12\relax}%
\providecommand \@@startlink[1]{}%
\providecommand \@@endlink[0]{}%
\providecommand \url  [0]{\begingroup\@sanitize@url \@url }%
\providecommand \@url [1]{\endgroup\@href {#1}{\urlprefix }}%
\providecommand \urlprefix  [0]{URL }%
\providecommand \Eprint [0]{\href }%
\providecommand \doibase [0]{http://dx.doi.org/}%
\providecommand \selectlanguage [0]{\@gobble}%
\providecommand \bibinfo  [0]{\@secondoftwo}%
\providecommand \bibfield  [0]{\@secondoftwo}%
\providecommand \translation [1]{[#1]}%
\providecommand \BibitemOpen [0]{}%
\providecommand \bibitemStop [0]{}%
\providecommand \bibitemNoStop [0]{.\EOS\space}%
\providecommand \EOS [0]{\spacefactor3000\relax}%
\providecommand \BibitemShut  [1]{\csname bibitem#1\endcsname}%
\let\auto@bib@innerbib\@empty
\bibitem [{\citenamefont {Rayleigh}(1879)}]{rayleigh}%
  \BibitemOpen
  \bibfield  {author} {\bibinfo {author} {\bibfnamefont {Lord}\ \bibnamefont
  {Rayleigh}},\ }\bibfield  {title} {\enquote {\bibinfo {title} {{XXXI.
  Investigations in optics, with special reference to the spectroscope}},}\
  }\href {\doibase 10.1080/14786447908639684} {\bibfield  {journal} {\bibinfo
  {journal} {Philosophical Magazine Series 5}\ }\textbf {\bibinfo {volume}
  {8}},\ \bibinfo {pages} {261--274} (\bibinfo {year} {1879})}\BibitemShut
  {NoStop}%
\bibitem [{\citenamefont {Born}\ and\ \citenamefont {Wolf}(1999)}]{born_wolf}%
  \BibitemOpen
  \bibfield  {author} {\bibinfo {author} {\bibfnamefont {Max}\ \bibnamefont
  {Born}}\ and\ \bibinfo {author} {\bibfnamefont {Emil}\ \bibnamefont {Wolf}},\
  }\href@noop {} {\emph {\bibinfo {title} {Principles of Optics:
  Electromagnetic Theory of Propagation, Interference and Diffraction of
  Light}}}\ (\bibinfo  {publisher} {Cambridge University Press},\ \bibinfo
  {address} {Cambridge},\ \bibinfo {year} {1999})\BibitemShut {NoStop}%
\bibitem [{\citenamefont {Hell}\ and\ \citenamefont {Wichmann}(1994)}]{hell94}%
  \BibitemOpen
  \bibfield  {author} {\bibinfo {author} {\bibfnamefont {Stefan~W.}\
  \bibnamefont {Hell}}\ and\ \bibinfo {author} {\bibfnamefont {Jan}\
  \bibnamefont {Wichmann}},\ }\bibfield  {title} {\enquote {\bibinfo {title}
  {Breaking the diffraction resolution limit by stimulated emission:
  stimulated-emission-depletion fluorescence microscopy},}\ }\href {\doibase
  10.1364/OL.19.000780} {\bibfield  {journal} {\bibinfo  {journal} {Opt.
  Lett.}\ }\textbf {\bibinfo {volume} {19}},\ \bibinfo {pages} {780--782}
  (\bibinfo {year} {1994})}\BibitemShut {NoStop}%
\bibitem [{\citenamefont {Betzig}(1995)}]{betzig95}%
  \BibitemOpen
  \bibfield  {author} {\bibinfo {author} {\bibfnamefont {Eric}\ \bibnamefont
  {Betzig}},\ }\bibfield  {title} {\enquote {\bibinfo {title} {Proposed method
  for molecular optical imaging},}\ }\href {\doibase 10.1364/OL.20.000237}
  {\bibfield  {journal} {\bibinfo  {journal} {Opt. Lett.}\ }\textbf {\bibinfo
  {volume} {20}},\ \bibinfo {pages} {237--239} (\bibinfo {year}
  {1995})}\BibitemShut {NoStop}%
\bibitem [{\citenamefont {Hell}(2007)}]{hell}%
  \BibitemOpen
  \bibfield  {author} {\bibinfo {author} {\bibfnamefont {Stefan~W.}\
  \bibnamefont {Hell}},\ }\bibfield  {title} {\enquote {\bibinfo {title}
  {Far-field optical nanoscopy},}\ }\href {\doibase 10.1126/science.1137395}
  {\bibfield  {journal} {\bibinfo  {journal} {Science}\ }\textbf {\bibinfo
  {volume} {316}},\ \bibinfo {pages} {1153--1158} (\bibinfo {year}
  {2007})}\BibitemShut {NoStop}%
\bibitem [{\citenamefont {Betzig}\ \emph {et~al.}(2006)\citenamefont {Betzig},
  \citenamefont {Patterson}, \citenamefont {Sougrat}, \citenamefont
  {Lindwasser}, \citenamefont {Olenych}, \citenamefont {Bonifacino},
  \citenamefont {Davidson}, \citenamefont {Lippincott-Schwartz},\ and\
  \citenamefont {Hess}}]{betzig}%
  \BibitemOpen
  \bibfield  {author} {\bibinfo {author} {\bibfnamefont {Eric}\ \bibnamefont
  {Betzig}}, \bibinfo {author} {\bibfnamefont {George~H.}\ \bibnamefont
  {Patterson}}, \bibinfo {author} {\bibfnamefont {Rachid}\ \bibnamefont
  {Sougrat}}, \bibinfo {author} {\bibfnamefont {O.~Wolf}\ \bibnamefont
  {Lindwasser}}, \bibinfo {author} {\bibfnamefont {Scott}\ \bibnamefont
  {Olenych}}, \bibinfo {author} {\bibfnamefont {Juan~S.}\ \bibnamefont
  {Bonifacino}}, \bibinfo {author} {\bibfnamefont {Michael~W.}\ \bibnamefont
  {Davidson}}, \bibinfo {author} {\bibfnamefont {Jennifer}\ \bibnamefont
  {Lippincott-Schwartz}}, \ and\ \bibinfo {author} {\bibfnamefont {Harald~F.}\
  \bibnamefont {Hess}},\ }\bibfield  {title} {\enquote {\bibinfo {title}
  {Imaging intracellular fluorescent proteins at nanometer resolution},}\
  }\href {\doibase 10.1126/science.1127344} {\bibfield  {journal} {\bibinfo
  {journal} {Science}\ }\textbf {\bibinfo {volume} {313}},\ \bibinfo {pages}
  {1642--1645} (\bibinfo {year} {2006})}\BibitemShut {NoStop}%
\bibitem [{\citenamefont {Moerner}(2007)}]{moerner}%
  \BibitemOpen
  \bibfield  {author} {\bibinfo {author} {\bibfnamefont {William~E.}\
  \bibnamefont {Moerner}},\ }\bibfield  {title} {\enquote {\bibinfo {title}
  {New directions in single-molecule imaging and analysis},}\ }\href {\doibase
  10.1073/pnas.0610081104} {\bibfield  {journal} {\bibinfo  {journal}
  {Proceedings of the National Academy of Sciences}\ }\textbf {\bibinfo
  {volume} {104}},\ \bibinfo {pages} {12596--12602} (\bibinfo {year}
  {2007})}\BibitemShut {NoStop}%
\bibitem [{\citenamefont {Bettens}\ \emph {et~al.}(1999)\citenamefont
  {Bettens}, \citenamefont {Van~Dyck}, \citenamefont {den Dekker},
  \citenamefont {Sijbers},\ and\ \citenamefont {van~den Bos}}]{bettens}%
  \BibitemOpen
  \bibfield  {author} {\bibinfo {author} {\bibfnamefont {E.}~\bibnamefont
  {Bettens}}, \bibinfo {author} {\bibfnamefont {D.}~\bibnamefont {Van~Dyck}},
  \bibinfo {author} {\bibfnamefont {A.~J.}\ \bibnamefont {den Dekker}},
  \bibinfo {author} {\bibfnamefont {J.}~\bibnamefont {Sijbers}}, \ and\
  \bibinfo {author} {\bibfnamefont {A.}~\bibnamefont {van~den Bos}},\
  }\bibfield  {title} {\enquote {\bibinfo {title} {Model-based two-object
  resolution from observations having counting statistics},}\ }\href {\doibase
  http://dx.doi.org/10.1016/S0304-3991(99)00006-6} {\bibfield  {journal}
  {\bibinfo  {journal} {Ultramicroscopy}\ }\textbf {\bibinfo {volume} {77}},\
  \bibinfo {pages} {37--48} (\bibinfo {year} {1999})}\BibitemShut {NoStop}%
\bibitem [{\citenamefont {Van~Aert}\ \emph {et~al.}(2002)\citenamefont
  {Van~Aert}, \citenamefont {den Dekker}, \citenamefont {Van~Dyck},\ and\
  \citenamefont {van~den Bos}}]{vanaert}%
  \BibitemOpen
  \bibfield  {author} {\bibinfo {author} {\bibfnamefont {S.}~\bibnamefont
  {Van~Aert}}, \bibinfo {author} {\bibfnamefont {A.~J.}\ \bibnamefont {den
  Dekker}}, \bibinfo {author} {\bibfnamefont {D.}~\bibnamefont {Van~Dyck}}, \
  and\ \bibinfo {author} {\bibfnamefont {A.}~\bibnamefont {van~den Bos}},\
  }\bibfield  {title} {\enquote {\bibinfo {title} {High-resolution electron
  microscopy and electron tomography: resolution versus precision},}\ }\href
  {\doibase http://dx.doi.org/10.1016/S1047-8477(02)00016-3} {\bibfield
  {journal} {\bibinfo  {journal} {Journal of Structural Biology}\ }\textbf
  {\bibinfo {volume} {138}},\ \bibinfo {pages} {21--33} (\bibinfo {year}
  {2002})}\BibitemShut {NoStop}%
\bibitem [{\citenamefont {Ram}\ \emph {et~al.}(2006)\citenamefont {Ram},
  \citenamefont {Ward},\ and\ \citenamefont {Ober}}]{ram}%
  \BibitemOpen
  \bibfield  {author} {\bibinfo {author} {\bibfnamefont {Sripad}\ \bibnamefont
  {Ram}}, \bibinfo {author} {\bibfnamefont {E.~Sally}\ \bibnamefont {Ward}}, \
  and\ \bibinfo {author} {\bibfnamefont {Raimund~J.}\ \bibnamefont {Ober}},\
  }\bibfield  {title} {\enquote {\bibinfo {title} {Beyond {R}ayleigh's
  criterion: A resolution measure with application to single-molecule
  microscopy},}\ }\href {\doibase 10.1073/pnas.0508047103} {\bibfield
  {journal} {\bibinfo  {journal} {Proceedings of the National Academy of
  Sciences of the United States of America}\ }\textbf {\bibinfo {volume}
  {103}},\ \bibinfo {pages} {4457--4462} (\bibinfo {year} {2006})}\BibitemShut
  {NoStop}%
\bibitem [{\citenamefont {Van~Trees}(2001)}]{vantrees}%
  \BibitemOpen
  \bibfield  {author} {\bibinfo {author} {\bibfnamefont {Harry~L.}\
  \bibnamefont {Van~Trees}},\ }\href@noop {} {\emph {\bibinfo {title}
  {Detection, Estimation, and Modulation Theory, Part I.}}}\ (\bibinfo
  {publisher} {John Wiley \& Sons},\ \bibinfo {address} {New York},\ \bibinfo
  {year} {2001})\BibitemShut {NoStop}%
\bibitem [{\citenamefont {Pawley}(2006)}]{pawley}%
  \BibitemOpen
  \bibinfo {editor} {\bibfnamefont {James~B.}\ \bibnamefont {Pawley}},\ ed.,\
  \href {\doibase 10.1007/978-0-387-45524-2} {\emph {\bibinfo {title} {Handbook
  of Biological Confocal Microscopy}}}\ (\bibinfo  {publisher} {Springer},\
  \bibinfo {address} {New York},\ \bibinfo {year} {2006})\BibitemShut {NoStop}%
\bibitem [{\citenamefont {Zmuidzinas}(2003)}]{zmuidzinas03}%
  \BibitemOpen
  \bibfield  {author} {\bibinfo {author} {\bibfnamefont {Jonas}\ \bibnamefont
  {Zmuidzinas}},\ }\bibfield  {title} {\enquote {\bibinfo {title}
  {{C}ram\'{e}r--{R}ao sensitivity limits for astronomical instruments:
  implications for interferometer design},}\ }\href {\doibase
  10.1364/JOSAA.20.000218} {\bibfield  {journal} {\bibinfo  {journal} {J. Opt.
  Soc. Am. A}\ }\textbf {\bibinfo {volume} {20}},\ \bibinfo {pages} {218--233}
  (\bibinfo {year} {2003})}\BibitemShut {NoStop}%
\bibitem [{\citenamefont {Labeyrie}\ \emph {et~al.}(2006)\citenamefont
  {Labeyrie}, \citenamefont {Lipson},\ and\ \citenamefont
  {Nisenson}}]{labeyrie}%
  \BibitemOpen
  \bibfield  {author} {\bibinfo {author} {\bibfnamefont {Antoine}\ \bibnamefont
  {Labeyrie}}, \bibinfo {author} {\bibfnamefont {Stephen~G.}\ \bibnamefont
  {Lipson}}, \ and\ \bibinfo {author} {\bibfnamefont {Peter}\ \bibnamefont
  {Nisenson}},\ }\href {http://dx.doi.org/10.1017/CBO9780511617638} {\emph
  {\bibinfo {title} {An Introduction to Optical Stellar Interferometry}}}\
  (\bibinfo  {publisher} {Cambridge University Press},\ \bibinfo {address}
  {Cambridge},\ \bibinfo {year} {2006})\BibitemShut {NoStop}%
\bibitem [{\citenamefont {Howell}(2006)}]{howell06}%
  \BibitemOpen
  \bibfield  {author} {\bibinfo {author} {\bibfnamefont {Steve~B.}\
  \bibnamefont {Howell}},\ }\href@noop {} {\emph {\bibinfo {title} {Handbook of
  CCD Astronomy}}}\ (\bibinfo  {publisher} {Cambridge University Press},\
  \bibinfo {address} {Cambridge},\ \bibinfo {year} {2006})\BibitemShut
  {NoStop}%
\bibitem [{\citenamefont {Huber}\ \emph {et~al.}(2013)\citenamefont {Huber},
  \citenamefont {Pauluhn}, \citenamefont {Culhane}, \citenamefont {Timothy},
  \citenamefont {Wilhelm},\ and\ \citenamefont {Zehnder}}]{huber}%
  \BibitemOpen
  \bibinfo {editor} {\bibfnamefont {Martin C.~E.}\ \bibnamefont {Huber}},
  \bibinfo {editor} {\bibfnamefont {Anuschka}\ \bibnamefont {Pauluhn}},
  \bibinfo {editor} {\bibfnamefont {J.~Len}\ \bibnamefont {Culhane}}, \bibinfo
  {editor} {\bibfnamefont {J.~Gethyn}\ \bibnamefont {Timothy}}, \bibinfo
  {editor} {\bibfnamefont {Klaus}\ \bibnamefont {Wilhelm}}, \ and\ \bibinfo
  {editor} {\bibfnamefont {Alex}\ \bibnamefont {Zehnder}},\ eds.,\ \href
  {\doibase 10.1007/978-1-4614-7804-1_16} {\emph {\bibinfo {title} {Observing
  Photons in Space: A Guide to Experimental Space Astronomy}}}\ (\bibinfo
  {publisher} {Springer},\ \bibinfo {address} {New York},\ \bibinfo {year}
  {2013})\ Chap.\ \bibinfo {chapter} {High-accuracy positioning: astrometry},
  pp.\ \bibinfo {pages} {299--311}\BibitemShut {NoStop}%
\bibitem [{\citenamefont {Helstrom}(1976)}]{helstrom}%
  \BibitemOpen
  \bibfield  {author} {\bibinfo {author} {\bibfnamefont {Carl~W.}\ \bibnamefont
  {Helstrom}},\ }\href@noop {} {\emph {\bibinfo {title} {Quantum Detection and
  Estimation Theory}}}\ (\bibinfo  {publisher} {Academic Press},\ \bibinfo
  {address} {New York},\ \bibinfo {year} {1976})\BibitemShut {NoStop}%
\bibitem [{\citenamefont {Giovannetti}\ \emph {et~al.}(2011)\citenamefont
  {Giovannetti}, \citenamefont {Lloyd},\ and\ \citenamefont
  {Maccone}}]{glm2011}%
  \BibitemOpen
  \bibfield  {author} {\bibinfo {author} {\bibfnamefont {Vittorio}\
  \bibnamefont {Giovannetti}}, \bibinfo {author} {\bibfnamefont {Seth}\
  \bibnamefont {Lloyd}}, \ and\ \bibinfo {author} {\bibfnamefont {Lorenzo}\
  \bibnamefont {Maccone}},\ }\bibfield  {title} {\enquote {\bibinfo {title}
  {Advances in quantum metrology},}\ }\href
  {http://dx.doi.org/10.1038/nphoton.2011.35} {\bibfield  {journal} {\bibinfo
  {journal} {Nature Photon.}\ }\textbf {\bibinfo {volume} {5}},\ \bibinfo
  {pages} {222--229} (\bibinfo {year} {2011})}\BibitemShut {NoStop}%
\bibitem [{\citenamefont {Tsang}(2015)}]{localization}%
  \BibitemOpen
  \bibfield  {author} {\bibinfo {author} {\bibfnamefont {Mankei}\ \bibnamefont
  {Tsang}},\ }\bibfield  {title} {\enquote {\bibinfo {title} {Quantum limits to
  optical point-source localization},}\ }\href {\doibase
  10.1364/OPTICA.2.000646} {\bibfield  {journal} {\bibinfo  {journal} {Optica}\
  }\textbf {\bibinfo {volume} {2}},\ \bibinfo {pages} {646--653} (\bibinfo
  {year} {2015})}\BibitemShut {NoStop}%
\bibitem [{\citenamefont {Hayashi}(2005)}]{hayashi05}%
  \BibitemOpen
  \bibinfo {editor} {\bibfnamefont {Masahito}\ \bibnamefont {Hayashi}},\ ed.,\
  \href {\doibase 10.1142/9789812563071} {\emph {\bibinfo {title} {Asymptotic
  Theory of Quantum Statistical Inference: Selected Papers}}}\ (\bibinfo
  {publisher} {World Scientific},\ \bibinfo {address} {Singapore},\ \bibinfo
  {year} {2005})\BibitemShut {NoStop}%
\bibitem [{\citenamefont {Fujiwara}(2006)}]{fujiwara2006}%
  \BibitemOpen
  \bibfield  {author} {\bibinfo {author} {\bibfnamefont {Akio}\ \bibnamefont
  {Fujiwara}},\ }\bibfield  {title} {\enquote {\bibinfo {title} {Strong
  consistency and asymptotic efficiency for adaptive quantum estimation
  problems},}\ }\href {http://stacks.iop.org/0305-4470/39/i=40/a=014}
  {\bibfield  {journal} {\bibinfo  {journal} {Journal of Physics A:
  Mathematical and General}\ }\textbf {\bibinfo {volume} {39}},\ \bibinfo
  {pages} {12489} (\bibinfo {year} {2006})}\BibitemShut {NoStop}%
\bibitem [{\citenamefont {Helstrom}(1973)}]{helstrom73b}%
  \BibitemOpen
  \bibfield  {author} {\bibinfo {author} {\bibfnamefont {Carl~W.}\ \bibnamefont
  {Helstrom}},\ }\bibfield  {title} {\enquote {\bibinfo {title} {Resolution of
  point sources of light as analyzed by quantum detection theory},}\ }\href
  {\doibase 10.1109/TIT.1973.1055052} {\bibfield  {journal} {\bibinfo
  {journal} {IEEE Transactions on Information Theory}\ }\textbf {\bibinfo
  {volume} {19}},\ \bibinfo {pages} {389--398} (\bibinfo {year}
  {1973})}\BibitemShut {NoStop}%
\bibitem [{\citenamefont {Goodman}(1985)}]{goodman_stat}%
  \BibitemOpen
  \bibfield  {author} {\bibinfo {author} {\bibfnamefont {Joseph~W.}\
  \bibnamefont {Goodman}},\ }\href@noop {} {\emph {\bibinfo {title}
  {Statistical Optics}}}\ (\bibinfo  {publisher} {Wiley},\ \bibinfo {address}
  {New York},\ \bibinfo {year} {1985})\BibitemShut {NoStop}%
\bibitem [{\citenamefont {Mandel}\ and\ \citenamefont {Wolf}(1995)}]{mandel}%
  \BibitemOpen
  \bibfield  {author} {\bibinfo {author} {\bibfnamefont {Leonard}\ \bibnamefont
  {Mandel}}\ and\ \bibinfo {author} {\bibfnamefont {Emil}\ \bibnamefont
  {Wolf}},\ }\href {\doibase 10.1017/CBO9781139644105} {\emph {\bibinfo {title}
  {Optical Coherence and Quantum Optics}}}\ (\bibinfo  {publisher} {Cambridge
  University Press},\ \bibinfo {address} {Cambridge},\ \bibinfo {year}
  {1995})\BibitemShut {NoStop}%
\bibitem [{\citenamefont {Mandel}(1959)}]{mandel59}%
  \BibitemOpen
  \bibfield  {author} {\bibinfo {author} {\bibfnamefont {Leonard}\ \bibnamefont
  {Mandel}},\ }\bibfield  {title} {\enquote {\bibinfo {title} {Fluctuations of
  photon beams: The distribution of the photo-electrons},}\ }\href
  {http://stacks.iop.org/0370-1328/74/i=3/a=301} {\bibfield  {journal}
  {\bibinfo  {journal} {Proceedings of the Physical Society}\ }\textbf
  {\bibinfo {volume} {74}},\ \bibinfo {pages} {233} (\bibinfo {year}
  {1959})}\BibitemShut {NoStop}%
\bibitem [{\citenamefont {Gottesman}\ \emph {et~al.}(2012)\citenamefont
  {Gottesman}, \citenamefont {Jennewein},\ and\ \citenamefont
  {Croke}}]{gottesman}%
  \BibitemOpen
  \bibfield  {author} {\bibinfo {author} {\bibfnamefont {Daniel}\ \bibnamefont
  {Gottesman}}, \bibinfo {author} {\bibfnamefont {Thomas}\ \bibnamefont
  {Jennewein}}, \ and\ \bibinfo {author} {\bibfnamefont {Sarah}\ \bibnamefont
  {Croke}},\ }\bibfield  {title} {\enquote {\bibinfo {title} {Longer-baseline
  telescopes using quantum repeaters},}\ }\href {\doibase
  10.1103/PhysRevLett.109.070503} {\bibfield  {journal} {\bibinfo  {journal}
  {Phys. Rev. Lett.}\ }\textbf {\bibinfo {volume} {109}},\ \bibinfo {pages}
  {070503} (\bibinfo {year} {2012})}\BibitemShut {NoStop}%
\bibitem [{\citenamefont {Tsang}(2011)}]{stellar}%
  \BibitemOpen
  \bibfield  {author} {\bibinfo {author} {\bibfnamefont {Mankei}\ \bibnamefont
  {Tsang}},\ }\bibfield  {title} {\enquote {\bibinfo {title} {Quantum
  nonlocality in weak-thermal-light interferometry},}\ }\href {\doibase
  10.1103/PhysRevLett.107.270402} {\bibfield  {journal} {\bibinfo  {journal}
  {Phys. Rev. Lett.}\ }\textbf {\bibinfo {volume} {107}},\ \bibinfo {pages}
  {270402} (\bibinfo {year} {2011})}\BibitemShut {NoStop}%
\bibitem [{\citenamefont {Goodman}(2004)}]{goodman}%
  \BibitemOpen
  \bibfield  {author} {\bibinfo {author} {\bibfnamefont {Joseph~W.}\
  \bibnamefont {Goodman}},\ }\href@noop {} {\emph {\bibinfo {title}
  {Introduction to Fourier Optics}}}\ (\bibinfo  {publisher}
  {Mc{G}raw-{H}ill},\ \bibinfo {address} {New York},\ \bibinfo {year}
  {2004})\BibitemShut {NoStop}%
\bibitem [{\citenamefont {Yuen}\ and\ \citenamefont
  {Shapiro}(1978)}]{yuen_shapiro1}%
  \BibitemOpen
  \bibfield  {author} {\bibinfo {author} {\bibfnamefont {Horace~P.}\
  \bibnamefont {Yuen}}\ and\ \bibinfo {author} {\bibfnamefont {Jeffrey~H.}\
  \bibnamefont {Shapiro}},\ }\bibfield  {title} {\enquote {\bibinfo {title}
  {Optical communication with two-photon coherent states--{P}art {I}:
  Quantum-state propagation and quantum-noise},}\ }\href {\doibase
  10.1109/TIT.1978.1055958} {\bibfield  {journal} {\bibinfo  {journal} {IEEE
  Transactions on Information Theory}\ }\textbf {\bibinfo {volume} {24}},\
  \bibinfo {pages} {657--668} (\bibinfo {year} {1978})}\BibitemShut {NoStop}%
\bibitem [{\citenamefont {Shapiro}(2009)}]{shapiro09}%
  \BibitemOpen
  \bibfield  {author} {\bibinfo {author} {\bibfnamefont {Jeffrey~H.}\
  \bibnamefont {Shapiro}},\ }\bibfield  {title} {\enquote {\bibinfo {title}
  {The quantum theory of optical communications},}\ }\href {\doibase
  10.1109/JSTQE.2009.2024959} {\bibfield  {journal} {\bibinfo  {journal} {IEEE
  Journal of Selected Topics in Quantum Electronics}\ }\textbf {\bibinfo
  {volume} {15}},\ \bibinfo {pages} {1547--1569} (\bibinfo {year}
  {2009})}\BibitemShut {NoStop}%
\bibitem [{\citenamefont {Ober}\ \emph {et~al.}(2004)\citenamefont {Ober},
  \citenamefont {Ram},\ and\ \citenamefont {Ward}}]{ober}%
  \BibitemOpen
  \bibfield  {author} {\bibinfo {author} {\bibfnamefont {Raimund~J.}\
  \bibnamefont {Ober}}, \bibinfo {author} {\bibfnamefont {Sripad}\ \bibnamefont
  {Ram}}, \ and\ \bibinfo {author} {\bibfnamefont {E.~Sally}\ \bibnamefont
  {Ward}},\ }\bibfield  {title} {\enquote {\bibinfo {title} {Localization
  accuracy in single-molecule microscopy},}\ }\href {\doibase
  10.1016/S0006-3495(04)74193-4} {\bibfield  {journal} {\bibinfo  {journal}
  {Biophysical Journal}\ }\textbf {\bibinfo {volume} {86}},\ \bibinfo {pages}
  {1185--1200} (\bibinfo {year} {2004})}\BibitemShut {NoStop}%
\bibitem [{\citenamefont {Deschout}\ \emph {et~al.}(2014)\citenamefont
  {Deschout}, \citenamefont {Zanacchi}, \citenamefont {Mlodzianoski},
  \citenamefont {Diaspro}, \citenamefont {Bewersdorf}, \citenamefont {Hess},\
  and\ \citenamefont {Braeckmans}}]{deschout}%
  \BibitemOpen
  \bibfield  {author} {\bibinfo {author} {\bibfnamefont {Hendrik}\ \bibnamefont
  {Deschout}}, \bibinfo {author} {\bibfnamefont {Francesca~Cella}\ \bibnamefont
  {Zanacchi}}, \bibinfo {author} {\bibfnamefont {Michael}\ \bibnamefont
  {Mlodzianoski}}, \bibinfo {author} {\bibfnamefont {Alberto}\ \bibnamefont
  {Diaspro}}, \bibinfo {author} {\bibfnamefont {Joerg}\ \bibnamefont
  {Bewersdorf}}, \bibinfo {author} {\bibfnamefont {Samuel~T.}\ \bibnamefont
  {Hess}}, \ and\ \bibinfo {author} {\bibfnamefont {Kevin}\ \bibnamefont
  {Braeckmans}},\ }\bibfield  {title} {\enquote {\bibinfo {title} {Precisely
  and accurately localizing single emitters in fluorescence microscopy},}\
  }\href {http://dx.doi.org/10.1038/nmeth.2843} {\bibfield  {journal} {\bibinfo
   {journal} {Nature Methods}\ }\textbf {\bibinfo {volume} {11}},\ \bibinfo
  {pages} {253--266} (\bibinfo {year} {2014})}\BibitemShut {NoStop}%
\bibitem [{\citenamefont {Chao}\ \emph {et~al.}(2016)\citenamefont {Chao},
  \citenamefont {Sally~Ward},\ and\ \citenamefont {Ober}}]{chao16}%
  \BibitemOpen
  \bibfield  {author} {\bibinfo {author} {\bibfnamefont {Jerry}\ \bibnamefont
  {Chao}}, \bibinfo {author} {\bibfnamefont {E.}~\bibnamefont {Sally~Ward}}, \
  and\ \bibinfo {author} {\bibfnamefont {Raimund~J.}\ \bibnamefont {Ober}},\
  }\bibfield  {title} {\enquote {\bibinfo {title} {Fisher information theory
  for parameter estimation in single molecule microscopy: tutorial},}\ }\href
  {\doibase 10.1364/JOSAA.33.000B36} {\bibfield  {journal} {\bibinfo  {journal}
  {Journal of the Optical Society of America A}\ }\textbf {\bibinfo {volume}
  {33}},\ \bibinfo {pages} {B36} (\bibinfo {year} {2016})}\BibitemShut
  {NoStop}%
\bibitem [{imp()}]{implicit}%
  \BibitemOpen
  \href@noop {} {}\bibinfo {note} {For notational simplicity, the dependence of
  mathematical quantities on {$\theta$} is not written out
  explicitly.}\BibitemShut {Stop}%
\bibitem [{\citenamefont {{Lindegren}}(1978)}]{lindegren78}%
  \BibitemOpen
  \bibfield  {author} {\bibinfo {author} {\bibfnamefont {Lennart}\ \bibnamefont
  {{Lindegren}}},\ }\bibfield  {title} {\enquote {\bibinfo {title}
  {{Photoelectric astrometry - A comparison of methods for precise image
  location}},}\ }in\ \href@noop {} {\emph {\bibinfo {booktitle} {IAU Colloq.
  48: Modern Astrometry}}},\ \bibinfo {editor} {edited by\ \bibinfo {editor}
  {\bibfnamefont {F.~V.}\ \bibnamefont {{Prochazka}}}\ and\ \bibinfo {editor}
  {\bibfnamefont {R.~H.}\ \bibnamefont {{Tucker}}}}\ (\bibinfo {year} {1978})\
  pp.\ \bibinfo {pages} {197--217}\BibitemShut {NoStop}%
\bibitem [{\citenamefont {King}(1983)}]{king83}%
  \BibitemOpen
  \bibfield  {author} {\bibinfo {author} {\bibfnamefont {Ivan~R.}\ \bibnamefont
  {King}},\ }\bibfield  {title} {\enquote {\bibinfo {title} {Accuracy of
  measurement of star images on a pixel array},}\ }\href
  {http://www.jstor.org/stable/40678139} {\bibfield  {journal} {\bibinfo
  {journal} {Publications of the Astronomical Society of the Pacific}\ }\textbf
  {\bibinfo {volume} {95}},\ \bibinfo {pages} {163--168} (\bibinfo {year}
  {1983})}\BibitemShut {NoStop}%
\bibitem [{\citenamefont {Yariv}(1989)}]{yariv}%
  \BibitemOpen
  \bibfield  {author} {\bibinfo {author} {\bibfnamefont {Amnon}\ \bibnamefont
  {Yariv}},\ }\href@noop {} {\emph {\bibinfo {title} {Quantum Electronics}}}\
  (\bibinfo  {publisher} {Wiley},\ \bibinfo {address} {New York},\ \bibinfo
  {year} {1989})\BibitemShut {NoStop}%
\bibitem [{\citenamefont {Yariv}\ and\ \citenamefont {Yeh}(2007)}]{yariv_yeh}%
  \BibitemOpen
  \bibfield  {author} {\bibinfo {author} {\bibfnamefont {Amnon}\ \bibnamefont
  {Yariv}}\ and\ \bibinfo {author} {\bibfnamefont {Pochi}\ \bibnamefont
  {Yeh}},\ }\href@noop {} {\emph {\bibinfo {title} {Photonics: Optical
  Electronics in Modern Communications}}}\ (\bibinfo  {publisher} {Oxford
  University Press},\ \bibinfo {address} {New York},\ \bibinfo {year}
  {2007})\BibitemShut {NoStop}%
\bibitem [{\citenamefont {Sorin}\ \emph {et~al.}(1986)\citenamefont {Sorin},
  \citenamefont {Kim},\ and\ \citenamefont {Shaw}}]{sorin}%
  \BibitemOpen
  \bibfield  {author} {\bibinfo {author} {\bibfnamefont {W.~V.}\ \bibnamefont
  {Sorin}}, \bibinfo {author} {\bibfnamefont {B.~Y.}\ \bibnamefont {Kim}}, \
  and\ \bibinfo {author} {\bibfnamefont {H.~J.}\ \bibnamefont {Shaw}},\
  }\bibfield  {title} {\enquote {\bibinfo {title} {Highly selective evanescent
  modal filter for two-mode optical fibers},}\ }\href {\doibase
  10.1364/OL.11.000581} {\bibfield  {journal} {\bibinfo  {journal} {Optics
  Letters}\ }\textbf {\bibinfo {volume} {11}},\ \bibinfo {pages} {581--583}
  (\bibinfo {year} {1986})}\BibitemShut {NoStop}%
\bibitem [{\citenamefont {Townes}(2000)}]{townes}%
  \BibitemOpen
  \bibfield  {author} {\bibinfo {author} {\bibfnamefont {Charles~H.}\
  \bibnamefont {Townes}},\ }\enquote {\bibinfo {title} {Noise and sensitivity
  in interferometry},}\ in\ \href@noop {} {\emph {\bibinfo {booktitle}
  {Principles of Long Baseline Stellar Interferometry}}},\ \bibinfo {editor}
  {edited by\ \bibinfo {editor} {\bibfnamefont {Peter~R.}\ \bibnamefont
  {Lawson}}}\ (\bibinfo {year} {2000})\ Chap.~\bibinfo {chapter} {4}, pp.\
  \bibinfo {pages} {59--70}\BibitemShut {NoStop}%
\bibitem [{\citenamefont {Nair}()}]{nair_unpublished}%
  \BibitemOpen
  \bibfield  {author} {\bibinfo {author} {\bibfnamefont {Ranjith}\ \bibnamefont
  {Nair}},\ }\href@noop {} {}\bibinfo {note} {Unpublished}\BibitemShut
  {NoStop}%
\bibitem [{\citenamefont {Wasserman}(2004)}]{wasserman}%
  \BibitemOpen
  \bibfield  {author} {\bibinfo {author} {\bibfnamefont {Larry~A.}\
  \bibnamefont {Wasserman}},\ }\href@noop {} {\emph {\bibinfo {title} {All of
  Statistics}}}\ (\bibinfo  {publisher} {Springer},\ \bibinfo {address} {New
  York},\ \bibinfo {year} {2004})\BibitemShut {NoStop}%
\bibitem [{\citenamefont {{Ang}}\ \emph {et~al.}(2016)\citenamefont {{Ang}},
  \citenamefont {{Nair}},\ and\ \citenamefont {{Tsang}}}]{ant}%
  \BibitemOpen
  \bibfield  {author} {\bibinfo {author} {\bibfnamefont {Shan~Zheng}\
  \bibnamefont {{Ang}}}, \bibinfo {author} {\bibfnamefont {Ranjith}\
  \bibnamefont {{Nair}}}, \ and\ \bibinfo {author} {\bibfnamefont {Mankei}\
  \bibnamefont {{Tsang}}},\ }\bibfield  {title} {\enquote {\bibinfo {title}
  {{Quantum limit for two-dimensional resolution of two incoherent optical
  point sources}},}\ }\href@noop {} {\bibfield  {journal} {\bibinfo  {journal}
  {ArXiv e-prints}\ } (\bibinfo {year} {2016})},\ \Eprint
  {http://arxiv.org/abs/1606.00603} {arXiv:1606.00603 [quant-ph]} \BibitemShut
  {NoStop}%
\bibitem [{\citenamefont {Becker}(2009)}]{becker}%
  \BibitemOpen
  \bibinfo {editor} {\bibfnamefont {Werner}\ \bibnamefont {Becker}},\ ed.,\
  \href {\doibase 10.1007/978-3-540-76965-1} {\emph {\bibinfo {title} {Neutron
  Stars and Pulsars}}}\ (\bibinfo  {publisher} {Springer},\ \bibinfo {address}
  {Berlin},\ \bibinfo {year} {2009})\BibitemShut {NoStop}%
\bibitem [{\citenamefont {Michalet}\ \emph {et~al.}(2006)\citenamefont
  {Michalet}, \citenamefont {Weiss},\ and\ \citenamefont
  {J{\"a}ger}}]{michalet06}%
  \BibitemOpen
  \bibfield  {author} {\bibinfo {author} {\bibfnamefont {Xavier}\ \bibnamefont
  {Michalet}}, \bibinfo {author} {\bibfnamefont {Shimon}\ \bibnamefont
  {Weiss}}, \ and\ \bibinfo {author} {\bibfnamefont {Marcus}\ \bibnamefont
  {J{\"a}ger}},\ }\bibfield  {title} {\enquote {\bibinfo {title}
  {Single-molecule fluorescence studies of protein folding and conformational
  dynamics},}\ }\href {\doibase 10.1021/cr0404343} {\bibfield  {journal}
  {\bibinfo  {journal} {Chemical Reviews}\ }\textbf {\bibinfo {volume} {106}},\
  \bibinfo {pages} {1785--1813} (\bibinfo {year} {2006})}\BibitemShut {NoStop}%
\bibitem [{\citenamefont {{Tsang}}\ \emph {et~al.}(2015)\citenamefont
  {{Tsang}}, \citenamefont {{Nair}},\ and\ \citenamefont {{Lu}}}]{tnlv1}%
  \BibitemOpen
  \bibfield  {author} {\bibinfo {author} {\bibfnamefont {Mankei}\ \bibnamefont
  {{Tsang}}}, \bibinfo {author} {\bibfnamefont {Ranjith}\ \bibnamefont
  {{Nair}}}, \ and\ \bibinfo {author} {\bibfnamefont {Xiao-Ming}\ \bibnamefont
  {{Lu}}},\ }\bibfield  {title} {\enquote {\bibinfo {title} {{Quantum theory of
  superresolution for two incoherent optical point sources}},}\ }\href@noop {}
  {\bibfield  {journal} {\bibinfo  {journal} {ArXiv e-prints}\ } (\bibinfo
  {year} {2015})},\ \Eprint {http://arxiv.org/abs/1511.00552v1}
  {arXiv:1511.00552v1 [quant-ph]} \BibitemShut {NoStop}%
\bibitem [{\citenamefont {{Tsang}}\ \emph {et~al.}(2016)\citenamefont
  {{Tsang}}, \citenamefont {{Nair}},\ and\ \citenamefont {{Lu}}}]{tnl2}%
  \BibitemOpen
  \bibfield  {author} {\bibinfo {author} {\bibfnamefont {Mankei}\ \bibnamefont
  {{Tsang}}}, \bibinfo {author} {\bibfnamefont {Ranjith}\ \bibnamefont
  {{Nair}}}, \ and\ \bibinfo {author} {\bibfnamefont {Xiao-Ming}\ \bibnamefont
  {{Lu}}},\ }\bibfield  {title} {\enquote {\bibinfo {title} {{Semiclassical
  Theory of Superresolution for Two Incoherent Optical Point Sources}},}\
  }\href@noop {} {\bibfield  {journal} {\bibinfo  {journal} {ArXiv e-prints}\ }
  (\bibinfo {year} {2016})},\ \Eprint {http://arxiv.org/abs/1602.04655}
  {arXiv:1602.04655 [quant-ph]} \BibitemShut {NoStop}%
\bibitem [{\citenamefont {Nair}\ and\ \citenamefont {Tsang}(2016)}]{sliver}%
  \BibitemOpen
  \bibfield  {author} {\bibinfo {author} {\bibfnamefont {Ranjith}\ \bibnamefont
  {Nair}}\ and\ \bibinfo {author} {\bibfnamefont {Mankei}\ \bibnamefont
  {Tsang}},\ }\bibfield  {title} {\enquote {\bibinfo {title} {Interferometric
  superlocalization of two incoherent optical point sources},}\ }\href
  {\doibase 10.1364/OE.24.003684} {\bibfield  {journal} {\bibinfo  {journal}
  {Opt. Express}\ }\textbf {\bibinfo {volume} {24}},\ \bibinfo {pages}
  {3684--3701} (\bibinfo {year} {2016})}\BibitemShut {NoStop}%
\bibitem [{\citenamefont {{Nair}}\ and\ \citenamefont
  {{Tsang}}(2016)}]{nair_tsang16}%
  \BibitemOpen
  \bibfield  {author} {\bibinfo {author} {\bibfnamefont {Ranjith}\ \bibnamefont
  {{Nair}}}\ and\ \bibinfo {author} {\bibfnamefont {Mankei}\ \bibnamefont
  {{Tsang}}},\ }\bibfield  {title} {\enquote {\bibinfo {title} {{Far-field
  Super-resolution of Thermal Electromagnetic Sources at the Quantum Limit}},}\
  }\href@noop {} {\bibfield  {journal} {\bibinfo  {journal} {ArXiv e-prints}\ }
  (\bibinfo {year} {2016})},\ \Eprint {http://arxiv.org/abs/1604.00937}
  {arXiv:1604.00937 [quant-ph]} \BibitemShut {NoStop}%
\bibitem [{\citenamefont {{Lupo}}\ and\ \citenamefont
  {{Pirandola}}(2016)}]{lupo}%
  \BibitemOpen
  \bibfield  {author} {\bibinfo {author} {\bibfnamefont {C.}~\bibnamefont
  {{Lupo}}}\ and\ \bibinfo {author} {\bibfnamefont {S.}~\bibnamefont
  {{Pirandola}}},\ }\bibfield  {title} {\enquote {\bibinfo {title} {{Ultimate
  precision bound of quantum and sub-wavelength imaging}},}\ }\href@noop {}
  {\bibfield  {journal} {\bibinfo  {journal} {ArXiv e-prints}\ } (\bibinfo
  {year} {2016})},\ \Eprint {http://arxiv.org/abs/1604.07367} {arXiv:1604.07367
  [quant-ph]} \BibitemShut {NoStop}%
\bibitem [{\citenamefont {Tang}\ \emph {et~al.}(2016)\citenamefont {Tang},
  \citenamefont {{Durak}},\ and\ \citenamefont {{Ling}}}]{tang16}%
  \BibitemOpen
  \bibfield  {author} {\bibinfo {author} {\bibfnamefont {Zong~Sheng}\
  \bibnamefont {Tang}}, \bibinfo {author} {\bibfnamefont {Kadir}\ \bibnamefont
  {{Durak}}}, \ and\ \bibinfo {author} {\bibfnamefont {Alexander}\ \bibnamefont
  {{Ling}}},\ }\bibfield  {title} {\enquote {\bibinfo {title} {{Fault-tolerant
  and finite-error localization for point emitters within the diffraction
  limit}},}\ }\href@noop {} {\bibfield  {journal} {\bibinfo  {journal} {ArXiv
  e-prints}\ } (\bibinfo {year} {2016})},\ \Eprint
  {http://arxiv.org/abs/1605.07297} {arXiv:1605.07297 [physics.optics]}
  \BibitemShut {NoStop}%
\bibitem [{\citenamefont {Yang}\ \emph {et~al.}(2016)\citenamefont {Yang},
  \citenamefont {Taschilina}, \citenamefont {Moiseev}, \citenamefont {Simon},\
  and\ \citenamefont {Lvovsky}}]{lvovsky16}%
  \BibitemOpen
  \bibfield  {author} {\bibinfo {author} {\bibfnamefont {Fan}\ \bibnamefont
  {Yang}}, \bibinfo {author} {\bibfnamefont {Arina}\ \bibnamefont
  {Taschilina}}, \bibinfo {author} {\bibfnamefont {Eugene~S.}\ \bibnamefont
  {Moiseev}}, \bibinfo {author} {\bibfnamefont {Christoph}\ \bibnamefont
  {Simon}}, \ and\ \bibinfo {author} {\bibfnamefont {Alexander~I.}\
  \bibnamefont {Lvovsky}},\ }\bibfield  {title} {\enquote {\bibinfo {title}
  {Far-field linear optical superresolution via heterodyne detection in a
  higher-order local oscillator mode},}\ }\href@noop {} {\bibfield  {journal}
  {\bibinfo  {journal} {ArXiv e-prints}\ } (\bibinfo {year} {2016})},\ \Eprint
  {http://arxiv.org/abs/1606.02662} {arXiv:1606.02662 [physics.optics]}
  \BibitemShut {NoStop}%
\bibitem [{\citenamefont {{Tham}}\ \emph {et~al.}(2016)\citenamefont {{Tham}},
  \citenamefont {{Ferretti}},\ and\ \citenamefont {{Steinberg}}}]{steinberg16}%
  \BibitemOpen
  \bibfield  {author} {\bibinfo {author} {\bibfnamefont {Weng~Kian}\
  \bibnamefont {{Tham}}}, \bibinfo {author} {\bibfnamefont {Hugo}\ \bibnamefont
  {{Ferretti}}}, \ and\ \bibinfo {author} {\bibfnamefont {Aephraim~M.}\
  \bibnamefont {{Steinberg}}},\ }\bibfield  {title} {\enquote {\bibinfo {title}
  {{Beating Rayleigh's Curse by Imaging Using Phase Information}},}\
  }\href@noop {} {\bibfield  {journal} {\bibinfo  {journal} {ArXiv e-prints}\ }
  (\bibinfo {year} {2016})},\ \Eprint {http://arxiv.org/abs/1606.02666}
  {arXiv:1606.02666 [physics.optics]} \BibitemShut {NoStop}%
\bibitem [{\citenamefont {Pa{\'u}r}\ \emph {et~al.}(2016)\citenamefont
  {Pa{\'u}r}, \citenamefont {Stoklasa}, \citenamefont {Hradil}, \citenamefont
  {S{\'a}nchez-Soto},\ and\ \citenamefont {Rehacek}}]{paur16}%
  \BibitemOpen
  \bibfield  {author} {\bibinfo {author} {\bibfnamefont {Martin}\ \bibnamefont
  {Pa{\'u}r}}, \bibinfo {author} {\bibfnamefont {Bohumil}\ \bibnamefont
  {Stoklasa}}, \bibinfo {author} {\bibfnamefont {Zdenek}\ \bibnamefont
  {Hradil}}, \bibinfo {author} {\bibfnamefont {Luis~L.}\ \bibnamefont
  {S{\'a}nchez-Soto}}, \ and\ \bibinfo {author} {\bibfnamefont {Jaroslav}\
  \bibnamefont {Rehacek}},\ }\bibfield  {title} {\enquote {\bibinfo {title}
  {Achieving quantum-limited optical resolution},}\ }\href@noop {} {\bibfield
  {journal} {\bibinfo  {journal} {ArXiv e-prints}\ } (\bibinfo {year}
  {2016})},\ \Eprint {http://arxiv.org/abs/1606.08332} {arXiv:1606.08332
  [quant-ph]} \BibitemShut {NoStop}%
\bibitem [{\citenamefont {Helstrom}(1970{\natexlab{a}})}]{helstrom70}%
  \BibitemOpen
  \bibfield  {author} {\bibinfo {author} {\bibfnamefont {Carl~W.}\ \bibnamefont
  {Helstrom}},\ }\bibfield  {title} {\enquote {\bibinfo {title} {Estimation of
  object parameters by a quantum-limited optical system},}\ }\href {\doibase
  10.1364/JOSA.60.000233} {\bibfield  {journal} {\bibinfo  {journal} {J. Opt.
  Soc. Am.}\ }\textbf {\bibinfo {volume} {60}},\ \bibinfo {pages} {233--239}
  (\bibinfo {year} {1970}{\natexlab{a}})}\BibitemShut {NoStop}%
\bibitem [{\citenamefont {Bobroff}(1986)}]{bobroff}%
  \BibitemOpen
  \bibfield  {author} {\bibinfo {author} {\bibfnamefont {Norman}\ \bibnamefont
  {Bobroff}},\ }\bibfield  {title} {\enquote {\bibinfo {title} {Position
  measurement with a resolution and noise-limited instrument},}\ }\href
  {\doibase 10.1063/1.1138619} {\bibfield  {journal} {\bibinfo  {journal}
  {Review of Scientific Instruments}\ }\textbf {\bibinfo {volume} {57}},\
  \bibinfo {pages} {1152--1157} (\bibinfo {year} {1986})}\BibitemShut {NoStop}%
\bibitem [{\citenamefont {Delaubert}\ \emph {et~al.}(2008)\citenamefont
  {Delaubert}, \citenamefont {Treps}, \citenamefont {Fabre}, \citenamefont
  {Bachor},\ and\ \citenamefont {R\'efr\'egier}}]{delaubert}%
  \BibitemOpen
  \bibfield  {author} {\bibinfo {author} {\bibfnamefont {V.}~\bibnamefont
  {Delaubert}}, \bibinfo {author} {\bibfnamefont {N.}~\bibnamefont {Treps}},
  \bibinfo {author} {\bibfnamefont {C.}~\bibnamefont {Fabre}}, \bibinfo
  {author} {\bibfnamefont {H.~A.}\ \bibnamefont {Bachor}}, \ and\ \bibinfo
  {author} {\bibfnamefont {P.}~\bibnamefont {R\'efr\'egier}},\ }\bibfield
  {title} {\enquote {\bibinfo {title} {Quantum limits in image processing},}\
  }\href {\doibase 10.1209/0295-5075/81/44001} {\bibfield  {journal} {\bibinfo
  {journal} {EPL (Europhysics Letters)}\ }\textbf {\bibinfo {volume} {81}},\
  \bibinfo {pages} {44001} (\bibinfo {year} {2008})}\BibitemShut {NoStop}%
\bibitem [{\citenamefont {Nair}\ and\ \citenamefont {Yen}(2011)}]{nair_yen}%
  \BibitemOpen
  \bibfield  {author} {\bibinfo {author} {\bibfnamefont {Ranjith}\ \bibnamefont
  {Nair}}\ and\ \bibinfo {author} {\bibfnamefont {Brent~J.}\ \bibnamefont
  {Yen}},\ }\bibfield  {title} {\enquote {\bibinfo {title} {Optimal quantum
  states for image sensing in loss},}\ }\href {\doibase
  10.1103/PhysRevLett.107.193602} {\bibfield  {journal} {\bibinfo  {journal}
  {Phys. Rev. Lett.}\ }\textbf {\bibinfo {volume} {107}},\ \bibinfo {pages}
  {193602} (\bibinfo {year} {2011})}\BibitemShut {NoStop}%
\bibitem [{\citenamefont {P\'erez-Delgado}\ \emph {et~al.}(2012)\citenamefont
  {P\'erez-Delgado}, \citenamefont {Pearce},\ and\ \citenamefont
  {Kok}}]{perez12}%
  \BibitemOpen
  \bibfield  {author} {\bibinfo {author} {\bibfnamefont {Carlos~A.}\
  \bibnamefont {P\'erez-Delgado}}, \bibinfo {author} {\bibfnamefont {Mark~E.}\
  \bibnamefont {Pearce}}, \ and\ \bibinfo {author} {\bibfnamefont {Pieter}\
  \bibnamefont {Kok}},\ }\bibfield  {title} {\enquote {\bibinfo {title}
  {Fundamental limits of classical and quantum imaging},}\ }\href {\doibase
  10.1103/PhysRevLett.109.123601} {\bibfield  {journal} {\bibinfo  {journal}
  {Phys. Rev. Lett.}\ }\textbf {\bibinfo {volume} {109}},\ \bibinfo {pages}
  {123601} (\bibinfo {year} {2012})}\BibitemShut {NoStop}%
\bibitem [{\citenamefont {{Fabre}}\ \emph {et~al.}(2000)\citenamefont
  {{Fabre}}, \citenamefont {{Fouet}},\ and\ \citenamefont
  {{Ma{\^i}tre}}}]{fabre}%
  \BibitemOpen
  \bibfield  {author} {\bibinfo {author} {\bibfnamefont {C.}~\bibnamefont
  {{Fabre}}}, \bibinfo {author} {\bibfnamefont {J.~B.}\ \bibnamefont
  {{Fouet}}}, \ and\ \bibinfo {author} {\bibfnamefont {A.}~\bibnamefont
  {{Ma{\^i}tre}}},\ }\bibfield  {title} {\enquote {\bibinfo {title} {{Quantum
  limits in the measurement of very small displacements in optical images}},}\
  }\href {\doibase 10.1364/OL.25.000076} {\bibfield  {journal} {\bibinfo
  {journal} {Optics Letters}\ }\textbf {\bibinfo {volume} {25}},\ \bibinfo
  {pages} {76--78} (\bibinfo {year} {2000})}\BibitemShut {NoStop}%
\bibitem [{\citenamefont {Barnett}\ \emph {et~al.}(2003)\citenamefont
  {Barnett}, \citenamefont {Fabre},\ and\ \citenamefont
  {Ma{\^{i}}tre}}]{barnett}%
  \BibitemOpen
  \bibfield  {author} {\bibinfo {author} {\bibfnamefont {S.~M.}\ \bibnamefont
  {Barnett}}, \bibinfo {author} {\bibfnamefont {C.}~\bibnamefont {Fabre}}, \
  and\ \bibinfo {author} {\bibfnamefont {A.}~\bibnamefont {Ma{\^{i}}tre}},\
  }\bibfield  {title} {\enquote {\bibinfo {title} {Ultimate quantum limits for
  resolution of beam displacements},}\ }\href {\doibase
  10.1140/epjd/e2003-00003-3} {\bibfield  {journal} {\bibinfo  {journal} {The
  European Physical Journal D - Atomic, Molecular, Optical and Plasma Physics}\
  }\textbf {\bibinfo {volume} {22}},\ \bibinfo {pages} {513--519} (\bibinfo
  {year} {2003})}\BibitemShut {NoStop}%
\bibitem [{\citenamefont {Treps}\ \emph {et~al.}(2003)\citenamefont {Treps},
  \citenamefont {Grosse}, \citenamefont {Bowen}, \citenamefont {Fabre},
  \citenamefont {Bachor},\ and\ \citenamefont {Lam}}]{treps}%
  \BibitemOpen
  \bibfield  {author} {\bibinfo {author} {\bibfnamefont {Nicolas}\ \bibnamefont
  {Treps}}, \bibinfo {author} {\bibfnamefont {Nicolai}\ \bibnamefont {Grosse}},
  \bibinfo {author} {\bibfnamefont {Warwick~P.}\ \bibnamefont {Bowen}},
  \bibinfo {author} {\bibfnamefont {Claude}\ \bibnamefont {Fabre}}, \bibinfo
  {author} {\bibfnamefont {Hans-A.}\ \bibnamefont {Bachor}}, \ and\ \bibinfo
  {author} {\bibfnamefont {Ping~Koy}\ \bibnamefont {Lam}},\ }\bibfield  {title}
  {\enquote {\bibinfo {title} {A quantum laser pointer},}\ }\href {\doibase
  10.1126/science.1086489} {\bibfield  {journal} {\bibinfo  {journal}
  {Science}\ }\textbf {\bibinfo {volume} {301}},\ \bibinfo {pages} {940--943}
  (\bibinfo {year} {2003})}\BibitemShut {NoStop}%
\bibitem [{\citenamefont {Taylor}\ \emph {et~al.}(2013)\citenamefont {Taylor},
  \citenamefont {Janousek}, \citenamefont {Daria}, \citenamefont {Knittel},
  \citenamefont {Hage}, \citenamefont {Bachor},\ and\ \citenamefont
  {Bowen}}]{taylor2013}%
  \BibitemOpen
  \bibfield  {author} {\bibinfo {author} {\bibfnamefont {Michael~A.}\
  \bibnamefont {Taylor}}, \bibinfo {author} {\bibfnamefont {Jiri}\ \bibnamefont
  {Janousek}}, \bibinfo {author} {\bibfnamefont {Vincent}\ \bibnamefont
  {Daria}}, \bibinfo {author} {\bibfnamefont {Joachim}\ \bibnamefont
  {Knittel}}, \bibinfo {author} {\bibfnamefont {Boris}\ \bibnamefont {Hage}},
  \bibinfo {author} {\bibfnamefont {Hans-A.}\ \bibnamefont {Bachor}}, \ and\
  \bibinfo {author} {\bibfnamefont {Warwick~P.}\ \bibnamefont {Bowen}},\
  }\bibfield  {title} {\enquote {\bibinfo {title} {Biological measurement
  beyond the quantum limit},}\ }\href {\doibase 10.1038/nphoton.2012.346}
  {\bibfield  {journal} {\bibinfo  {journal} {Nature Photonics}\ }\textbf
  {\bibinfo {volume} {7}},\ \bibinfo {pages} {229--233} (\bibinfo {year}
  {2013})}\BibitemShut {NoStop}%
\bibitem [{\citenamefont {Kolobov}(1999)}]{kolobov}%
  \BibitemOpen
  \bibfield  {author} {\bibinfo {author} {\bibfnamefont {Mikhail~I.}\
  \bibnamefont {Kolobov}},\ }\bibfield  {title} {\enquote {\bibinfo {title}
  {The spatial behavior of nonclassical light},}\ }\href {\doibase
  10.1103/RevModPhys.71.1539} {\bibfield  {journal} {\bibinfo  {journal} {Rev.
  Mod. Phys.}\ }\textbf {\bibinfo {volume} {71}},\ \bibinfo {pages}
  {1539--1589} (\bibinfo {year} {1999})}\BibitemShut {NoStop}%
\bibitem [{\citenamefont {Kolobov}(2007)}]{kolobov07}%
  \BibitemOpen
  \bibinfo {editor} {\bibfnamefont {Mikhail~I.}\ \bibnamefont {Kolobov}},\
  ed.,\ \href@noop {} {\emph {\bibinfo {title} {Quantum Imaging}}}\ (\bibinfo
  {publisher} {Springer},\ \bibinfo {address} {New York},\ \bibinfo {year}
  {2007})\BibitemShut {NoStop}%
\bibitem [{\citenamefont {Helstrom}(1970{\natexlab{b}})}]{helstrom70b}%
  \BibitemOpen
  \bibfield  {author} {\bibinfo {author} {\bibfnamefont {Carl~W.}\ \bibnamefont
  {Helstrom}},\ }\bibfield  {title} {\enquote {\bibinfo {title} {Resolvability
  of objects from the standpoint of statistical parameter estimation},}\ }\href
  {\doibase 10.1364/JOSA.60.000659} {\bibfield  {journal} {\bibinfo  {journal}
  {J. Opt. Soc. Am.}\ }\textbf {\bibinfo {volume} {60}},\ \bibinfo {pages}
  {659--666} (\bibinfo {year} {1970}{\natexlab{b}})}\BibitemShut {NoStop}%
\bibitem [{\citenamefont {Kolobov}\ and\ \citenamefont
  {Fabre}(2000)}]{kolobov_fabre}%
  \BibitemOpen
  \bibfield  {author} {\bibinfo {author} {\bibfnamefont {Mikhail~I.}\
  \bibnamefont {Kolobov}}\ and\ \bibinfo {author} {\bibfnamefont {Claude}\
  \bibnamefont {Fabre}},\ }\bibfield  {title} {\enquote {\bibinfo {title}
  {Quantum limits on optical resolution},}\ }\href {\doibase
  10.1103/PhysRevLett.85.3789} {\bibfield  {journal} {\bibinfo  {journal}
  {Phys. Rev. Lett.}\ }\textbf {\bibinfo {volume} {85}},\ \bibinfo {pages}
  {3789--3792} (\bibinfo {year} {2000})}\BibitemShut {NoStop}%
\bibitem [{\citenamefont {Beskrovnyy}\ and\ \citenamefont
  {Kolobov}(2005)}]{beskrovnyy05}%
  \BibitemOpen
  \bibfield  {author} {\bibinfo {author} {\bibfnamefont {Vladislav~N.}\
  \bibnamefont {Beskrovnyy}}\ and\ \bibinfo {author} {\bibfnamefont
  {Mikhail~I.}\ \bibnamefont {Kolobov}},\ }\bibfield  {title} {\enquote
  {\bibinfo {title} {Quantum limits of super-resolution in reconstruction of
  optical objects},}\ }\href {\doibase 10.1103/PhysRevA.71.043802} {\bibfield
  {journal} {\bibinfo  {journal} {Phys. Rev. A}\ }\textbf {\bibinfo {volume}
  {71}},\ \bibinfo {pages} {043802} (\bibinfo {year} {2005})}\BibitemShut
  {NoStop}%
\bibitem [{\citenamefont {Beskrovny}\ and\ \citenamefont
  {Kolobov}(2008)}]{beskrovny08}%
  \BibitemOpen
  \bibfield  {author} {\bibinfo {author} {\bibfnamefont {Vladislav~N.}\
  \bibnamefont {Beskrovny}}\ and\ \bibinfo {author} {\bibfnamefont
  {Mikhail~I.}\ \bibnamefont {Kolobov}},\ }\bibfield  {title} {\enquote
  {\bibinfo {title} {Quantum-statistical analysis of superresolution for
  optical systems with circular symmetry},}\ }\href {\doibase
  10.1103/PhysRevA.78.043824} {\bibfield  {journal} {\bibinfo  {journal} {Phys.
  Rev. A}\ }\textbf {\bibinfo {volume} {78}},\ \bibinfo {pages} {043824}
  (\bibinfo {year} {2008})}\BibitemShut {NoStop}%
\bibitem [{\citenamefont {Pich\'{e}}\ \emph {et~al.}(2012)\citenamefont
  {Pich\'{e}}, \citenamefont {Leach}, \citenamefont {Johnson}, \citenamefont
  {Salvail}, \citenamefont {Kolobov},\ and\ \citenamefont {Boyd}}]{piche}%
  \BibitemOpen
  \bibfield  {author} {\bibinfo {author} {\bibfnamefont {Kevin}\ \bibnamefont
  {Pich\'{e}}}, \bibinfo {author} {\bibfnamefont {Jonathan}\ \bibnamefont
  {Leach}}, \bibinfo {author} {\bibfnamefont {Allan~S.}\ \bibnamefont
  {Johnson}}, \bibinfo {author} {\bibfnamefont {Jeff~Z.}\ \bibnamefont
  {Salvail}}, \bibinfo {author} {\bibfnamefont {Mikhail~I.}\ \bibnamefont
  {Kolobov}}, \ and\ \bibinfo {author} {\bibfnamefont {Robert~W.}\ \bibnamefont
  {Boyd}},\ }\bibfield  {title} {\enquote {\bibinfo {title} {Experimental
  realization of optical eigenmode super-resolution},}\ }\href {\doibase
  10.1364/OE.20.026424} {\bibfield  {journal} {\bibinfo  {journal} {Opt.
  Express}\ }\textbf {\bibinfo {volume} {20}},\ \bibinfo {pages} {26424--26433}
  (\bibinfo {year} {2012})}\BibitemShut {NoStop}%
\bibitem [{\citenamefont {Horstmeyer}\ \emph {et~al.}(2016)\citenamefont
  {Horstmeyer}, \citenamefont {Heintzmann}, \citenamefont {Popescu},
  \citenamefont {Waller},\ and\ \citenamefont {Yang}}]{horstmeyer}%
  \BibitemOpen
  \bibfield  {author} {\bibinfo {author} {\bibfnamefont {Roarke}\ \bibnamefont
  {Horstmeyer}}, \bibinfo {author} {\bibfnamefont {Rainer}\ \bibnamefont
  {Heintzmann}}, \bibinfo {author} {\bibfnamefont {Gabriel}\ \bibnamefont
  {Popescu}}, \bibinfo {author} {\bibfnamefont {Laura}\ \bibnamefont {Waller}},
  \ and\ \bibinfo {author} {\bibfnamefont {Changhuei}\ \bibnamefont {Yang}},\
  }\bibfield  {title} {\enquote {\bibinfo {title} {Standardizing the resolution
  claims for coherent microscopy},}\ }\href
  {http://dx.doi.org/10.1038/nphoton.2015.279} {\bibfield  {journal} {\bibinfo
  {journal} {Nature Photonics}\ }\textbf {\bibinfo {volume} {10}},\ \bibinfo
  {pages} {68--71} (\bibinfo {year} {2016})}\BibitemShut {NoStop}%
\bibitem [{\citenamefont {Tsang}(2009)}]{centroid}%
  \BibitemOpen
  \bibfield  {author} {\bibinfo {author} {\bibfnamefont {Mankei}\ \bibnamefont
  {Tsang}},\ }\bibfield  {title} {\enquote {\bibinfo {title} {Quantum imaging
  beyond the diffraction limit by optical centroid measurements},}\ }\href
  {\doibase 10.1103/PhysRevLett.102.253601} {\bibfield  {journal} {\bibinfo
  {journal} {Phys. Rev. Lett.}\ }\textbf {\bibinfo {volume} {102}},\ \bibinfo
  {pages} {253601} (\bibinfo {year} {2009})}\BibitemShut {NoStop}%
\bibitem [{\citenamefont {Giovannetti}\ \emph {et~al.}(2009)\citenamefont
  {Giovannetti}, \citenamefont {Lloyd}, \citenamefont {Maccone},\ and\
  \citenamefont {Shapiro}}]{glm_imaging}%
  \BibitemOpen
  \bibfield  {author} {\bibinfo {author} {\bibfnamefont {Vittorio}\
  \bibnamefont {Giovannetti}}, \bibinfo {author} {\bibfnamefont {Seth}\
  \bibnamefont {Lloyd}}, \bibinfo {author} {\bibfnamefont {Lorenzo}\
  \bibnamefont {Maccone}}, \ and\ \bibinfo {author} {\bibfnamefont
  {Jeffrey~H.}\ \bibnamefont {Shapiro}},\ }\bibfield  {title} {\enquote
  {\bibinfo {title} {Sub-{R}ayleigh-diffraction-bound quantum imaging},}\
  }\href {\doibase 10.1103/PhysRevA.79.013827} {\bibfield  {journal} {\bibinfo
  {journal} {Phys. Rev. A}\ }\textbf {\bibinfo {volume} {79}},\ \bibinfo
  {pages} {013827} (\bibinfo {year} {2009})}\BibitemShut {NoStop}%
\bibitem [{\citenamefont {Shin}\ \emph {et~al.}(2011)\citenamefont {Shin},
  \citenamefont {Chan}, \citenamefont {Chang},\ and\ \citenamefont
  {Boyd}}]{shin}%
  \BibitemOpen
  \bibfield  {author} {\bibinfo {author} {\bibfnamefont {Heedeuk}\ \bibnamefont
  {Shin}}, \bibinfo {author} {\bibfnamefont {Kam Wai~Clifford}\ \bibnamefont
  {Chan}}, \bibinfo {author} {\bibfnamefont {Hye~Jeong}\ \bibnamefont {Chang}},
  \ and\ \bibinfo {author} {\bibfnamefont {Robert~W.}\ \bibnamefont {Boyd}},\
  }\bibfield  {title} {\enquote {\bibinfo {title} {Quantum spatial
  superresolution by optical centroid measurements},}\ }\href {\doibase
  10.1103/PhysRevLett.107.083603} {\bibfield  {journal} {\bibinfo  {journal}
  {Phys. Rev. Lett.}\ }\textbf {\bibinfo {volume} {107}},\ \bibinfo {pages}
  {083603} (\bibinfo {year} {2011})}\BibitemShut {NoStop}%
\bibitem [{\citenamefont {Rozema}\ \emph {et~al.}(2014)\citenamefont {Rozema},
  \citenamefont {Bateman}, \citenamefont {Mahler}, \citenamefont {Okamoto},
  \citenamefont {Feizpour}, \citenamefont {Hayat},\ and\ \citenamefont
  {Steinberg}}]{rozema}%
  \BibitemOpen
  \bibfield  {author} {\bibinfo {author} {\bibfnamefont {Lee~A.}\ \bibnamefont
  {Rozema}}, \bibinfo {author} {\bibfnamefont {James~D.}\ \bibnamefont
  {Bateman}}, \bibinfo {author} {\bibfnamefont {Dylan~H.}\ \bibnamefont
  {Mahler}}, \bibinfo {author} {\bibfnamefont {Ryo}\ \bibnamefont {Okamoto}},
  \bibinfo {author} {\bibfnamefont {Amir}\ \bibnamefont {Feizpour}}, \bibinfo
  {author} {\bibfnamefont {Alex}\ \bibnamefont {Hayat}}, \ and\ \bibinfo
  {author} {\bibfnamefont {Aephraim~M.}\ \bibnamefont {Steinberg}},\ }\bibfield
   {title} {\enquote {\bibinfo {title} {Scalable spatial superresolution using
  entangled photons},}\ }\href {\doibase 10.1103/PhysRevLett.112.223602}
  {\bibfield  {journal} {\bibinfo  {journal} {Phys. Rev. Lett.}\ }\textbf
  {\bibinfo {volume} {112}},\ \bibinfo {pages} {223602} (\bibinfo {year}
  {2014})}\BibitemShut {NoStop}%
\bibitem [{\citenamefont {Oppel}\ \emph {et~al.}(2012)\citenamefont {Oppel},
  \citenamefont {B\"uttner}, \citenamefont {Kok},\ and\ \citenamefont {von
  Zanthier}}]{oppel}%
  \BibitemOpen
  \bibfield  {author} {\bibinfo {author} {\bibfnamefont {S.}~\bibnamefont
  {Oppel}}, \bibinfo {author} {\bibfnamefont {T.}~\bibnamefont {B\"uttner}},
  \bibinfo {author} {\bibfnamefont {P.}~\bibnamefont {Kok}}, \ and\ \bibinfo
  {author} {\bibfnamefont {J.}~\bibnamefont {von Zanthier}},\ }\bibfield
  {title} {\enquote {\bibinfo {title} {Superresolving multiphoton interferences
  with independent light sources},}\ }\href {\doibase
  10.1103/PhysRevLett.109.233603} {\bibfield  {journal} {\bibinfo  {journal}
  {Phys. Rev. Lett.}\ }\textbf {\bibinfo {volume} {109}},\ \bibinfo {pages}
  {233603} (\bibinfo {year} {2012})}\BibitemShut {NoStop}%
\bibitem [{\citenamefont {Schwartz}\ \emph {et~al.}(2013)\citenamefont
  {Schwartz}, \citenamefont {Levitt}, \citenamefont {Tenne}, \citenamefont
  {Itzhakov}, \citenamefont {Deutsch},\ and\ \citenamefont
  {Oron}}]{schwartz13}%
  \BibitemOpen
  \bibfield  {author} {\bibinfo {author} {\bibfnamefont {Osip}\ \bibnamefont
  {Schwartz}}, \bibinfo {author} {\bibfnamefont {Jonathan~M.}\ \bibnamefont
  {Levitt}}, \bibinfo {author} {\bibfnamefont {Ron}\ \bibnamefont {Tenne}},
  \bibinfo {author} {\bibfnamefont {Stella}\ \bibnamefont {Itzhakov}}, \bibinfo
  {author} {\bibfnamefont {Zvicka}\ \bibnamefont {Deutsch}}, \ and\ \bibinfo
  {author} {\bibfnamefont {Dan}\ \bibnamefont {Oron}},\ }\bibfield  {title}
  {\enquote {\bibinfo {title} {Superresolution microscopy with quantum
  emitters},}\ }\href {\doibase 10.1021/nl402552m} {\bibfield  {journal}
  {\bibinfo  {journal} {Nano Letters}\ }\textbf {\bibinfo {volume} {13}},\
  \bibinfo {pages} {5832--5836} (\bibinfo {year} {2013})}\BibitemShut {NoStop}%
\bibitem [{\citenamefont {Cui}\ \emph {et~al.}(2013)\citenamefont {Cui},
  \citenamefont {Sun}, \citenamefont {Chen}, \citenamefont {Gong},\ and\
  \citenamefont {Guo}}]{cui13}%
  \BibitemOpen
  \bibfield  {author} {\bibinfo {author} {\bibfnamefont {Jin-Ming}\
  \bibnamefont {Cui}}, \bibinfo {author} {\bibfnamefont {Fang-Wen}\
  \bibnamefont {Sun}}, \bibinfo {author} {\bibfnamefont {Xiang-Dong}\
  \bibnamefont {Chen}}, \bibinfo {author} {\bibfnamefont {Zhao-Jun}\
  \bibnamefont {Gong}}, \ and\ \bibinfo {author} {\bibfnamefont {Guang-Can}\
  \bibnamefont {Guo}},\ }\bibfield  {title} {\enquote {\bibinfo {title}
  {Quantum statistical imaging of particles without restriction of the
  diffraction limit},}\ }\href {\doibase 10.1103/PhysRevLett.110.153901}
  {\bibfield  {journal} {\bibinfo  {journal} {Phys. Rev. Lett.}\ }\textbf
  {\bibinfo {volume} {110}},\ \bibinfo {pages} {153901} (\bibinfo {year}
  {2013})}\BibitemShut {NoStop}%
\bibitem [{\citenamefont {Gatto~Monticone}\ \emph {et~al.}(2014)\citenamefont
  {Gatto~Monticone}, \citenamefont {Katamadze}, \citenamefont {Traina},
  \citenamefont {Moreva}, \citenamefont {Forneris}, \citenamefont
  {Ruo-Berchera}, \citenamefont {Olivero}, \citenamefont {Degiovanni},
  \citenamefont {Brida},\ and\ \citenamefont {Genovese}}]{monticone}%
  \BibitemOpen
  \bibfield  {author} {\bibinfo {author} {\bibfnamefont {D.}~\bibnamefont
  {Gatto~Monticone}}, \bibinfo {author} {\bibfnamefont {K.}~\bibnamefont
  {Katamadze}}, \bibinfo {author} {\bibfnamefont {P.}~\bibnamefont {Traina}},
  \bibinfo {author} {\bibfnamefont {E.}~\bibnamefont {Moreva}}, \bibinfo
  {author} {\bibfnamefont {J.}~\bibnamefont {Forneris}}, \bibinfo {author}
  {\bibfnamefont {I.}~\bibnamefont {Ruo-Berchera}}, \bibinfo {author}
  {\bibfnamefont {P.}~\bibnamefont {Olivero}}, \bibinfo {author} {\bibfnamefont
  {I.~P.}\ \bibnamefont {Degiovanni}}, \bibinfo {author} {\bibfnamefont
  {G.}~\bibnamefont {Brida}}, \ and\ \bibinfo {author} {\bibfnamefont
  {M.}~\bibnamefont {Genovese}},\ }\bibfield  {title} {\enquote {\bibinfo
  {title} {Beating the abbe diffraction limit in confocal microscopy via
  nonclassical photon statistics},}\ }\href {\doibase
  10.1103/PhysRevLett.113.143602} {\bibfield  {journal} {\bibinfo  {journal}
  {Phys. Rev. Lett.}\ }\textbf {\bibinfo {volume} {113}},\ \bibinfo {pages}
  {143602} (\bibinfo {year} {2014})}\BibitemShut {NoStop}%
\bibitem [{\citenamefont {Boto}\ \emph {et~al.}(2000)\citenamefont {Boto},
  \citenamefont {Kok}, \citenamefont {Abrams}, \citenamefont {Braunstein},
  \citenamefont {Williams},\ and\ \citenamefont {Dowling}}]{boto}%
  \BibitemOpen
  \bibfield  {author} {\bibinfo {author} {\bibfnamefont {Agedi~N.}\
  \bibnamefont {Boto}}, \bibinfo {author} {\bibfnamefont {Pieter}\ \bibnamefont
  {Kok}}, \bibinfo {author} {\bibfnamefont {Daniel~S.}\ \bibnamefont {Abrams}},
  \bibinfo {author} {\bibfnamefont {Samuel~L.}\ \bibnamefont {Braunstein}},
  \bibinfo {author} {\bibfnamefont {Colin~P.}\ \bibnamefont {Williams}}, \ and\
  \bibinfo {author} {\bibfnamefont {Jonathan~P.}\ \bibnamefont {Dowling}},\
  }\bibfield  {title} {\enquote {\bibinfo {title} {Quantum interferometric
  optical lithography: Exploiting entanglement to beat the diffraction
  limit},}\ }\href {\doibase 10.1103/PhysRevLett.85.2733} {\bibfield  {journal}
  {\bibinfo  {journal} {Phys. Rev. Lett.}\ }\textbf {\bibinfo {volume} {85}},\
  \bibinfo {pages} {2733--2736} (\bibinfo {year} {2000})}\BibitemShut {NoStop}%
\bibitem [{\citenamefont {Shih}(2007)}]{shih07}%
  \BibitemOpen
  \bibfield  {author} {\bibinfo {author} {\bibfnamefont {Yanhua}\ \bibnamefont
  {Shih}},\ }\bibfield  {title} {\enquote {\bibinfo {title} {Quantum
  imaging},}\ }\href {\doibase 10.1109/JSTQE.2007.902724} {\bibfield  {journal}
  {\bibinfo  {journal} {IEEE Journal of Selected Topics in Quantum
  Electronics}\ }\textbf {\bibinfo {volume} {13}},\ \bibinfo {pages}
  {1016--1030} (\bibinfo {year} {2007})}\BibitemShut {NoStop}%
\bibitem [{\citenamefont {Boyd}\ and\ \citenamefont
  {Dowling}(2012)}]{boyd2012}%
  \BibitemOpen
  \bibfield  {author} {\bibinfo {author} {\bibfnamefont {Robert~W.}\
  \bibnamefont {Boyd}}\ and\ \bibinfo {author} {\bibfnamefont {Jonathan~P.}\
  \bibnamefont {Dowling}},\ }\bibfield  {title} {\enquote {\bibinfo {title}
  {Quantum lithography: status of the field},}\ }\href {\doibase
  10.1007/s11128-011-0253-y} {\bibfield  {journal} {\bibinfo  {journal}
  {Quantum Information Processing}\ }\textbf {\bibinfo {volume} {11}},\
  \bibinfo {pages} {891--901} (\bibinfo {year} {2012})}\BibitemShut {NoStop}%
\bibitem [{\citenamefont {Hemmer}\ and\ \citenamefont
  {Zapata}(2012)}]{hemmer12}%
  \BibitemOpen
  \bibfield  {author} {\bibinfo {author} {\bibfnamefont {Philip~R.}\
  \bibnamefont {Hemmer}}\ and\ \bibinfo {author} {\bibfnamefont {Todd}\
  \bibnamefont {Zapata}},\ }\bibfield  {title} {\enquote {\bibinfo {title} {The
  universal scaling laws that determine the achievable resolution in different
  schemes for super-resolution imaging},}\ }\href
  {http://stacks.iop.org/2040-8986/14/i=8/a=083002} {\bibfield  {journal}
  {\bibinfo  {journal} {Journal of Optics}\ }\textbf {\bibinfo {volume} {14}},\
  \bibinfo {pages} {083002} (\bibinfo {year} {2012})}\BibitemShut {NoStop}%
\bibitem [{\citenamefont {Pittman}\ \emph {et~al.}(1995)\citenamefont
  {Pittman}, \citenamefont {Shih}, \citenamefont {Strekalov},\ and\
  \citenamefont {Sergienko}}]{pittman}%
  \BibitemOpen
  \bibfield  {author} {\bibinfo {author} {\bibfnamefont {T.~B.}\ \bibnamefont
  {Pittman}}, \bibinfo {author} {\bibfnamefont {Y.~H.}\ \bibnamefont {Shih}},
  \bibinfo {author} {\bibfnamefont {D.~V.}\ \bibnamefont {Strekalov}}, \ and\
  \bibinfo {author} {\bibfnamefont {A.~V.}\ \bibnamefont {Sergienko}},\
  }\bibfield  {title} {\enquote {\bibinfo {title} {Optical imaging by means of
  two-photon quantum entanglement},}\ }\href {\doibase
  10.1103/PhysRevA.52.R3429} {\bibfield  {journal} {\bibinfo  {journal} {Phys.
  Rev. A}\ }\textbf {\bibinfo {volume} {52}},\ \bibinfo {pages} {R3429--R3432}
  (\bibinfo {year} {1995})}\BibitemShut {NoStop}%
\bibitem [{\citenamefont {Gatti}\ \emph {et~al.}(2004)\citenamefont {Gatti},
  \citenamefont {Brambilla}, \citenamefont {Bache},\ and\ \citenamefont
  {Lugiato}}]{gatti}%
  \BibitemOpen
  \bibfield  {author} {\bibinfo {author} {\bibfnamefont {A.}~\bibnamefont
  {Gatti}}, \bibinfo {author} {\bibfnamefont {E.}~\bibnamefont {Brambilla}},
  \bibinfo {author} {\bibfnamefont {M.}~\bibnamefont {Bache}}, \ and\ \bibinfo
  {author} {\bibfnamefont {L.~A.}\ \bibnamefont {Lugiato}},\ }\bibfield
  {title} {\enquote {\bibinfo {title} {Ghost imaging with thermal light:
  Comparing entanglement and classical correlation},}\ }\href {\doibase
  10.1103/PhysRevLett.93.093602} {\bibfield  {journal} {\bibinfo  {journal}
  {Phys. Rev. Lett.}\ }\textbf {\bibinfo {volume} {93}},\ \bibinfo {pages}
  {093602} (\bibinfo {year} {2004})}\BibitemShut {NoStop}%
\bibitem [{\citenamefont {Erkmen}\ and\ \citenamefont
  {Shapiro}(2010)}]{erkmen10}%
  \BibitemOpen
  \bibfield  {author} {\bibinfo {author} {\bibfnamefont {Baris~I.}\
  \bibnamefont {Erkmen}}\ and\ \bibinfo {author} {\bibfnamefont {Jeffrey~H.}\
  \bibnamefont {Shapiro}},\ }\bibfield  {title} {\enquote {\bibinfo {title}
  {Ghost imaging: from quantum to classical to computational},}\ }\href
  {\doibase 10.1364/AOP.2.000405} {\bibfield  {journal} {\bibinfo  {journal}
  {Adv. Opt. Photon.}\ }\textbf {\bibinfo {volume} {2}},\ \bibinfo {pages}
  {405--450} (\bibinfo {year} {2010})}\BibitemShut {NoStop}%
\bibitem [{\citenamefont {Braunstein}\ and\ \citenamefont
  {Caves}(1994)}]{braunstein}%
  \BibitemOpen
  \bibfield  {author} {\bibinfo {author} {\bibfnamefont {Samuel~L.}\
  \bibnamefont {Braunstein}}\ and\ \bibinfo {author} {\bibfnamefont
  {Carlton~M.}\ \bibnamefont {Caves}},\ }\bibfield  {title} {\enquote {\bibinfo
  {title} {Statistical distance and the geometry of quantum states},}\ }\href
  {\doibase 10.1103/PhysRevLett.72.3439} {\bibfield  {journal} {\bibinfo
  {journal} {Phys. Rev. Lett.}\ }\textbf {\bibinfo {volume} {72}},\ \bibinfo
  {pages} {3439--3443} (\bibinfo {year} {1994})}\BibitemShut {NoStop}%
\bibitem [{\citenamefont {Paris}(2009)}]{paris}%
  \BibitemOpen
  \bibfield  {author} {\bibinfo {author} {\bibfnamefont {Matteo G.~A.}\
  \bibnamefont {Paris}},\ }\bibfield  {title} {\enquote {\bibinfo {title}
  {Quantum estimation for quantum technology},}\ }\href {\doibase
  10.1142/S0219749909004839} {\bibfield  {journal} {\bibinfo  {journal}
  {International Journal of Quantum Information}\ }\textbf {\bibinfo {volume}
  {7}},\ \bibinfo {pages} {125--137} (\bibinfo {year} {2009})}\BibitemShut
  {NoStop}%
\bibitem [{\citenamefont {Wiseman}\ and\ \citenamefont
  {Milburn}(2010)}]{wiseman_milburn}%
  \BibitemOpen
  \bibfield  {author} {\bibinfo {author} {\bibfnamefont {Howard~M.}\
  \bibnamefont {Wiseman}}\ and\ \bibinfo {author} {\bibfnamefont {Gerard~J.}\
  \bibnamefont {Milburn}},\ }\href {\doibase 10.1017/CBO9780511813948} {\emph
  {\bibinfo {title} {Quantum Measurement and Control}}}\ (\bibinfo  {publisher}
  {Cambridge University Press},\ \bibinfo {address} {Cambridge},\ \bibinfo
  {year} {2010})\BibitemShut {NoStop}%
\bibitem [{\citenamefont {Cox}\ and\ \citenamefont {Hinkley}(1979)}]{cox79}%
  \BibitemOpen
  \bibfield  {author} {\bibinfo {author} {\bibfnamefont {D.~R.}\ \bibnamefont
  {Cox}}\ and\ \bibinfo {author} {\bibfnamefont {D.~V.}\ \bibnamefont
  {Hinkley}},\ }\href@noop {} {\emph {\bibinfo {title} {Theoretical
  Statistics}}}\ (\bibinfo  {publisher} {Chapman \& Hall/CRC},\ \bibinfo
  {address} {Boca Raton},\ \bibinfo {year} {1979})\BibitemShut {NoStop}%
\bibitem [{\citenamefont {van~der Vaart}(1997)}]{vaart97}%
  \BibitemOpen
  \bibfield  {author} {\bibinfo {author} {\bibfnamefont {A.~W.}\ \bibnamefont
  {van~der Vaart}},\ }\href {\doibase 10.1007/978-1-4612-1880-7_27} {\emph
  {\bibinfo {title} {Festschrift for {Lucien Le Cam}: Research Papers in
  Probability and Statistics}}},\ edited by\ \bibinfo {editor} {\bibfnamefont
  {David}\ \bibnamefont {Pollard}}, \bibinfo {editor} {\bibfnamefont {Erik}\
  \bibnamefont {Torgersen}}, \ and\ \bibinfo {editor} {\bibfnamefont
  {Grace~L.}\ \bibnamefont {Yang}}\ (\bibinfo  {publisher} {Springer},\
  \bibinfo {address} {New York},\ \bibinfo {year} {1997})\ Chap.\ \bibinfo
  {chapter} {27. Superefficiency}, pp.\ \bibinfo {pages} {397--410}\BibitemShut
  {NoStop}%
\bibitem [{\citenamefont {{Tsang}}(2016)}]{tsang16}%
  \BibitemOpen
  \bibfield  {author} {\bibinfo {author} {\bibfnamefont {Mankei}\ \bibnamefont
  {{Tsang}}},\ }\bibfield  {title} {\enquote {\bibinfo {title} {{Conservative
  error measures for classical and quantum metrology}},}\ }\href@noop {}
  {\bibfield  {journal} {\bibinfo  {journal} {ArXiv e-prints}\ } (\bibinfo
  {year} {2016})},\ \Eprint {http://arxiv.org/abs/1605.03799} {arXiv:1605.03799
  [quant-ph]} \BibitemShut {NoStop}%
\end{thebibliography}%

\end{document}